\documentclass[prc,twocolumn,preprintnumbers,showpaNSs,superscriptaddress]{revtex4}
\usepackage[T1]{fontenc}
\usepackage{amsmath,aas_macros}
\usepackage{natbib}
\usepackage{soul}
\usepackage{xcolor}
\usepackage{lipsum}
\usepackage{float}
\usepackage{graphicx} 
\usepackage{multirow,makecell}
\usepackage{placeins}
\usepackage{bm,amsfonts,hhline,amssymb,microtype,float,latexsym,epsf,mathtools,times,makecell,array}
\usepackage{tabularx}
\usepackage{epstopdf}
\usepackage{gensymb}
\usepackage{pifont}
\usepackage{hyperref}
\usepackage{lineno}
\hypersetup{colorlinks=true, linkcolor=magenta, filecolor=magenta, urlcolor=magenta, citecolor=blue}
\newcommand{\Qsat}{Q_{\text {sat }}} 
 
\newcommand{\bea}{\begin{eqnarray}}
\newcommand{\eea}{\end{eqnarray}}

\definecolor{ForestGreen}{HTML}{228B22}

\usepackage[normalem]{ulem}  

\definecolor{darkred}{rgb}{0.7,.1,.2}

\begin{document}
\title{Rapidly rotating hot nuclear and 
hypernuclear compact stars: \\Integral parameters and universal relations }

\author{Stefanos Tsiopelas}
\email{stefanos.tsiopelas2@uwr.edu.pl}

\author{Armen Sedrakian}
\email{armen.sedrakian@uwr.edu.pl}
\affiliation{Institute of Theoretical Physics, 
University of Wroc\l{}aw, 50-204 Wroc\l{}aw, Poland}
\affiliation{Frankfurt Institute for Advanced Studies, D-60438 Frankfurt am Main, Germany}

\author{Micaela Oertel}
\email{micaela.oertel@astro.unistra.fr}
\affiliation{Observatoire astronomique 
de Strasbourg, CNRS, Universit\'e de Strasbourg,  67000 Strasbourg, France}
\affiliation{LUX, Observatoire de Paris, CNRS, Universit\'e PSL, Sorbonnes Universités, 5 place Jules Janssen, 92195 Meudon, France}

\date{October 17, 2025}

\begin{abstract}    
In this work, we investigate hot, isentropic compact stars in the limiting cases of static and maximally rotating configurations, focusing on how variations in the symmetry energy of the equation of state derived from covariant density functional theory affect stellar properties. We consider both nucleonic and hyperonic matter with systematically varied symmetry energy slopes, fixed entropies per baryon  $s / k_B=1$ and 3, and electron fractions $Y_e=0.1$ and $Y_e=0.4$, representative of conditions in binary neutron star mergers and proto-neutron stars. We compute and analyze mass--radius and moment-of-inertia-mass relations, as well as the dependence of the Keplerian (mass-shedding) frequency on mass, angular momentum, and the ratio of kinetic to gravitational energy. Furthermore, we show that several universal relations between global properties remain valid across both nucleonic and hyperonic equations of state with varying symmetry energy, both in the static and Keplerian limit, and for various combinations of the fixed entropy and electron fraction.  
\end{abstract}
\maketitle

\section{Introduction}
\label{sec:intro}

Covariant density functionals (CDFs) provide a fast and reliable framework for incorporating physical constraints derived from both nuclear many-body systems and astrophysical observations. Based on baryon-meson Lagrangians, CDFs offer access to a wide range of microscopic quantities, such as self-energies (classified by their Lorentz structure), matter composition, chemical potentials, and effective masses, for reviews see~\cite{Lattimer:2015nhk,Oertel_RMP_2017,Sedrakian2023}. A key advantage of this approach is its flexibility: model parameters can be readily adjusted to accommodate new data or evolving constraints on the equation of state (EoS) and related properties across micro- and macrophysical regimes. On the astrophysical side, key observational constraints include the masses of heavy pulsars~\cite{NANOGrav:2019,Fonseca:2021}, the simultaneous measurements of masses and radii of both canonical and massive neutron stars (hereafter NS)~\cite{NICER:2021a,NICER:2021b}, as well as tidal deformabilities of medium-mass compact stars in the GW170817 event~\cite{Abbott_2017}. These observations place stringent limits on the NS EoS, in the regime where matter is  cold (i.e., Fermi energies are much larger than the temperature) and in $\beta$-equilibrium. On the nuclear physics side, we note in particular that measurements from parity-violating electron scattering experiments  on $^{208} \mathrm{Pb}$ and $^{40}\mathrm{Ca}$ provide  constraints on the symmetry energy, inferred through analyses of neutron skin thicknesses~\cite{PREX-II:2021,Reinhard:2021,CREX:2022,Lattimer:2023}. Additional constraints come from the analysis of charged-pion spectra at high transverse momenta~\cite{SpiRIT:2021} and from heavy-ion collisions at ultrarelativistic energies studied at the Large Hadron Collider~\cite{Giacalone:2023}.

The CDF  approach was applied to finite temperatures/entropies for both nucleonic~\cite{Shen1998,Hempel:2009mc,Hempel:2011mk,Steiner:2012rk,FURUSAWA2017} and hyperonic~\cite{Ishizuka:2008gr,Colucci:2013pya,Marques_2017,Fortin_PASA_2018,Raduta_MNRAS_2020,Stone:2019blq,Sedrakian:2022kgj,Sedrakian:2021qjw,Kochankovski:2022rid,Tsiopelas2024EPJA,Barman:2025qrm} matter in a number of recent works, thus allowing applications
in the context of supernovas, binary neutron star (BNS) mergers, and proto-neutron stars (PNS). In Ref.~\cite{Tsiopelas2024EPJA} (hereafter TSO24), we constructed 3D tables of finite-temperature EoS of nuclear and hypernuclear matter in the range of densities, temperatures, and electron fractions that are needed for numerical simulations of hot compact objects and made them available on the \href{https://compose.obspm.fr}{{\sc CompOSE}} database~\cite{Typel:2013rza,Dexheimer2022Parti,ComposeCoreTeam:2022ddl}. These EoS allow us to vary the underlying parametrization, designed in Ref.~\cite{Li2023ApJ}, in such a manner that three values of the slope of the symmetry energy $L_{\rm sym}=30, 50,$ and 70 MeV at a fixed value of skewness $\Qsat$ allow systematic studies of the effect of the symmetry energy on the observables. The chosen value of $\Qsat = 400$ MeV corresponds to the minimal value imposed by the requirement that the hyperonic stars achieve the two-solar mass lower bound on the maximum mass of a compact star. The low-density part of the EoS containing inhomogeneous matter was taken from Ref.~\cite{Hempel:2009mc}. In TSO24, the global properties of static and rapidly rotating compact objects--both nucleonic and hyperonic--were computed at zero temperature, in order to validate the EoS against multimessenger astrophysical constraints.

The objective of this work is to extend the TSO24 study and to carry out a similar analysis at finite temperatures and entropies, in particular in the case of rapidly (rigidly) rotating NSs. We aim to extract information on various global properties of such stars at the Keplerian frequency, which represents the maximum rotation rate before mass shedding sets in. Our analysis also offers a systematic study of how modifications of the symmetry energy affect the global properties of compact stars and their universal relations. This provides a more coherent perspective compared to earlier works that relied on heterogeneous collections of EoS and pursued objectives different from those considered here. 
Of course, the present work is of interest not only as an extension of the cold EoS studies, but it also provides insights that are directly relevant to the dynamical evolution of hot and rapidly rotating compact objects formed in core-collapse supernovae, binary neutron-star mergers, and proto-neutron stars. In these environments, finite-temperature effects, changes in the symmetry energy, and rapid rotation jointly determine key observables such as the maximum mass, radius, moment of inertia, and gravitational-wave signatures. Therefore, our investigation provides a more unified and physically consistent framework for confronting upcoming multimessenger data.

This regime is important in the interpretation and extraction of the maximum Tolman-Oppenheimer-Volkoff (TOV) mass of a compact star from the event GW170817. Indeed, several authors have argued~\cite{Margalit2017,Rezzolla2018ApJ,Shibata2019PhRvD,
   Khadkikar:2021} that the gravitational-wave event GW170817 (and similar BNS mergers) can be used to place an upper bound on the maximum mass, $M_{\mathrm{TOV}}^{\star}$, of a nonrotating (static) cold and $\beta$-equilibrated compact star. This inference relies on the scenario in which the merger leads to the formation of a hypermassive neutron star (HMNS) -- an object temporarily supported against collapse by rapid differential rotation. Such a remnant is expected to eventually lose angular momentum, for example, through viscous effects related to the magnetic field, gravitational wave, and neutrino emission, and collapse into a black hole once rotation becomes uniform and centrifugal support becomes insufficient. The observed absence of long-lived postmerger electromagnetic signals supports this collapse scenario. 

 This inference of $M_{\mathrm{TOV}}^{max}$ relies on the existence of quasiuniversal relations between nonrotating stars and stars rotating at the Kepler frequency. Within this study, we will examine more generally (quasi) universal relations among the global properties of static and rigidly maximally fast rotating, i.e., Keplerian, compact stars at finite temperature. In this context, universality implies that these relations are largely independent of the underlying EoS.  Although universal relations among various global properties of NSs have been known for decades, they gained significant attention after the discovery that the moment of inertia ($I$), tidal deformability ($\Lambda$), and quadrupole moment ($Q$) obey such nearly EoS-independent relations, see Refs.~\cite{Yagi:2013a,Yagi:2013b,Yagi:2017}. Since then, (quasi) universal relations have been extensively investigated in a variety of contexts: (a) rapidly rotating NSs~\cite{Doneva:2013,Pappas:2014,Chakrabarti:2014,Cipolletta:2015,Breu:2016,Riahi:2019,Riahi2019,Koliogiannis:2020,Konstantinou:2022,Largani:2022,Li2023PhRvC}; (b) NSs containing hyperons or other heavy baryons~\cite{Li2023PhRvC}; (c) hot PNS and postmerger remnants~\cite{Martinon:2014,Marques:2017,Lenka:2018,Stone:2019,Raduta_MNRAS_2020,Khadkikar:2021}; (d) gravitational wave data analysis~\cite{Yagi:2013a,Agathos2015,Chatziioannou2018,Carson2019,Suleiman:2024ztn}.  Given their significance, it is worthwhile to explore these relations for the class of EoS introduced in TSO24, which feature systematic variations in the symmetry energy slope and allow us to vary the composition from purely nucleonic and hyperon-admixed matter.  We also note that the universality across the variations of $L_{\rm sym}$ for cold static and Keplerian configurations was already demonstrated in Ref.~\cite{Yeasin2025}, where $Q_{\rm sat}$ value has been varied as well. 

As mentioned earlier, an interesting application of universal relations arises when one infers from the postmerger dynamics of NSs a constraint on the maximum mass of a static configuration as discussed in Refs.~\cite{Rezzolla2018ApJ,Shibata2019PhRvD,Khadkikar:2021}. The main step here is to employ universal relations that link the maximum mass of uniformly rotating (Keplerian) configurations to that of their static counterparts.  By estimating the total gravitational mass of the remnant, one can then infer a conservative upper limit on the maximum mass of a nonrotating NS. In that context, Ref.~\cite{Khadkikar:2021} showed that universality does not hold between the hot supramassive Keplerian and the cold TOV configurations,  because of uncertainties due to the unknown entropy of the hot remnant.  We will focus on this relation among other things in the following. 

This paper is organized as follows.  In Sec.~\ref{sec:Kepler} we discuss the properties of compact stars with varying $L_{\rm sym}$, focusing on the scaling of various global quantities of the stars on these input parameters at finite fixed entropies and electron fractions as well as in $\beta$-equilibrium. In Sec.~\ref{sec:Universal} we present evidence for the universality of the relations between the global parameters of the stars with respect to the input EoS with variations of $L_{\rm sym}$ in the case of nuclear and hypernuclear stars. Our conclusions are collected in Sec.~\ref{sec:Conclusions}. The appendix contains Tables of the fitting parameters for all universal relations studied in this work, along with the uncertainty quantification.
\begin{figure*}
\includegraphics[width=0.6\linewidth]{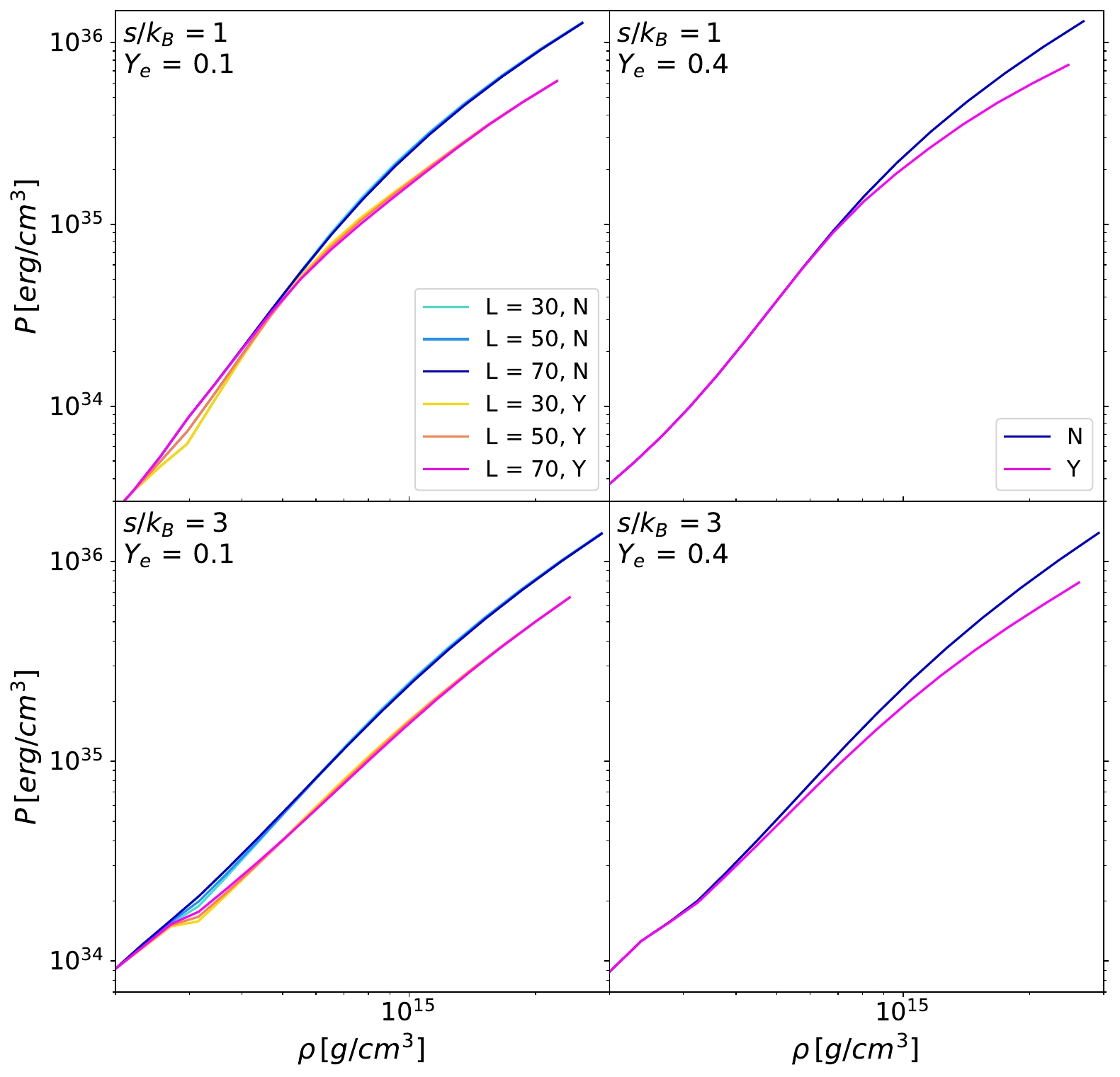}
\caption{Pressure versus energy for the nucleonic ($N$) and hyperonic $(Y)$ EoS models for different $L_{\rm sym}$, and for fixed values of $s/k_B = 1$ and $s/k_B = 3$ and electron fractions
$Y_{e}=0.1$ and 0.4.}
    \label{fig:EoSplots_fixed_sYe}
\end{figure*}    
\section{Static and rapidly rotating stars}
\label{sec:Kepler}

We begin by studying both static and maximally rotating (Keplerian) compact star configurations, which are isentropic and have a prescribed electron fraction. Such stars are representative of conditions found in the remnants of BNS mergers and  of core-collapse supernovae, i.e., PNS. 

The isentropic approximation is characteristic of the initial phases of these transients, each of which has its characteristic distinct physical scales. In PNS, convective effects lead--shortly after core bounce, on a timescale of less than 10s--the stellar interior to become almost isentropic and with a flat $Y_e$-profile, see e.g. Refs.~\cite{Roberts:2011yw,Pascal:2022qeg}.
In these early times, the entropy per baryon is in the range $s/k_B\sim 1-4$. As neutrinos escape, they carry away both energy and lepton number, leading to a decrease in the entropy per baryon over time. While the inner core may remain nearly isentropic for a while, entropy gradients begin to develop, particularly in the crust and outer layers. After about 30 seconds, the PNS begins its transition into a cold NS, and the average values drop below $s/k_B\sim 0.5$. The temperature is likewise nonuniform, with the outer layers and surface significantly cooler than the core. The situation after a BNS merger has some similarities, but also key differences to the PNS case, see Refs.~\cite{Radice2016,Sekiguchi2016,Radice2020,Sumiyoshi2021,Espino2024,Foucart2023}. The remnant, being initially an HMNS, quickly evolves into a supramassive NS (excluding the case of a prompt black hole formation). Immediately after the merger, at the center of the star, the neutrinos are trapped and maintain thermal and chemical equilibrium with matter. In this stage, the matter in these regions can be approximately modeled as isentropic, with entropy values typically in the range $s/k_B\sim 1-3$ in the core and higher in outer layers.   Unlike core-collapse supernovae, the strong differential rotation, intense shock heating, and violent oscillations lead to faster evolution: the timescale for neutrino diffusion and deleptonization is of the order of 10–100 milliseconds in the densest regions, depending on local density and temperature. As in the PNS, the remnant cools and expands, eventually becoming transparent to neutrinos after $\sim$ tens of milliseconds.  After this brief transition, the remnant develops significant entropy and temperature gradients, but the core remains hot and may still be approximately isentropic. Thus, while the isentropic approximation is reasonable for modeling the early, dense, and opaque phases of postmerger evolution, it gradually breaks down as neutrino decoupling progresses and temperature and composition gradients emerge. This discussion shows that considering isentropic configurations at constant $Y_e$ cannot describe PNS or BNS merger remnants realistically, but it represents a reasonable approximation allowing for studying physical effects, e.g., the effect of the nuclear symmetry energy, in a simple manner.

\subsection{Equations of state}

In this work, we use the EoS set for nuclear and hypernuclear matter given previously in Ref.~\cite{Tsiopelas2024EPJA}. There, a value of $\Qsat \ge 400$MeV was shown to be required to support hypernuclear compact stars with masses exceeding the well-established lower bound of $\sim 2 M_{\odot}$, as inferred from the mass measurement of PSR J0740+6620~\cite{NANOGrav:2019,Fonseca:2021}. Accordingly, throughout this work we fix ${\rm min}\,\Qsat = 400$~MeV to its minimum value required to produce a two-solar mass compact star with hyperons. As in Ref.~\cite{Tsiopelas2024EPJA}, we consider three representative values of the symmetry energy slope parameter: $L_{\rm sym} = 30$, 50, and 70~MeV.   Table \ref{tab:1} lists the parameters of  the CDFs used in this work, evaluated at saturation density.
The corresponding EoS are shown in Fig.~\ref{fig:EoSplots_fixed_sYe}.
\begin{table}[t]
\centering
\setlength{\tabcolsep}{4pt}
\caption{Parameters of the DDME2 functional and of the CDFs used in this work at saturation density $\rho_{\rm sat}$: energy per particle $E_{\rm sat}$, compressibility $K_{\rm sat}$, skewness $Q_{\rm sat}$, symmetry energy $J_{\rm sym}$ and its slope $L_{\rm sym}$, and effective Dirac mass $M_D^*$ in units of average vacuum mass of a nucleon $m_N$.}
\begin{tabular}{ccccccc}
\hline\hline
$\rho_{\rm sat}$ & $E_{\rm sat}$ & $K_{\rm sat}$ & $Q_{\rm sat}$ & 
$J_{\rm sym}$ & $L_{\rm sym}$ & $M_D^*/m_N$ \\
fm$^{-3}$ & MeV & MeV & MeV & 
MeV & MeV & \\
\hline
0.152 & $-16.14$ & 251.15 & 479.22 & 32.31 & 51.27 & 0.57 \\
0.152 & $-16.14$ & 251.15 & 400.00 & 30.08 & 30.00 & 0.57 \\
0.152 & $-16.14$ & 251.15 & 400.00 & 32.19 & 50.00 & 0.57 \\
0.152 & $-16.14$ & 251.15 & 400.00 & 33.99 & 70.00 & 0.57 \\
\hline\hline
\end{tabular}
\label{tab:1}
\end{table}

It is seen that high-density asymptotics of the EoS, which is controlled by $\Qsat$, is identical separately for nucleonic and hyperonic models. The hyperonic EoS is much softer due to the onset of additional degrees of freedom. The variations of the symmetry energy through $L_{\rm sym}$ are seen to affect the behavior of the EoS close to saturation density; the EoS models with smaller $L_{\rm sym}$ are softer. Furthermore, while the low-density asymptotics of all EoS are the same, the onset of hyperons manifests itself in softening of the EoS, which is more pronounced the smaller $L_{\rm sym}$ is. These variations are best seen in the case of $Y_e= 0.1$, whereas in the case $Y_e = 0.4$ (almost isospin symmetric matter), the nucleonic and hyperonic  EoS are separately indistinguishable in the figure. Finally, we note that for a larger entropy value, the softening of the EoS with the onset of hyperons is more pronounced. The reason is that larger entropies favor a larger population of hyperons, see e.g.~\cite{Oertel:2012qd,Oertel:2016xsn}.


\begin{figure*}[tbh]
\includegraphics[width=0.45\linewidth]{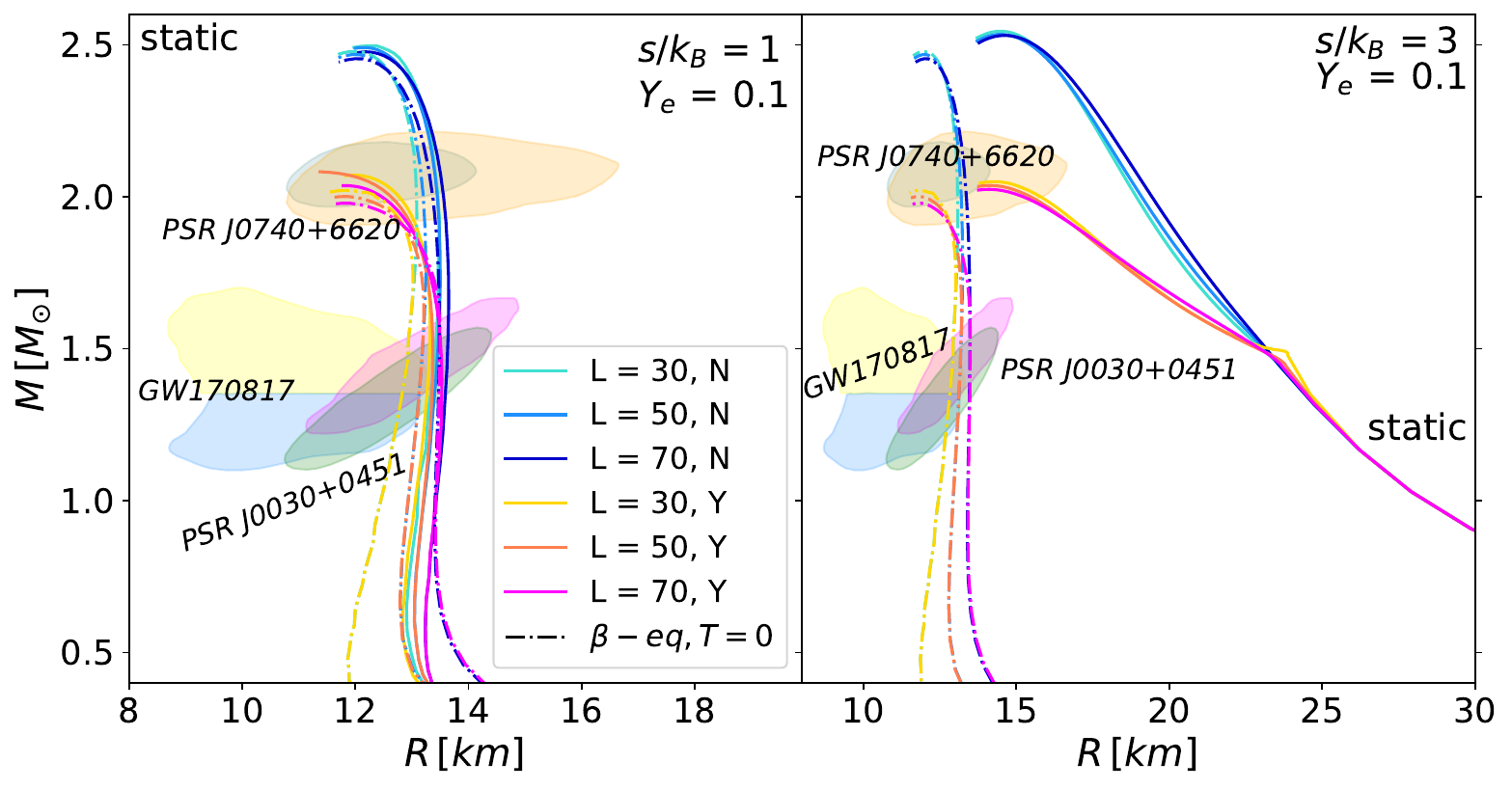}
\includegraphics[width=0.45\linewidth]{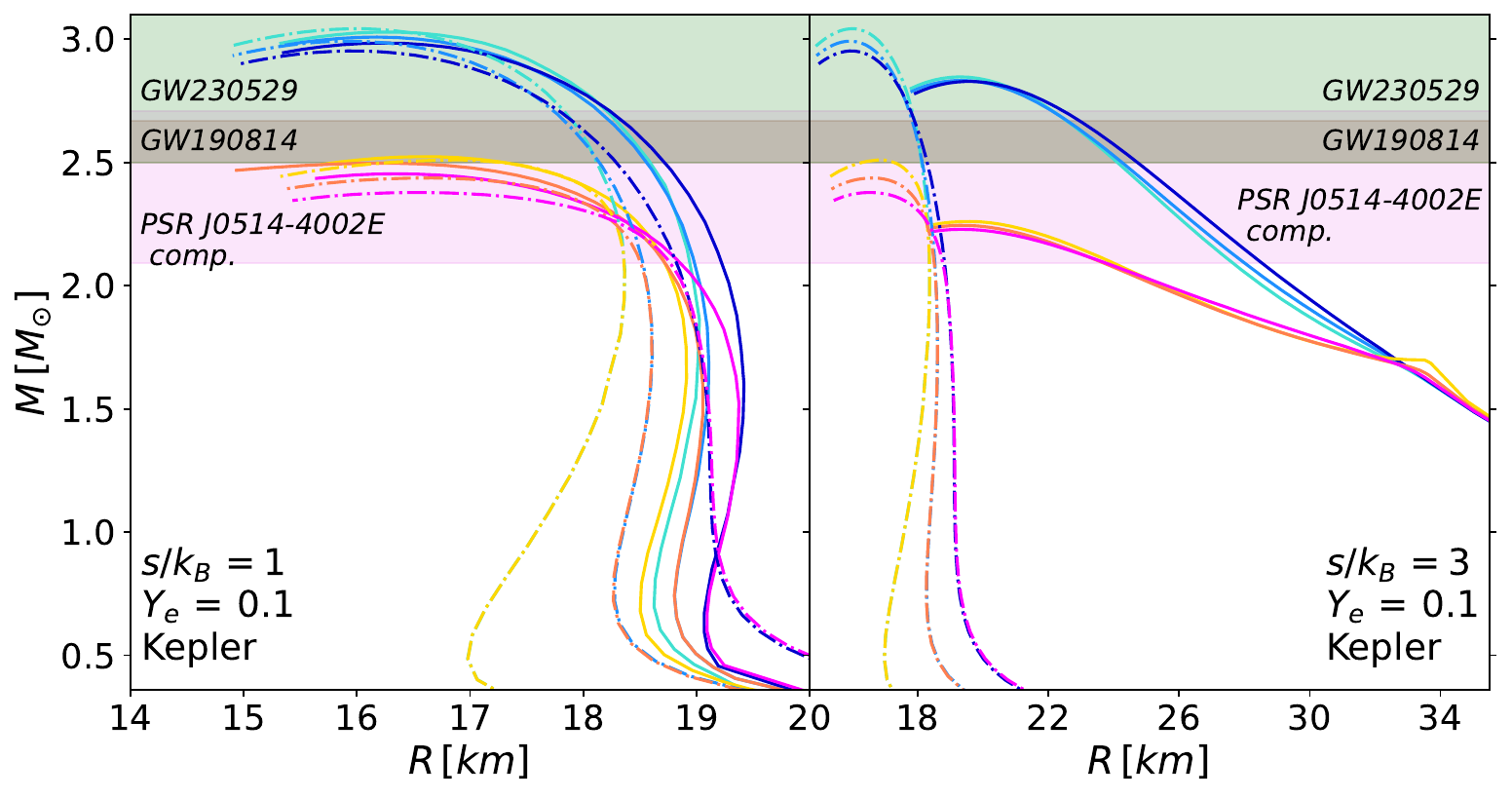}
\includegraphics[width=0.45\linewidth]{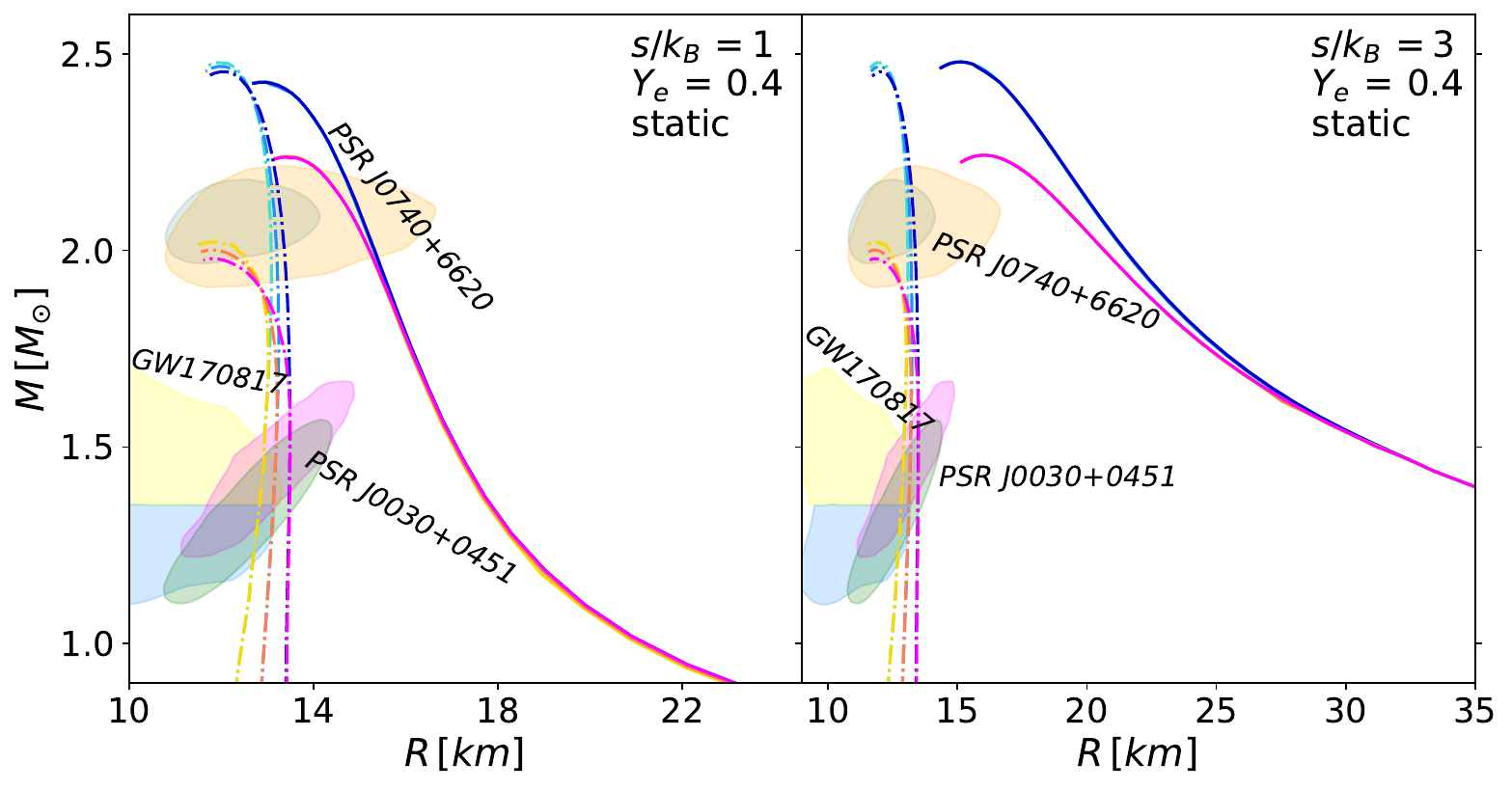}
\includegraphics[width=0.45\linewidth]{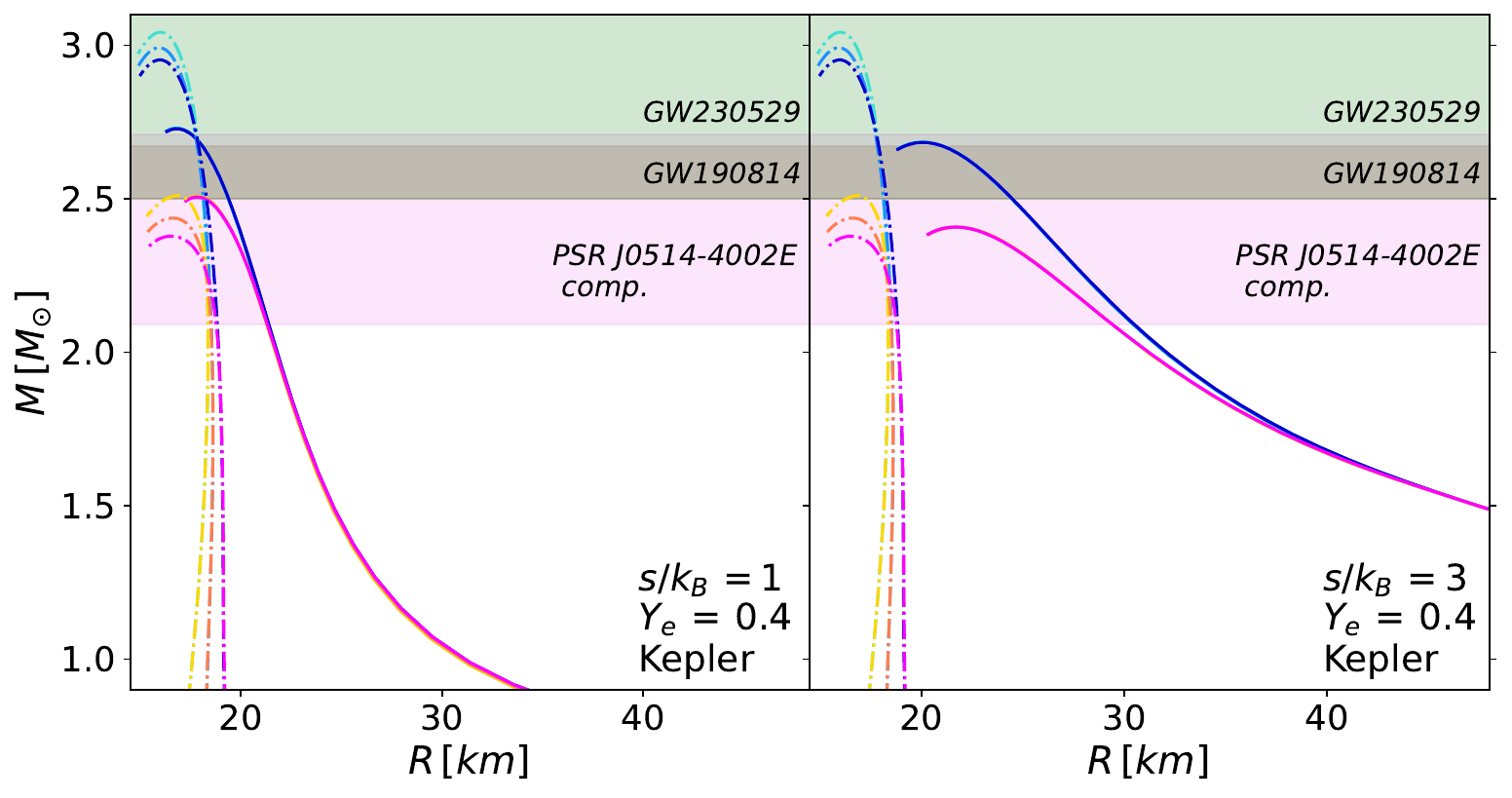}
\caption{Mass-radius relations for static and Keplerian stellar sequences defined through combinations of fixed values of entropy per baryon
  and electron fraction with nucleonic ($N$) and hyperonic ($Y$)compositions, and for various values of $L_{\rm sym}$.}
    \label{fig:MR_fixed_sYe}
\end{figure*}
\begin{figure*}[!]
\includegraphics[width=0.45\linewidth]{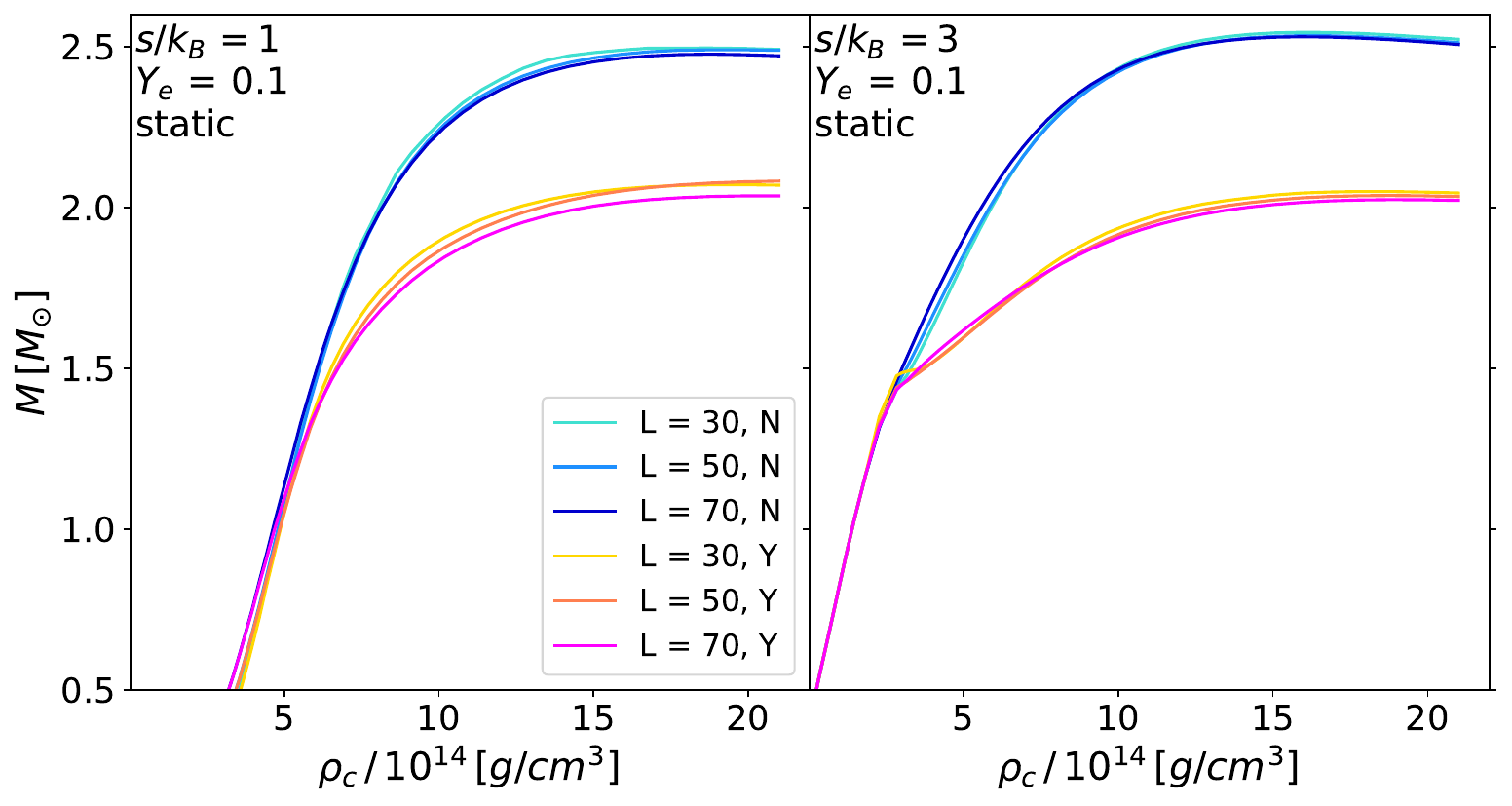}
\includegraphics[width=0.45\linewidth]{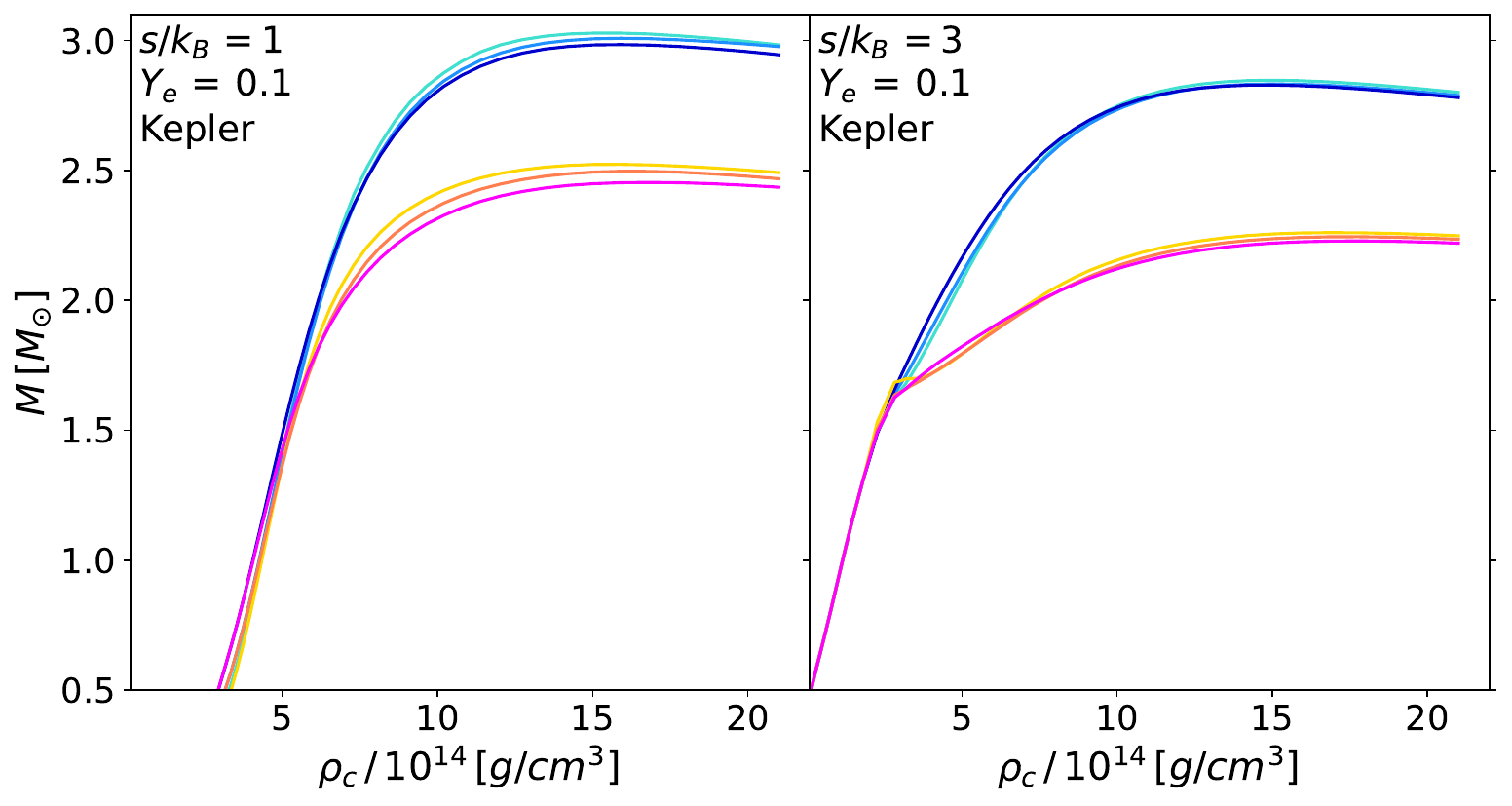}

\includegraphics[width=0.45\linewidth]{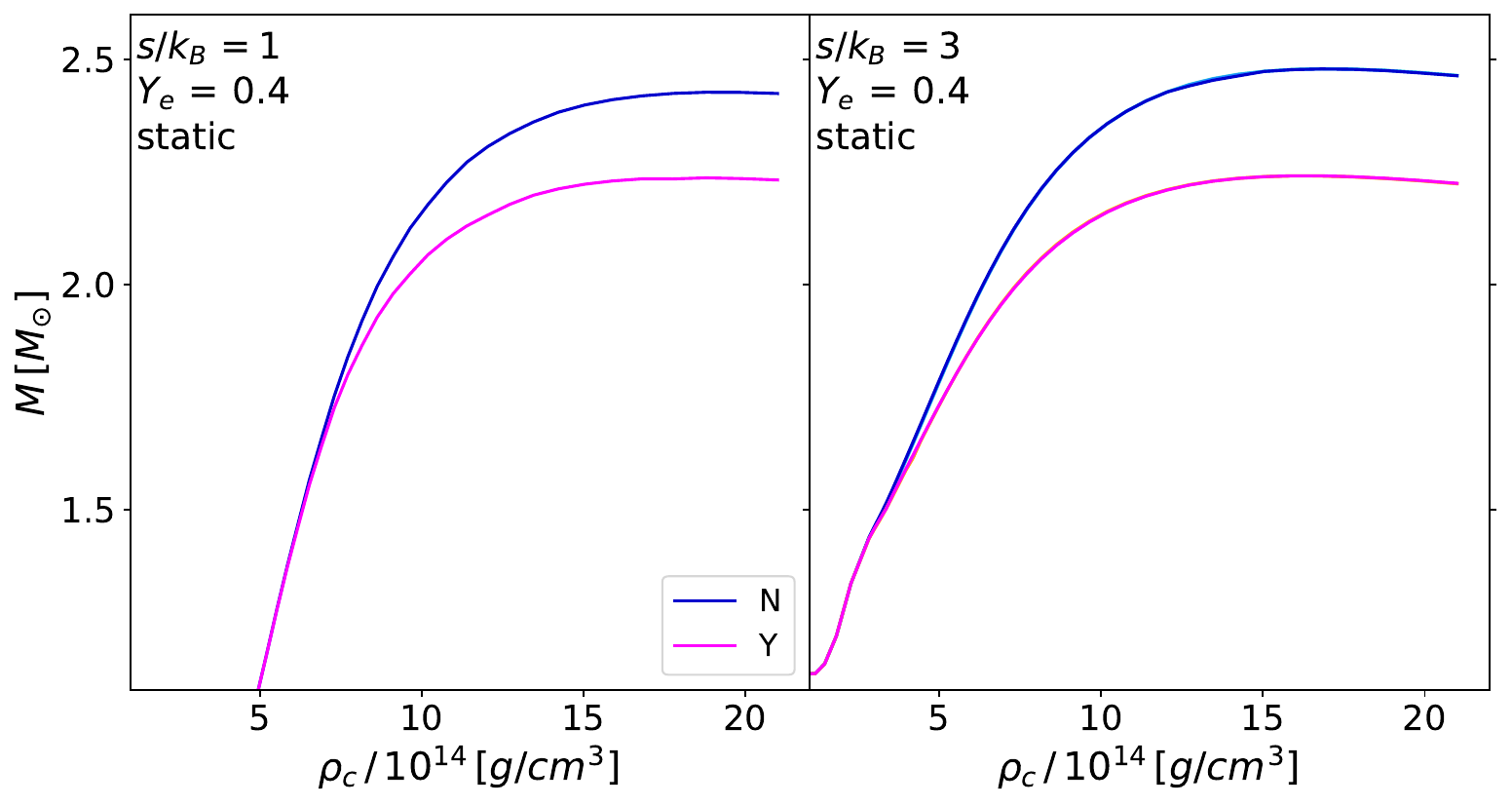}
\includegraphics[width=0.45\linewidth]{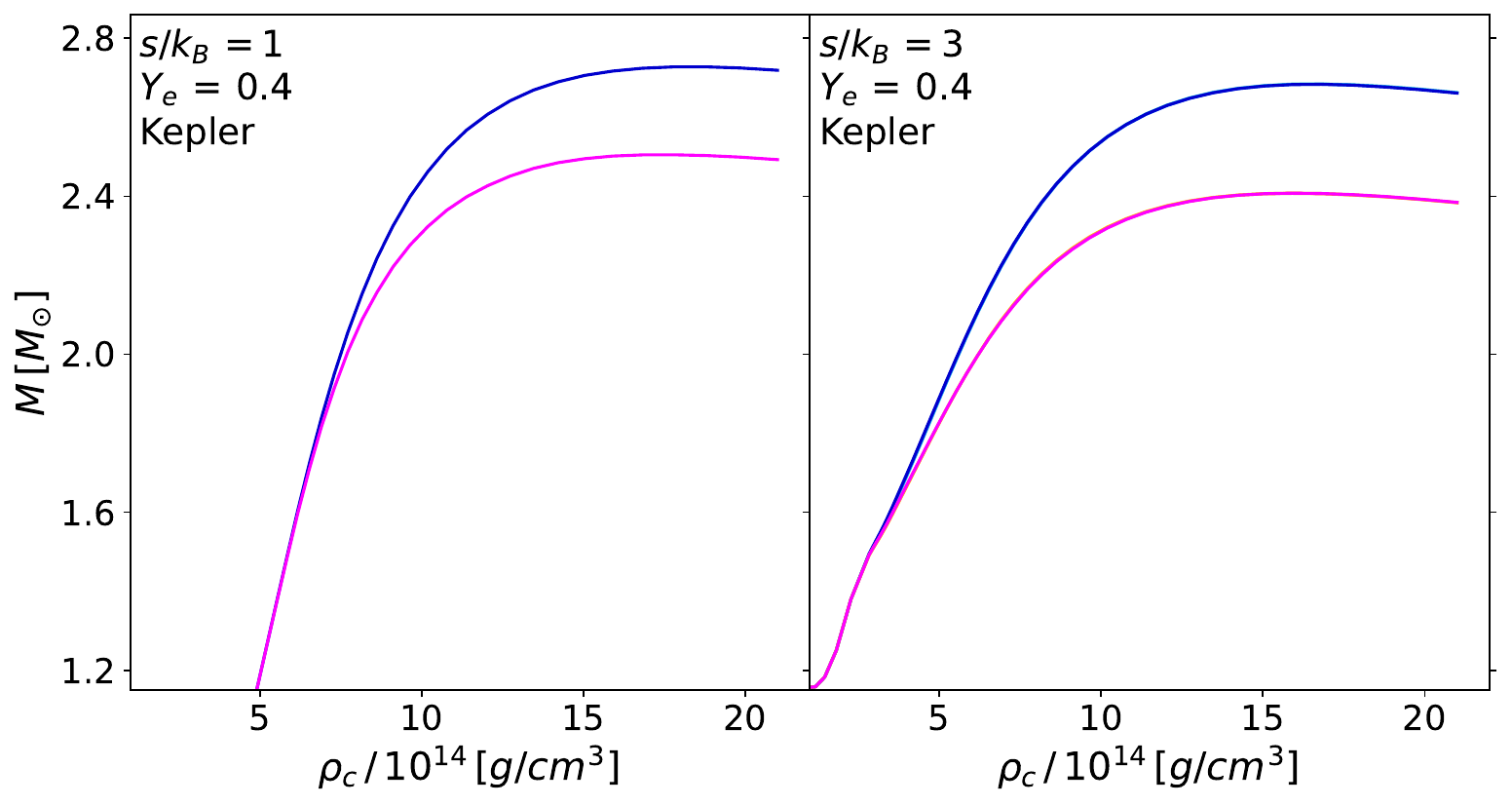}
\caption{Dependence of mass on central energy density for static and Keplerian sequences for fixed combinations of entropy per baryon
  and electron fractions and different values of $L_{\rm sym}$ in the case of nucleonic ($N$) and hyperonic ($Y$) EoS.
    }
    \label{fig:Mrhoa_fixed_sYe}
\end{figure*}

\subsection{Results for global NS properties}

To compute numerical models of hot, rapidly rotating stars, we employed the RNS code~(https://github.com/cgca/rns).  This code provides tools for constructing equilibrium configurations of relativistic rotating bodies~\cite{Stergioulas2003}, by solving the coupled Einstein field equations and equations of hydrostatic equilibrium. The solutions are obtained under the assumptions of stationarity and axisymmetry.  We input the isentropic EoS at fixed electron fraction into the RNS code to compute both static and Keplerian (maximally rotating) configurations. As discussed, e.g., in Refs.~\cite{Goussard1997,Marques:2017,Khadkikar:2021}, the formalism to solve for equilibrium configurations of the RNS code can be applied at finite temperature, too, under the above assumptions of constant entropy per baryon and electron fraction.

\begin{figure*}[!]
\includegraphics[width=0.7\linewidth]{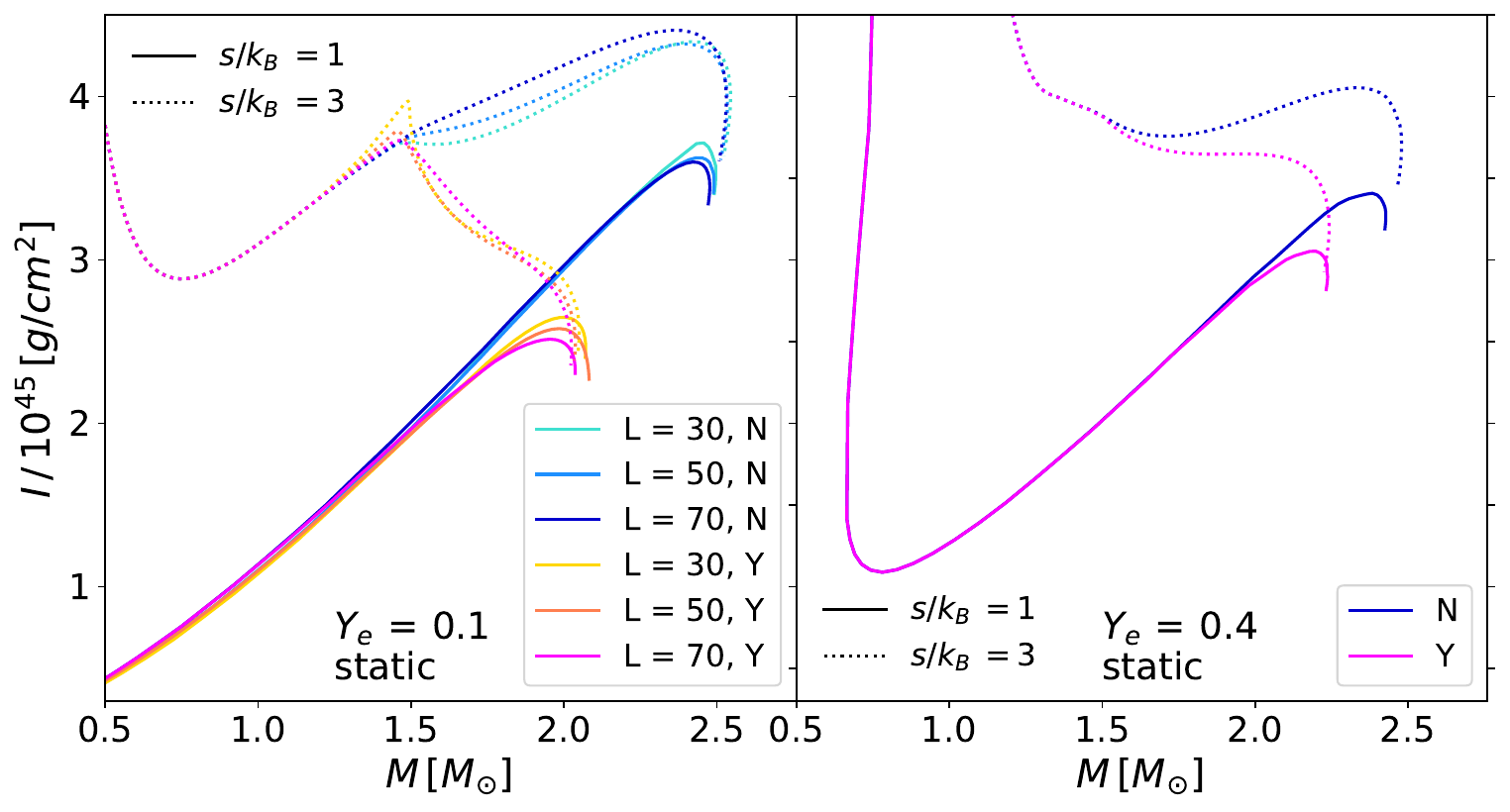}
\includegraphics[width=0.7\linewidth]{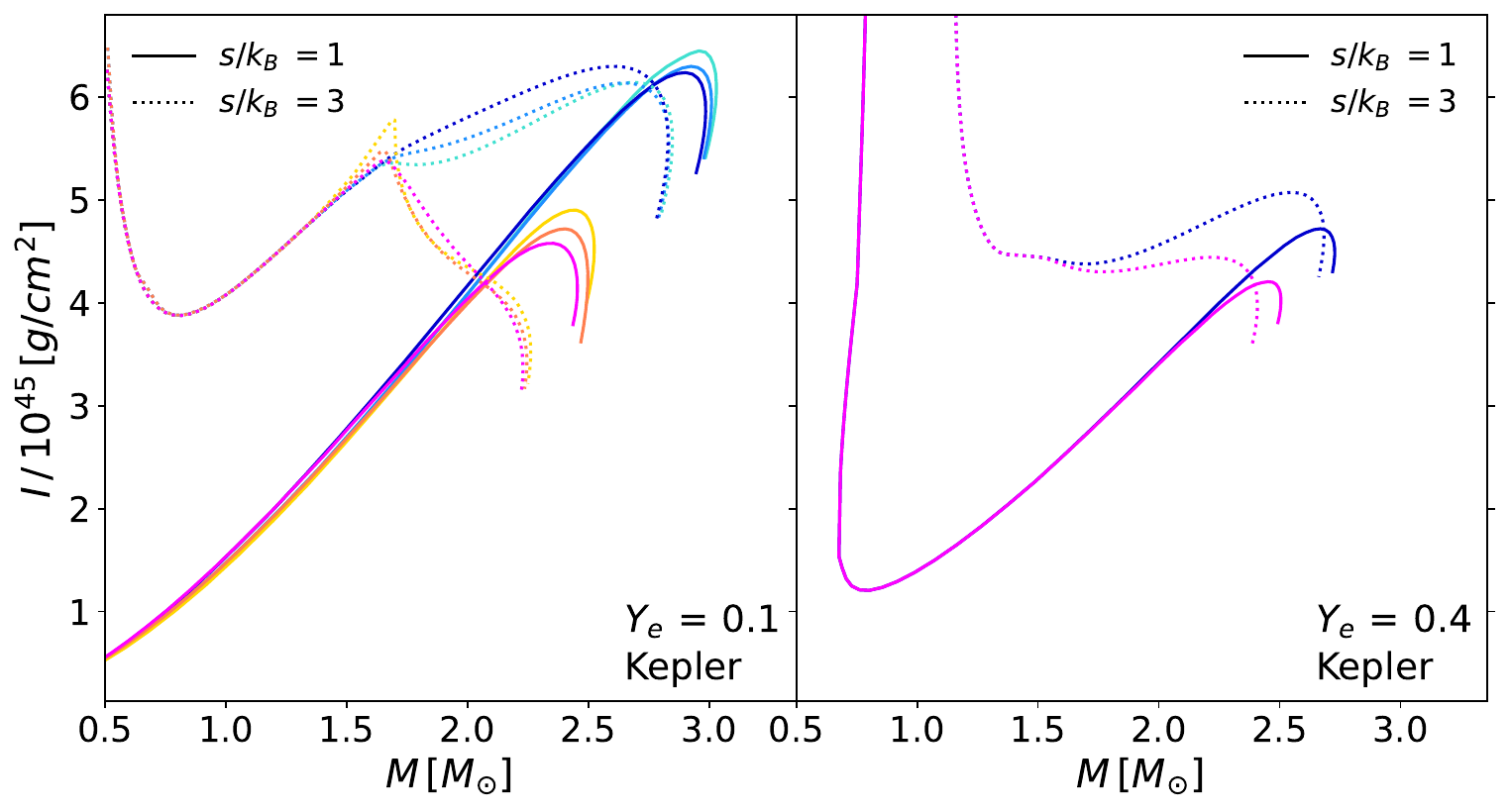}
    \caption{
      Dependence of moment of inertia on the gravitational mass for static and Keplerian sequences for fixed combinations of entropies per baryon
      and electron fractions and different values of $L_{\rm sym}$ in the case of nucleonic ($N$) and hyperonic ($Y$) EoS.
  }
    \label{fig:IM_fixed_sYe}
\end{figure*} 

We would like to caution that the determination of the radius of the PNS and BNS merger remnant is ambiguous because of the presence of an extended, low-density atmosphere. In our calculations, the outer edge of these transients is defined by a density cutoff corresponding to the lowest value available in the EOS table. However, the corresponding radius is somewhat arbitrary and does not reflect a physical surface in the usual sense. Consequently, our quoted radii should be interpreted with care, as they depend sensitively on the EOS's lowest density entry
$ 4.735\times 10^{-15}$ fm$^{-3}$  and on the thermodynamic and neutrino transport assumptions near the surface. Note that in Refs.~\cite{Raduta_MNRAS_2020,Khadkikar:2021} it was shown that the quantitative difference in radius values still remains small (below a few percent) for the considered range in entropy per baryon. In addition, we mainly show radii for model comparison purposes, and as discussed above, the low-density behavior of all EoS models is the same. We thus expect the model comparison to be meaningful.

  Our results for the mass-radius diagram are presented in Fig.~\ref{fig:MR_fixed_sYe}.  The left two columns show the results in the static limit for $s/k_B =1$ and $3$, respectively. The nucleonic stars are labeled as $N$, while the hypernuclear stars are labeled as $Y$. The following two columns show the same for Keplerian configurations, whereby $R$ refers to the equatorial circumferential radius.  The two upper panels correspond to electron fraction $Y_e=0.1$, representing typical conditions in BNS mergers, and the two lower panels--to fixed electron fraction $Y_e=0.4$, mimicking the environment in PNS; within each panel, we vary $L_{\rm sym}$. It is seen that the aforementioned softening of the EoS due to the onset of hyperons leads to lower maximum masses for both nonrotating and rapidly rotating stars. In the case of static, $\beta$-equilibrated stars, also shown in Fig.~\ref{fig:Mrhoa_fixed_sYe} the $M$-$R$ curves are constrained by the data: it is seen that these are consistent with NICER observations~\cite{NICER:2021a,NICER:2021b} for both canonical NSs ($M \sim 1.4 M_{\odot}$) and massive ones ($M \sim 2 M_{\odot}$), as discussed previously in Ref.~\cite{Tsiopelas2024EPJA}.  Furthermore, rapidly rotating, $\beta$-equilibrated stars with masses approaching three solar masses in the nucleonic ($N$) sequence are viable candidates for the so-called  ``mass-gap" compact objects, as suggested by gravitational wave events GW190814 and GW230529.

As expected, the nucleonic sequences differ from the hyperonic ones only once the central density exceeds the threshold for hyperon onset. The  $M$-$R$ relations for isentropic stars show similar behavior to those of cold stars; the following features are notable:
\begin{enumerate}
\item 
As a general trend, stars with lower values of ($L_{\rm sym}$) exhibit smaller radii and higher maximum masses, both in the cold, $\beta$-equilibrated case and for isentropic configurations. For a fixed $L_{\rm sym}$, isentropic stars reach somewhat larger maximum masses than their cold $\beta$-equilibrated counterparts, as seen for the cases $s/k_B = 1,3$ shown in the figure. It is worth noting, however, that when moving away from strict $\beta$-equilibrium toward small but finite values of $s/k_B$, the hyperon population increases in hypernuclear stars. This may initially reduce the maximum mass~\cite{Raduta_MNRAS_2020} before thermal pressure becomes dominant and drives the mass upward, as illustrated in the figure. Since our focus here is on the effects of varying $L_{\rm sym}$, we do not pursue this issue further and refer the reader to Ref.~\cite{Raduta_MNRAS_2020}. Finally, we note that for the values of $s/k_B$ shown, the increase in maximum mass is less pronounced for hyperonic EoS models than for nucleonic ones.

\item Isentropic stars generally have larger radii than their cold counterparts. The radius increases with entropy, primarily due to the expansion of the stellar envelope.

\item In isentropic stars, increasing the electron fraction from  $Y_e=0.1$ to $Y_e=0.4$
leads to a pronounced expansion of the envelope and a corresponding increase in the stellar radius. The effect of  $L_{\rm sym}$ becomes negligible in this regime, as the matter composition is close to the isospin symmetric limit. The maximum masses increase for the hyperonic stars, whereas they decrease slightly for the nucleonic ones. The reason is that a higher $Y_e$ disfavors hyperons, see also Ref.~\cite{Raduta_MNRAS_2020}.

\end{enumerate}
\begin{figure*}[hbt!]
\includegraphics[width=0.7\linewidth]{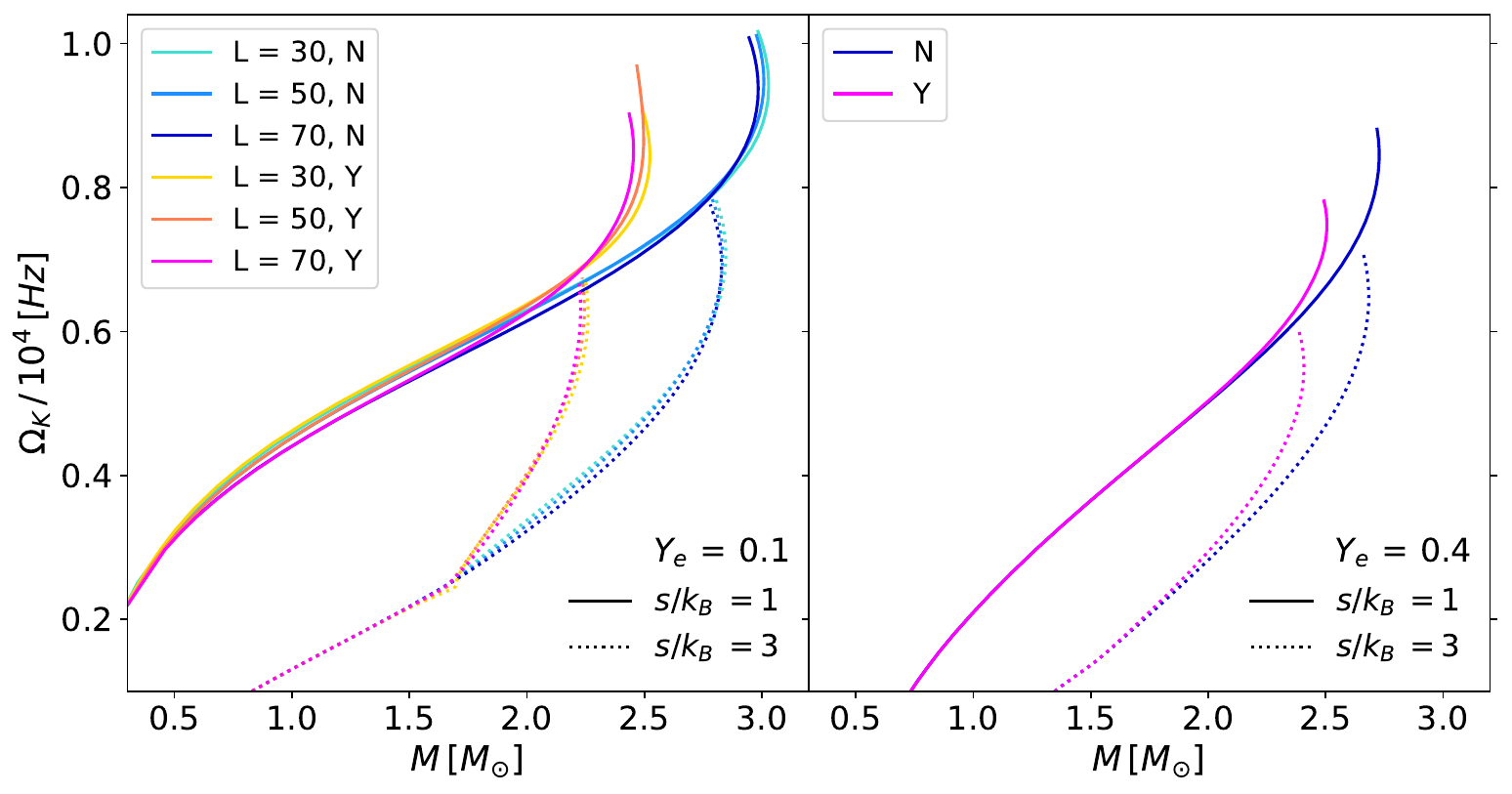}
\includegraphics[width=0.7\linewidth]{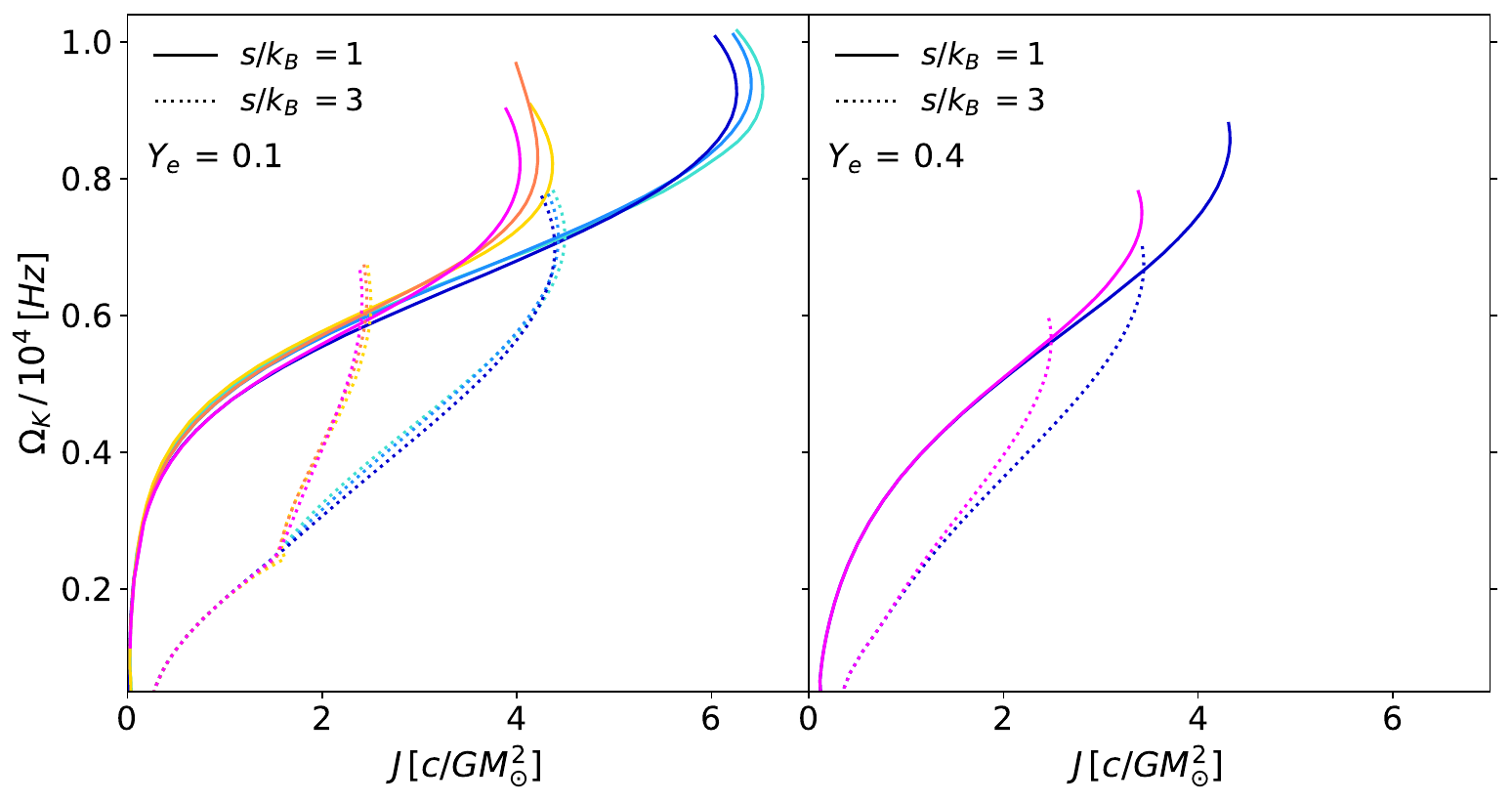}
\includegraphics[width=0.7\linewidth]{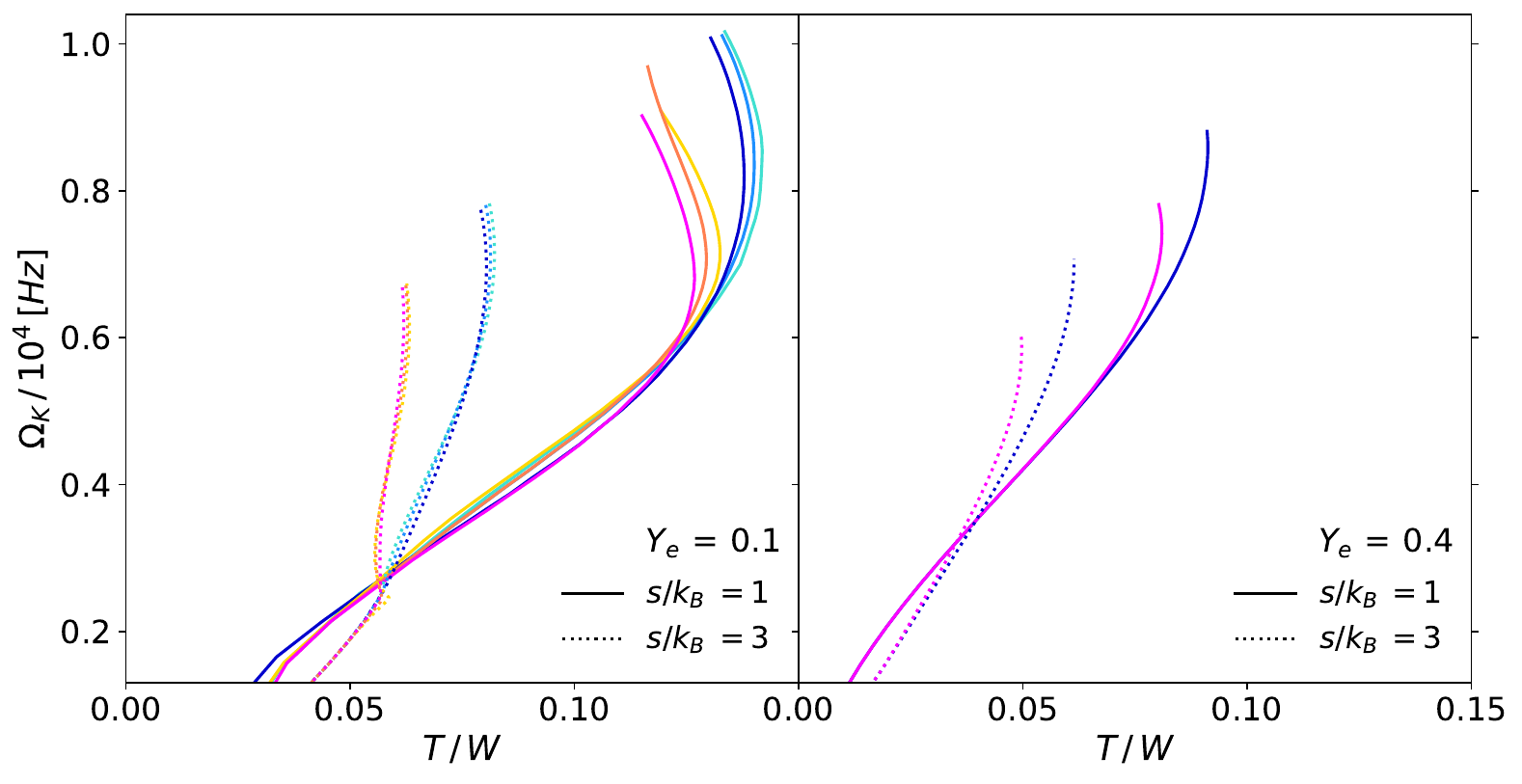}
    \caption{Keplerian frequency as a function of stellar mass (top panels), angular momentum (middle panels), and the ratio of kinetic to gravitational energies  $T/W$ (bottom panels) for nucleonic ($N$) and hyperonic ($Y$) EoS for fixed combinations of entropies per baryon and electron fractions.}
    \label{fig:OmegaM_fixed_sYe}
  \end{figure*}
  
The results for the mass as a function of central energy density are shown in Fig.~\ref{fig:Mrhoa_fixed_sYe}. 
The general trends observed in the $M$-$R$ diagrams persist here; however, the impact of varying entropy or electron fraction on the maximum mass becomes more clearly disentangled and easier to interpret.

The relationship between the maximum mass of Keplerian
 configurations, $M_K$, and the maximum mass of static stars, $M_S
 $, is well described by a linear fit of the form $M_K = a M_S + b
 $. The corresponding fit parameters for different values of $s/k_B$
 and electron fraction $Y_e$ are summarized in Table II. For each
 case, the table reports the best-fit coefficients along with their
 asymptotic standard errors, given both as absolute uncertainties and
 in units of the standard deviation $\sigma$, as well as the reduced
 $\chi^2$, providing a quantitative measure of the quality of the
 fits.

\begin{table}[hbt]
\centering
\begin{tabular}{|c|c|c|c|c|}
\hline
\multicolumn{5}{|c|}{} \\
\hline
$s/k_B$ & $Y_e$ & Parameter & Value & Asymptotic Standard Error \\
\hline
\multirow{8}{*}{1} 
  & 0.1 & a & 1.21560604 & $\pm$ 0.03120806 \\
  &     & b & $-0.01682827$ & $\pm$ 0.07132794 \\
  &     & $\sigma$ & \multicolumn{2}{c|}{0.016269} \\
  &     & $\chi^2$ & \multicolumn{2}{c|}{0.001059} \\
\cline{2-5}
  & 0.4 & a & 1.17054456 & $\pm$ 0.00328365 \\
  &     & b & $-0.11309495$ & $\pm$ 0.00766761 \\
  &     & $\sigma$ & \multicolumn{2}{c|}{0.000764} \\
  &     & $\chi^2$ & \multicolumn{2}{c|}{$2.336517 \times 10^{-6}$} \\
\hline
\multirow{5}{*}{3} 
  & 0.1 & a & 1.18459088 & $\pm $ 0.00179963 \\
  &     & b & $-0.16861531$ & 0.00414029 \\
  &     & $\sigma$ & \multicolumn{2}{c|}{0.001102} \\
  &     & $\chi^2$ & \multicolumn{2}{c|}{$4.853996\times10^{-6}$} \\
\cline{2-5}
  & 0.4 & a & 1.16216279 & $\pm $ 0.00017138 \\
  &     & b & $-0.19770778$ & $\pm $ 0.00040509 \\
  &     & $\sigma$ & \multicolumn{2}{c|}{$4.988388\times 10^{-5}$} \\
  &     & $\chi^2$ & \multicolumn{2}{c|}{$9.953607\times 10^{-9}$} \\
\hline
\end{tabular}
\caption{Fit parameters for the linear relation ($M_{K} = aM_{S}+b$) between the maximum mass of Keplerian stars and the maximum mass of static stars for different values of $s/k_B$ and $Y_e$. Each coefficient is listed with its asymptotic standard error, showing absolute $\pm$ values as well as $\sigma$ and reduced $\chi^2$.}
\label{tab:2}
\end{table}

Figure \ref{fig:IM_fixed_sYe} shows the moments of inertia of static (top row) and Keplerian (bottom row) stellar models with varying $L_{\rm sym}$ as a function of gravitational mass. The moment of inertia for a rotating star refers to the $I_{zz}$ component with rotation vector along the $z$ axis. Let us first discuss the static star. In the case $s/k_B=1$ and $Y_e = 0.1$, the moment of inertia of stable stars increases monotonically up to the maximum mass, after which it decreases abruptly, both for $N$ and $Y$ sequences. This is a direct consequence of the radius being almost fixed while the mass increases along the sequence, which is followed
by a segment of constant mass and decreasing radius in the region close to the maximum mass. The cases $s/k_B=3$ and $Y_e = 0.1$ as well as $s/k_B=3$
and $Y_e = 0.1$ and 0.4, show a different behavior, which can be traced to the much stronger variations in the radius: The moment of inertia exhibits a minimum with increasing mass, the minimum
being located at $M < M_\odot$ except for $s/A = 3$ and $Y_e = 0.4$, where the minimum is located around $ M\sim 1.5 M_\odot$. For $Y$ sequences, in addition, the moment of inertia starts decreasing at hyperon onset. This behavior results from the competition between an increase in the mass (which increases $I$) and the decrease in equatorial circumferential radius, which reduces $I$. 
The observed tendencies remain valid for Keplerian sequences, albeit with shifted scales, so we do not repeat the discussion here.
The variation of the moment of inertia with $L_{\rm sym}$ in the case $s/k_B = 1$ and $Y_e = 0.1$ for static stars exhibits an inversion: for low masses $M/M_{\odot} \leq 1.7$ for hyperonic configurations and $M/M_{\odot} \leq 2$ for nucleonic ones, the moment of inertia is larger for smaller $L_{\rm sym}$, whereas the opposite holds for higher masses. This behavior reflects the dependence of both mass and radius on $L_{\rm sym}$, which is most clearly visible near the maximum mass, where larger stellar masses correspond to larger moments of inertia. In the case $s/k_B = 3$ and $Y_e = 0.1$, the moment
of inertia is larger for smaller $L_{\rm sym}$ at the point of bifurcation of hyperonic sequences with increasing mass; this
reflects a similar kink in the mass-radius diagram. The remaining high-mass regime shows a complex, nonmonotonous behavior with two inversions
in the ordering of $I$ as a function of $L_{\rm sym}$. Because of the weak dependence of the EoS on $L_{\rm sym}$ for $Y_e = 0.4$, the variations
in the moment of inertia with  $L_{\rm sym}$ are insignificant. Again, in the case of Keplerian configurations, the variations with $L_{\rm sym}$  are similar to those for static stars, and we do not discuss them.

Figure~\ref{fig:OmegaM_fixed_sYe} shows the dependence of the Keplerian frequency on the mass, angular momentum, and kinetic to gravitational energy $T/W$ ratio for nucleonic and hyperonic isentropic and constant $Y_e$ configurations for varying $L_{\rm sym}$.  It is useful to recall the fit formula for cold NSs, which relates the Keplerian angular frequency and the nonrotating star's mass and radius~\cite{Cook:1994,Haensel2009,Riahi2019,Li2023PhRvC}
\bea\label{eq:Omega_K_scaling}
\Omega_K = 2\pi f_K \approx 2\pi  f_0
\left(\frac{M}{M_{\odot}}\right)^{1 / 2}\left(\frac{R}{10 \mathrm{~km}}\right)^{-3 / 2}~\rm{kHz},
\eea
where $f_0 = 1.04-1.08$, which reflects the general scaling of $\Omega_K$ with mass {\it and} radius. Formula \eqref{eq:Omega_K_scaling} is written down here for a qualitative understanding of the scalings, without an attempt to fit it to isentropic stars to determine the dependence of the $f_0$ parameter on the input $Y_e$ and $s/k_B$; this issue is left for future work.  The slow rise for masses $M/M_{\odot}\le 2$ (low-mass region) reflects the small variation in the radius with rapidly rising mass, whereas the steep rise for $M/M_{\odot} > 2$ (high-mass region) reflects the rapid reduction of the radius at almost constant mass.  In the low-mass region, the smaller $L_{\rm sym}$ is the larger $\Omega_K$ for constant mass. The ordering of curves reverses in the high-mass limit. This behavior is consistent with the $M(R)$ diagram, implying that more compact stars can rotate faster and have larger $\Omega_K$. A comparison of the stars with different entropies shows that the larger the entropy, the lower is $\Omega_K$; again, this result is well understood from the fact that larger entropy stars are more extended (i.e., have larger radius) and, therefore, less compact stars reach mass-shedding limit at smaller masses. Comparing the results for nucleonic and hyperonic EoS in the mass domain, where hyperonic sequences have already bifurcated from the nucleonic ones, one observes that the hyperonic stars have larger $\Omega_K$ for fixed same mass--a consequence of being more compact than the nucleonic counterparts.  However, note that they have a lower peak frequency, marked by the condition $d\Omega_K/dM = \infty$, as a consequence of lower maximal mass. Finally, the cases $Y_e = 0.1$ and 0.4 do not differ qualitatively, i.e., all scalings described above remain intact. For $Y_e=0.4$, the $\Omega_K(M)$ values by a factor of a few are lower for fixed mass, and the maximum frequency is lower as well; both effects can be traced again to the $M(R)$ relation, and show that more compact stars can rotate faster.

The middle panel of Fig.~\ref{fig:OmegaM_fixed_sYe} shows $\Omega_K(J)$, where $J$ is the angular momentum. This relation directly encodes the moment of inertia through the relation $J = I\Omega$; therefore, the behavior seen in this panel is a direct reflection of the variation of the moment of inertia with structure and EoS, already discussed in the context of Fig.~\ref{fig:IM_fixed_sYe}.  Furthermore, the discussion of the $\Omega_K(M)$ relation applies to the $\Omega_K(J)$ relation: specifically, it is seen that at fixed angular momentum, lower values of $L_{\rm sym} $ correspond to higher Keplerian frequencies $\Omega_K$ in the low-angular-momentum regime. This trend reverses at high angular momentum: more compact stars rotate faster before reaching the mass-shedding limit.  For fixed angular momentum, hyperonic stars exhibit larger $\Omega_K$ than nucleonic ones once their sequences diverge from each other, due again to the fact that they are more compact. However, their peak frequency (where $d \Omega_K / d J=0$ ) is smaller, which is tied to the lower maximum mass. The cases with $Y_e=0.1$ and $Y_e=0.4$ show the same qualitative behavior. For $Y_e=0.4$, the values of $\Omega_K$ are overall reduced, as seen the middle-right panel of Fig.~\ref{fig:OmegaM_fixed_sYe}  and as explained
above. 

The lower panel of Fig.~\ref{fig:OmegaM_fixed_sYe}  shows the dependence of $\Omega_K$ on the kinetic-to-gravitational energy ratio $T/W$ for isentropic nucleonic and hyperonic stars. The range of $T/W$ covered by stationary sequences of nucleonic and hyperonic stars is located far from
the dynamical bar-mode instability region, which is well-known for cold EoS to be above $T/W \simeq 0.27$~\cite{Chandrasekhar1969,Shibata2000ApJ,BaiottiPhysRevD}.
The same range  is, however, close to the value of $T/W\ge 0.14$  where secular (dissipation-driven) bar-mode instability   for uniformly rotating
cold NSs sets in~\cite{Gondek2002,Shibata2004,Saijo2006}. The latter effectively provides the maximum value of $T/W$ that may be realizable in nature. We observe the following trends: 
\begin{enumerate}
\item For fixed $T / W$, a smaller symmetry energy slope $L_{\rm sym}$ leads to a higher Keplerian frequency if $0\le T / W\le 0.125$ and the inverse is true  for higher $T / W$ for $s/k_B=1$ and $Y_e=0.1$.  The transition takes place close to the point of the onset of secular instability, with a limit of 0.14 for cold stars. The same inversion is observed for $s/k_B=3$ and  $Y_e=0.1$, but at lower values of $T/W\sim 0.06$ for hyperonic configurations and $T/W\sim 0.08$ for nucleonic ones.  The inversion is absent for $Y_e=0.4$ with hyperonic configurations showing larger Kepler frequency for fixed  $T/W$.
\item  For fixed $T / W$, the Kepler frequency is (slightly) higher for hyperonic than for the nucleonic models, with the exception of
  $Y_e=0.1$, $s/k_B = 1$ and $T/W\le 0.06$ regime. 
However, nucleonic models achieve larger frequencies before the stars become unstable at the maximum mass. 
\item While both $Y_e=0.1$ and $Y_e=0.4$ cases follow the same qualitative pattern, increasing the electron fraction from 0.1 to 0.4 leads to a systematic reduction in $\Omega_K$ values, which again can be explained by the stars being less compact for larger $Y_e$.
\end{enumerate}
As isentropic stars are less compact than their same-mass low-temperature counterparts, it is expected that the secular instability appears at lower values of $T/W$. The secular instabilities in hot compact stars have been discussed elsewhere~\cite{Lai2001,Camelio2021}.

\section{Universal relations}
\label{sec:Universal} 

Overall, mass, radius, and other macroscopic properties of compact stars are largely determined by the underlying EoS; a feature that has long been used to shed light on the properties of dense matter based on astrophysical observations. However, it has  been established that combinations of certain macroscopic properties comprise a set of universal relations that hold with remarkably weak dependence on the employed EoS, for a review see Ref.~\cite{Yagi:2017}. 
This universality, despite its still elusive origin, can for mature NS likewise be used to extract information from observational data. To name a few examples, reduction of EoS-related uncertainties in data analysis, constraints on otherwise inaccessible NS properties, and lifting of degeneracies between quadrupole moment and spin in gravitational waveforms from binary inspirals can be achieved by applying those universal relations. 

In this section, we examine the validity of universal relations in the case of isentropic stars with $s/k_B=1$ and $s/k_B=3$, both in the static and in the Keplerian limit, for matter with fixed electron fractions of $Y_e=0.1$ and $0.4$, for all the EoS models presented in Sec.~\ref{sec:Kepler}. The universal relations in question involve the normalized moment of inertia, the normalized quadrupole moment,  the stellar compactness, and the Kepler frequency.

In the following discussion, we consider NS sequences with central densities larger than $\rho_{c} \geq 2.8~10^{14}$ g~cm$^{-3}$ for all combinations of entropy and electron fraction. In addition, we take into account a small number of stars lying beyond the maximum mass on the unstable branch, with masses a few percent below the maximum mass of each sequence. This allows for a more accurate fitting of the high-mass end of the sequences.

The first universal relation examined in this work characterizes the normalized moment of inertia of a star, denoted as $I^{<} = I/(MR^2)$, through a polynomial expansion in stellar compactness following the approach of 
Ref.~\cite{Lattimer:2005}:
\bea\label{eq:tildeI-C}
I^{<}(C) = \sum_{j=0}^{m} a_j C^j.
\eea
Here, the compactness parameter is defined as $C = M/R$, where $M$ represents the stellar mass and $R$ the stellar radius. It is important to note that the compactness $C$ exhibits complex dependencies on several physical parameters, including the rotational velocity, entropy (for isentropic stellar models), and the composition of stellar matter. This polynomial representation was originally developed in Ref.~\cite{Lattimer:2005} for the specific case of cold, nonrotating NSs in $\beta$-equilibrium. In this regime, the relation demonstrates remarkable independence from the underlying EoS.
    
\begin{figure*}[!]
\includegraphics[width=0.37\linewidth]{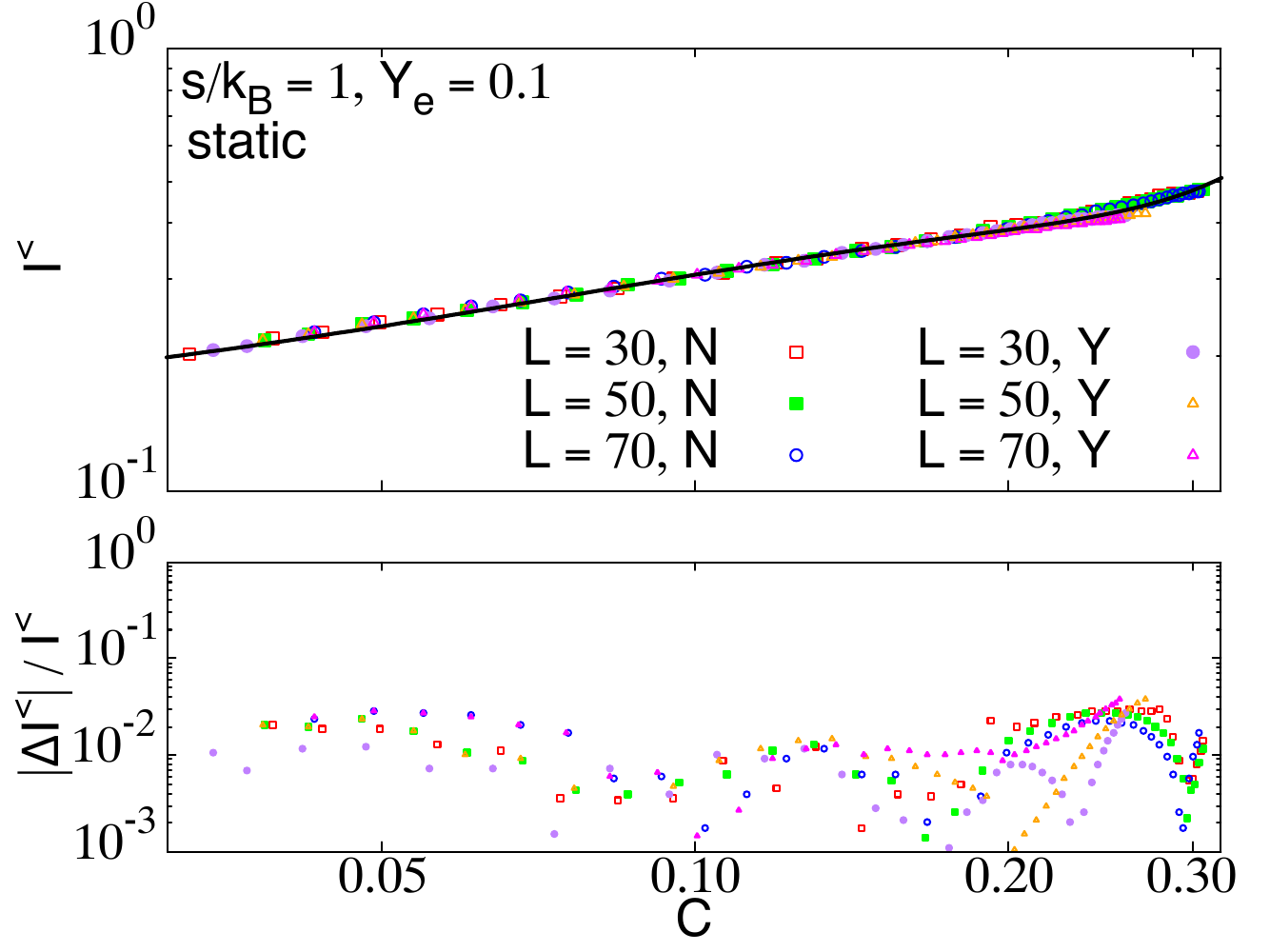}
\includegraphics[width=0.37\linewidth]{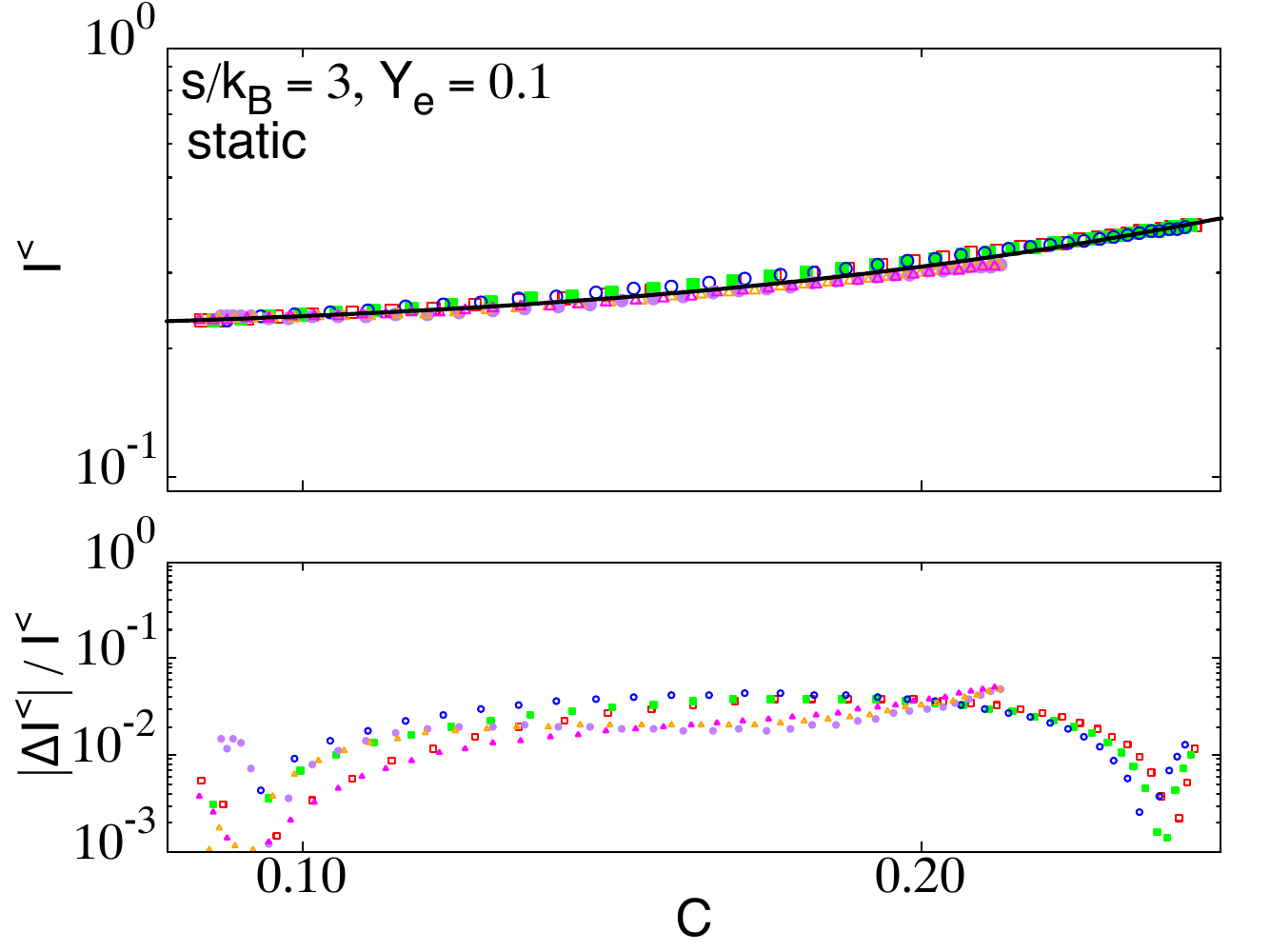}

\includegraphics[width=0.37\linewidth]{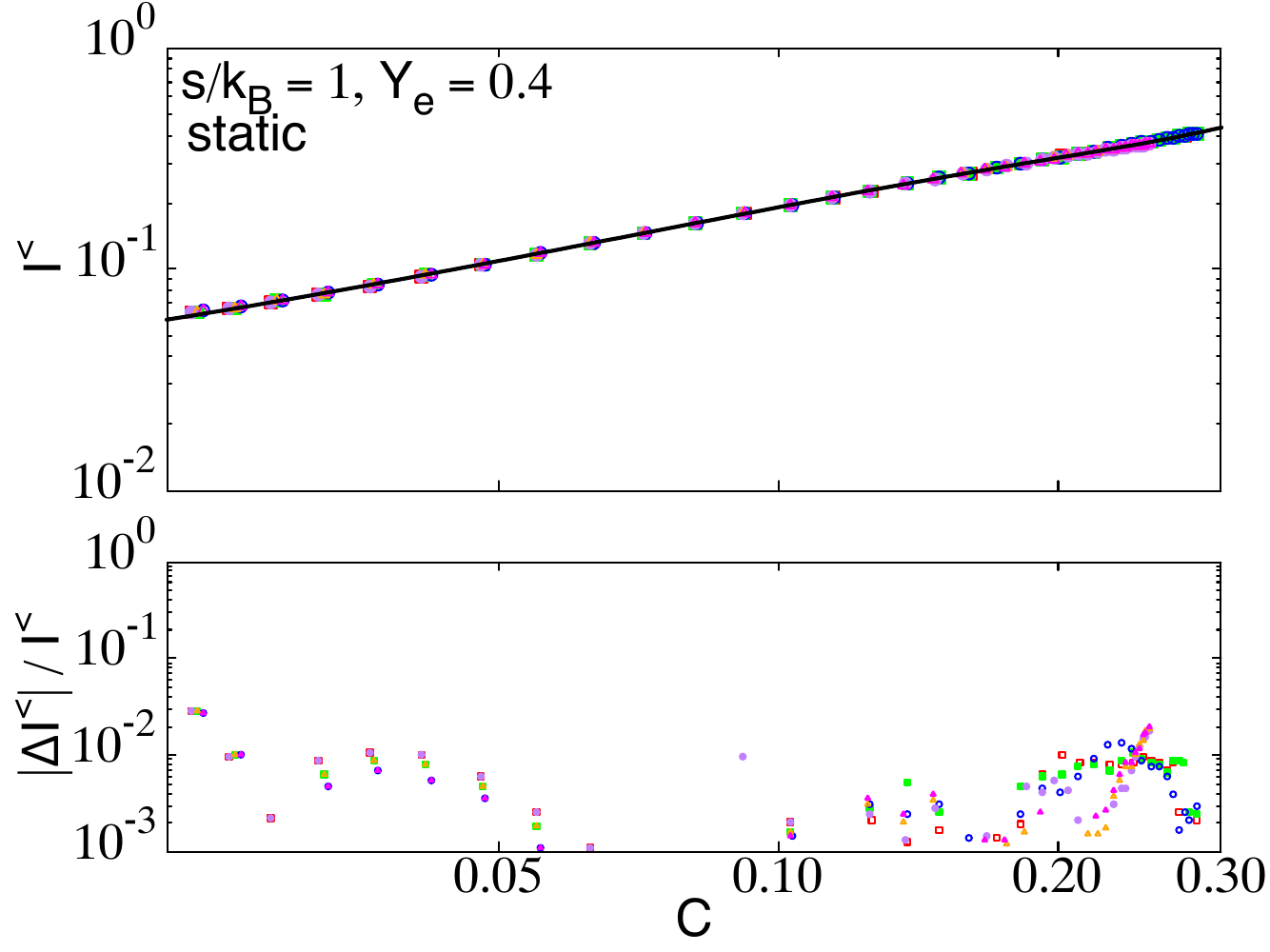}
\includegraphics[width=0.37\linewidth]{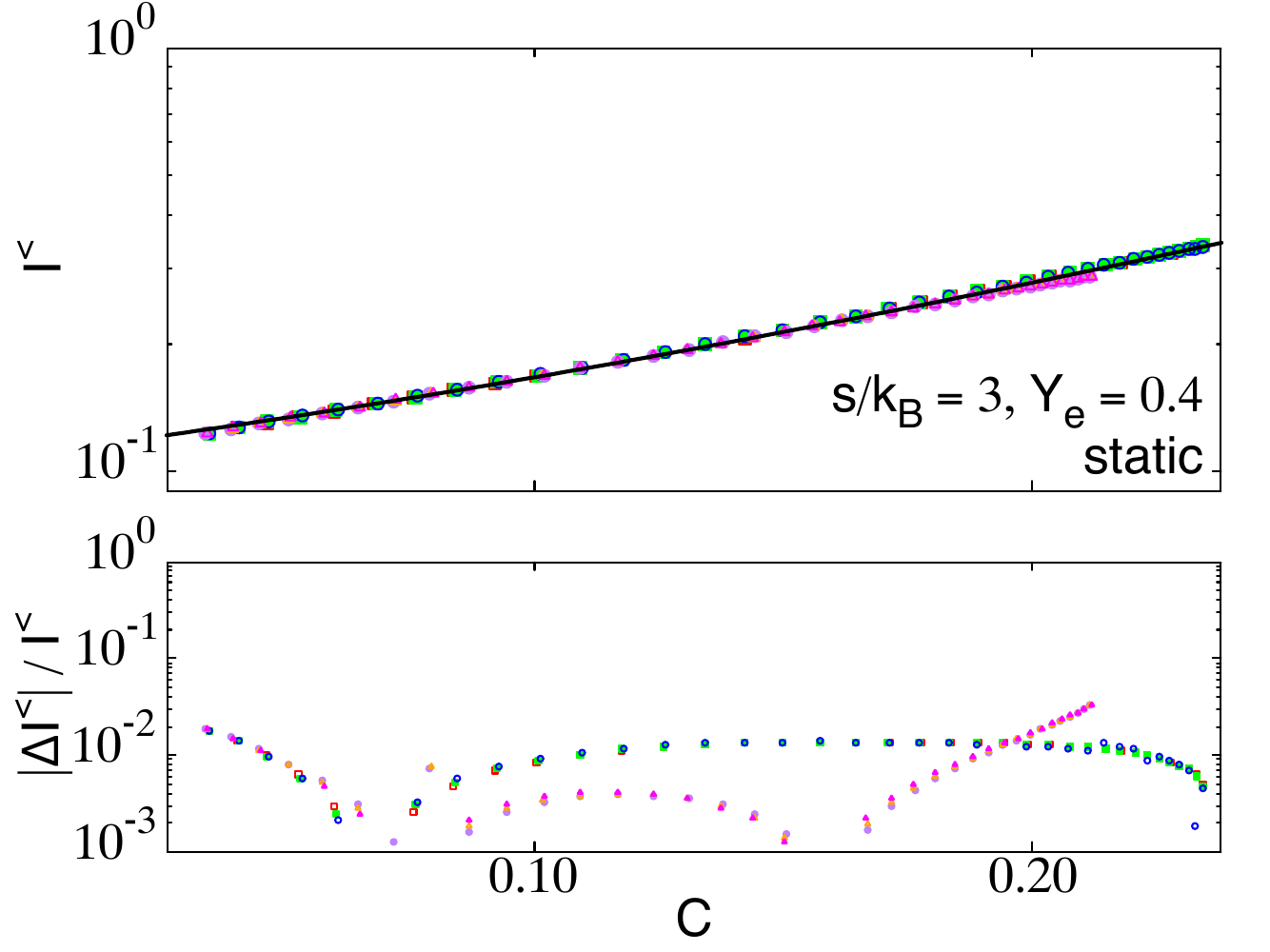}
    \caption{Log-log plot of the dimensionless moment of inertia $I^{<}$ versus compactness $C$ and relative residual errors ${\vert \Delta I^{<}\vert}/{I^{<\,fit}}$, where 
    $\Delta I^{<} = I^{<}-I^{< \, fit}$
    with respect to the fit by  Eq.~\eqref{eq:tildeI-C} for sequences of static stars of four different $(s/k_B, Y_e)$ combinations. In each subimage, the upper panel show the data points by varying the parameter $L_{\rm sym} \in [30, 50, 70]$ MeV for both nucleonic ($N$) and hyperonic ($Y$) compositions and the best-fit curve based on a polynomial expansion in $C$, while the lower panel displays the relative deviation between the numerical data and the fit.
 }
    \label{fig:I_tilde_a_Fit_Error}
\end{figure*}

Fig.~\ref{fig:I_tilde_a_Fit_Error} shows the data points, numerical fits, and relative errors $|\Delta I^{<}|/I^{<}$ of the same polynomial \eqref{eq:tildeI-C} for static stars of constant entropy per baryon $s/k_{B}=1$ (left column) and $s/k_{B}=3$ (right column) and two fixed electron fractions $Y_{e}=0.1$ (top) and $Y_{e}=0.4$ (bottom).  We observe that, independent of the value of $L_{\rm sym}$ and of the composition (nucleonic or hyperonic), the universality remains intact as long as the thermodynamic conditions—namely the values of $Y_e$ and $s/k_B$—are kept fixed. This observation is consistent with previous finite-temperature studies~\cite{Lenka:2018,Raduta_MNRAS_2020}, which established quasiuniversal relations for a given thermodynamic condition, with relative deviations across their heterogeneous EoS sets not exceeding 10\%.  The values of the coefficients of the polynomial obtained from our fit of Eq.~\eqref{eq:tildeI-C} to the data are listed in Table~\ref{tab:a_stat} of the Appendix. 
    
\begin{figure*}[!]
\includegraphics[width=0.37\linewidth]{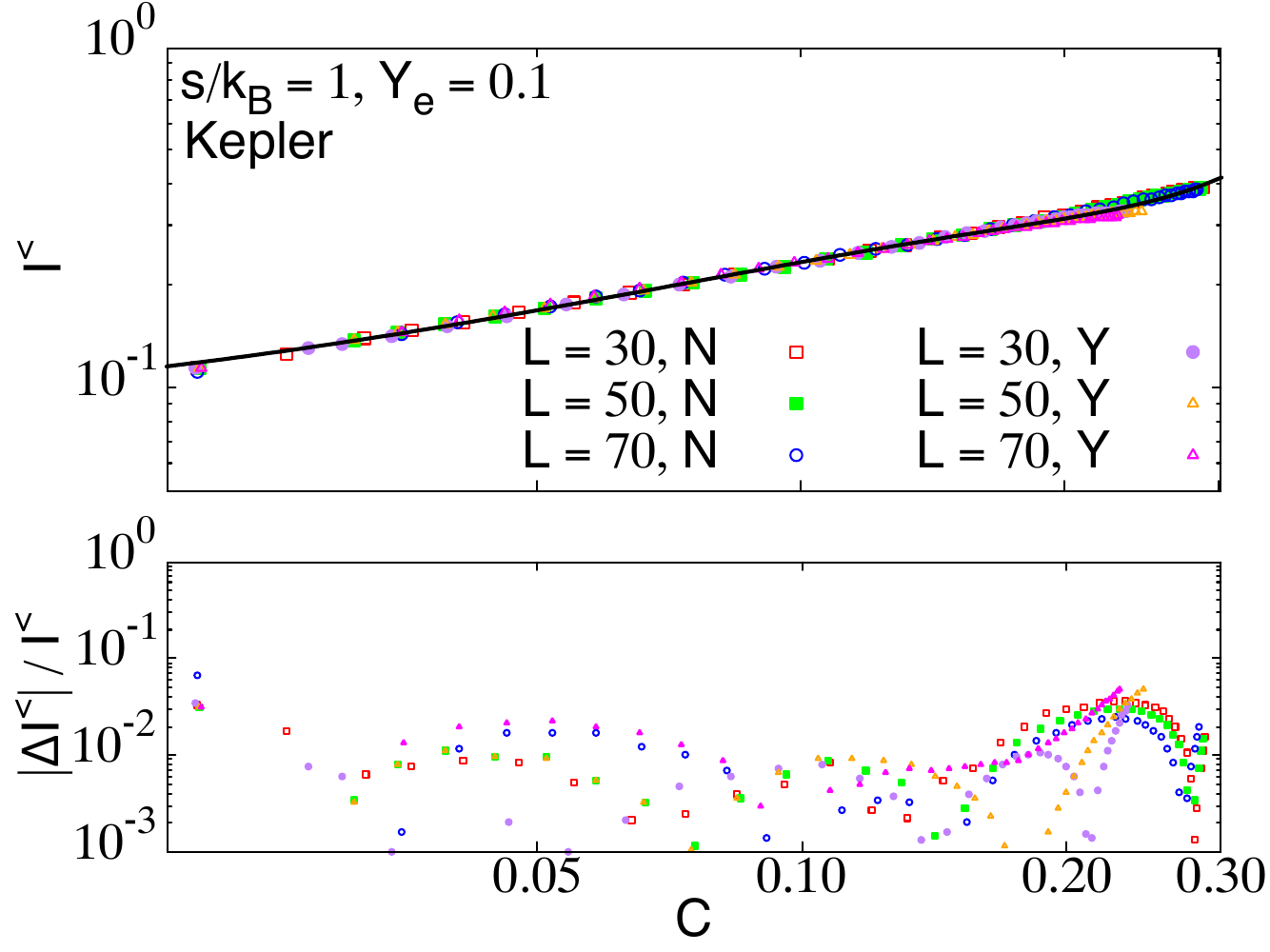}
\includegraphics[width=0.37\linewidth]{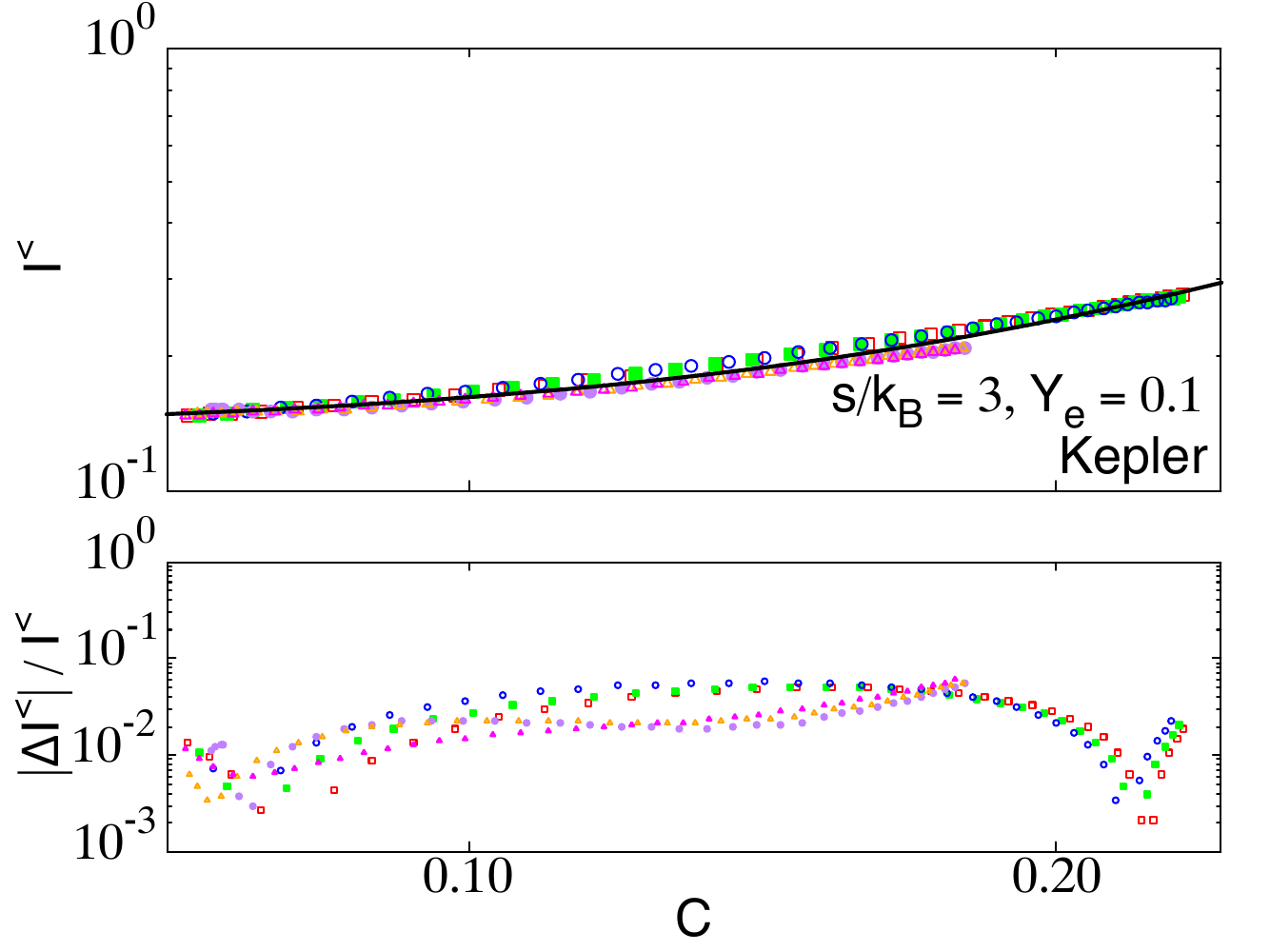}

\includegraphics[width=0.37\linewidth]{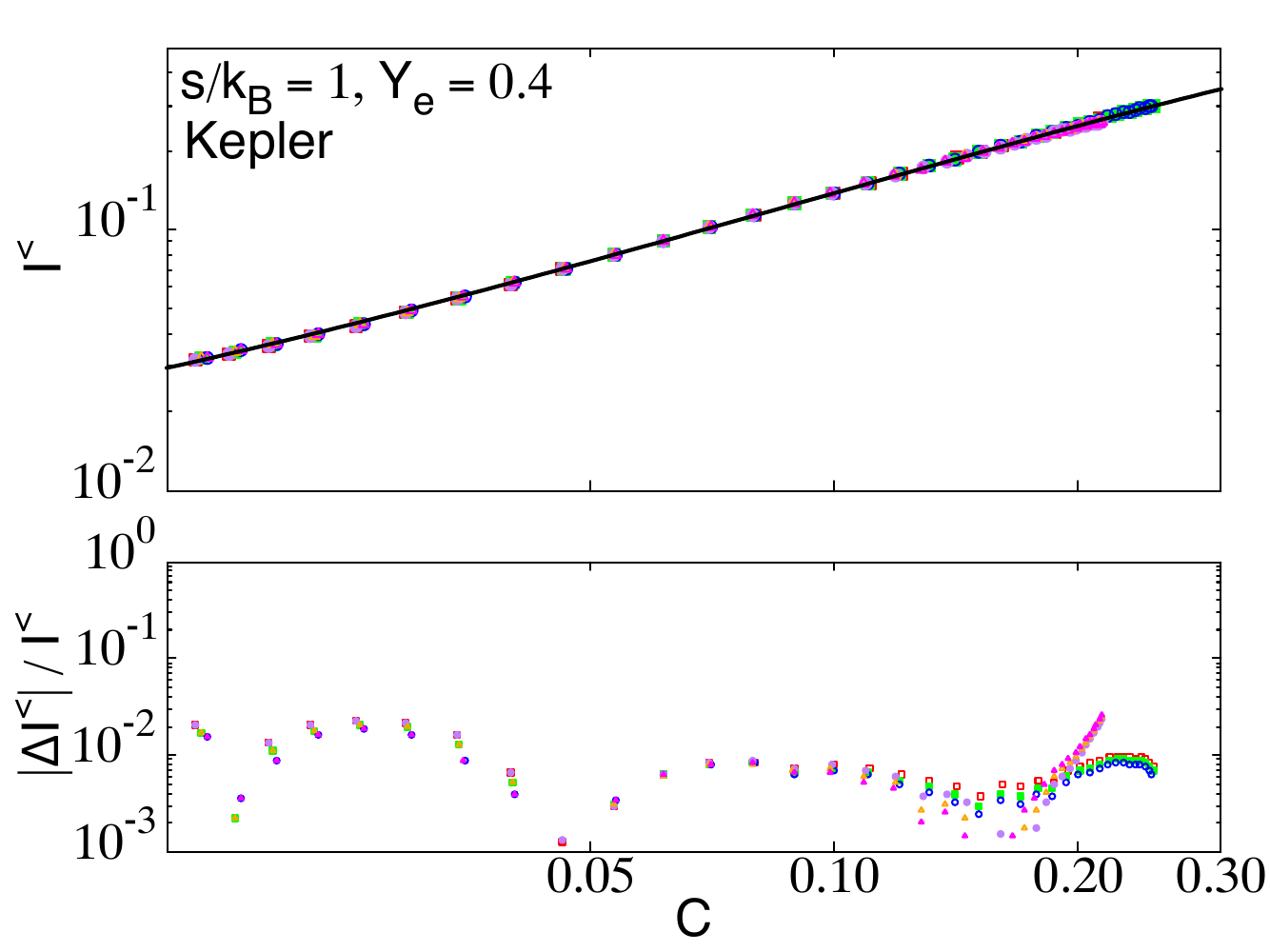}
\includegraphics[width=0.37\linewidth]{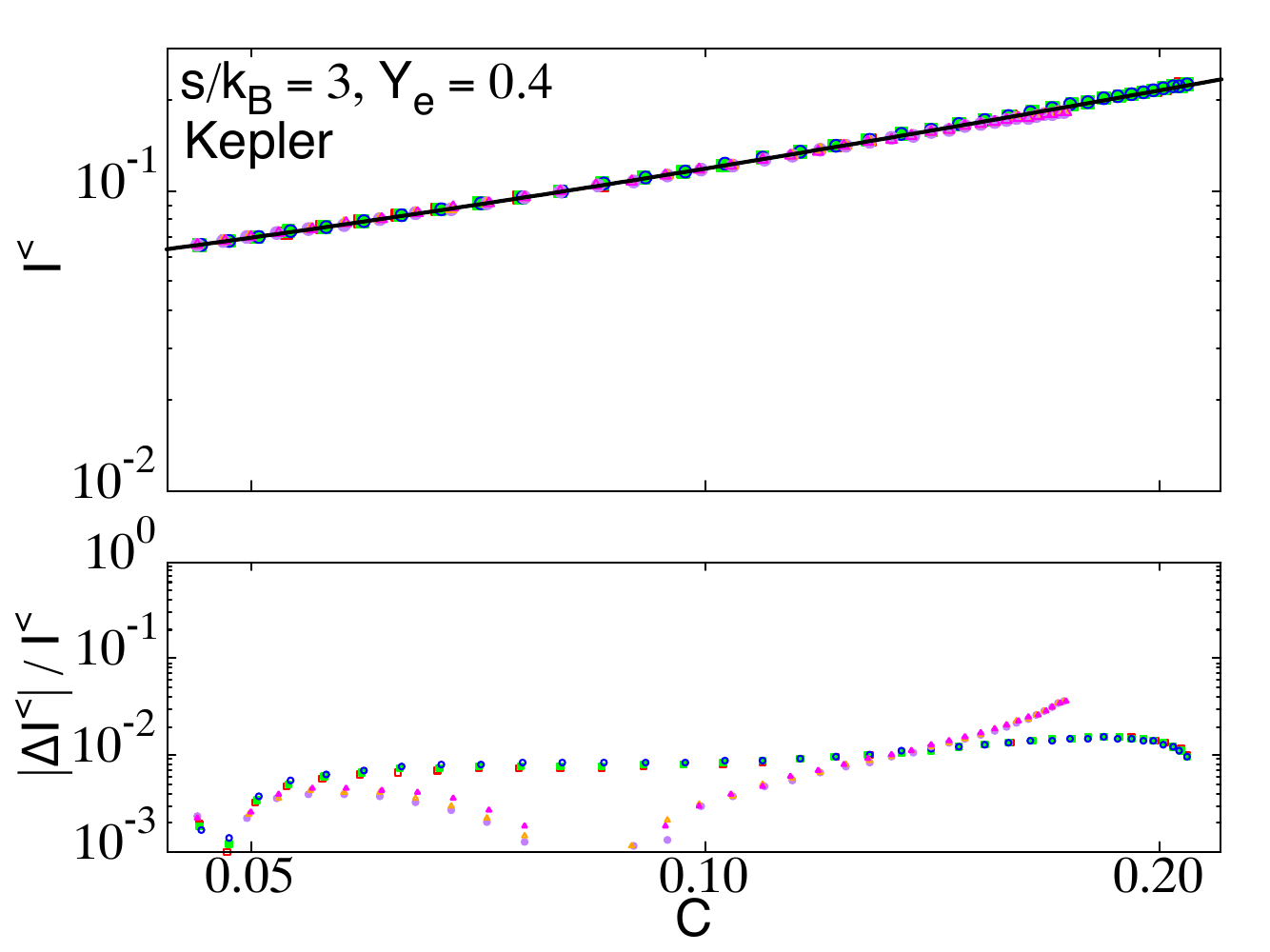}
    \caption{Same as in Fig.\ref{fig:I_tilde_a_Fit_Error} but for the case of rotating stars at the mass-shedding limit.}
    \label{fig:I_tilde_a_Fit_Error_kep}
\end{figure*}

The same procedure was repeated for the case of maximally rotating (Keplerian) configurations, and the results are presented in Fig.~\ref{fig:I_tilde_a_Fit_Error_kep}. The corresponding coefficients of the polynomial  are given in Table~\ref{tab:a_Kepler}. It is worth noting that in the high-entropy case, two polynomial terms are sufficient to find an accurate fit to the universal relation, for both values of electron fraction. In the low-entropy cases, on the other hand, the number of terms required is no more than four.

\begin{figure*}[!]
\includegraphics[width=0.37\linewidth]{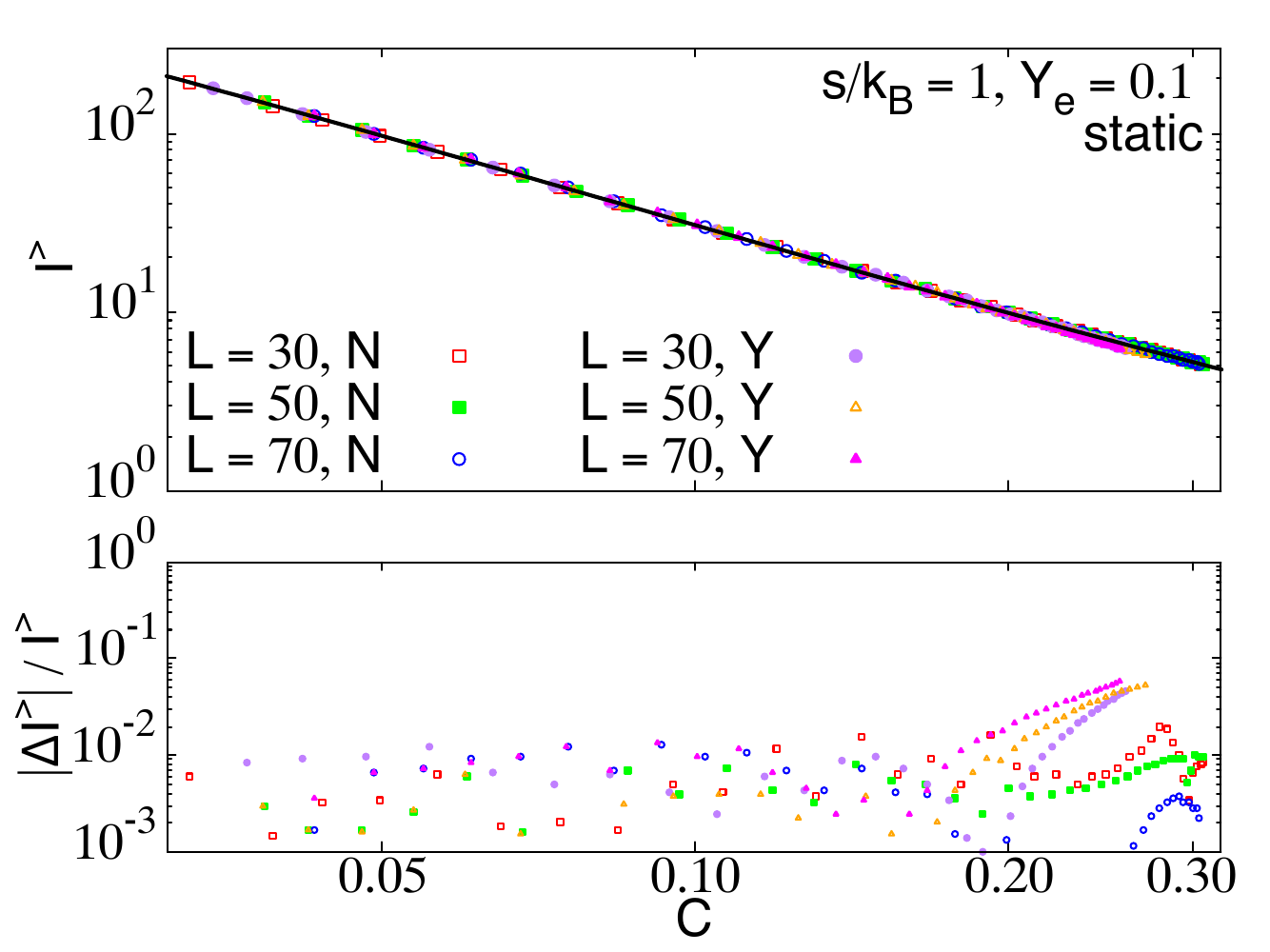}
\includegraphics[width=0.37\linewidth]{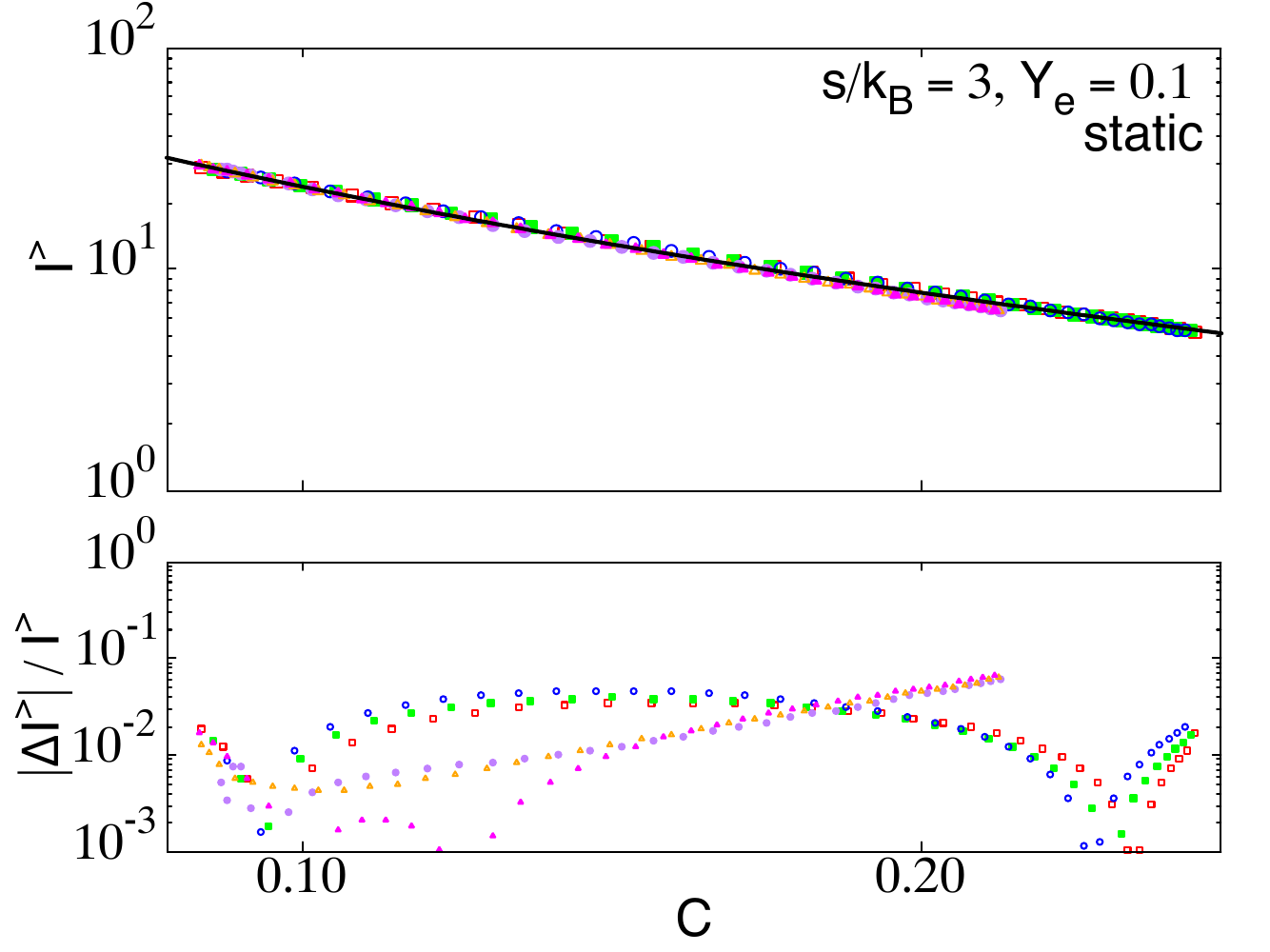}

\includegraphics[width=0.37\linewidth]{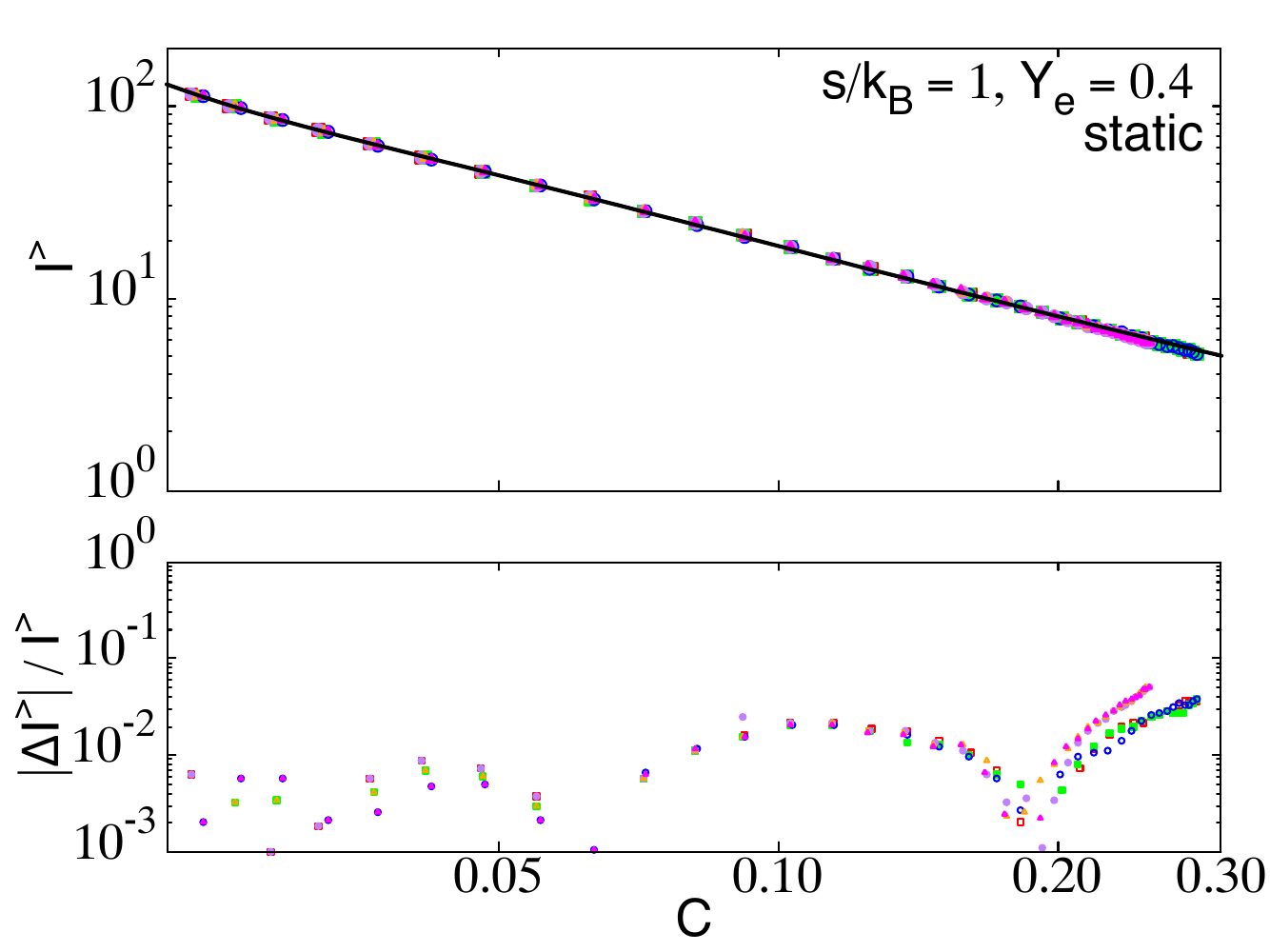}
\includegraphics[width=0.37\linewidth]{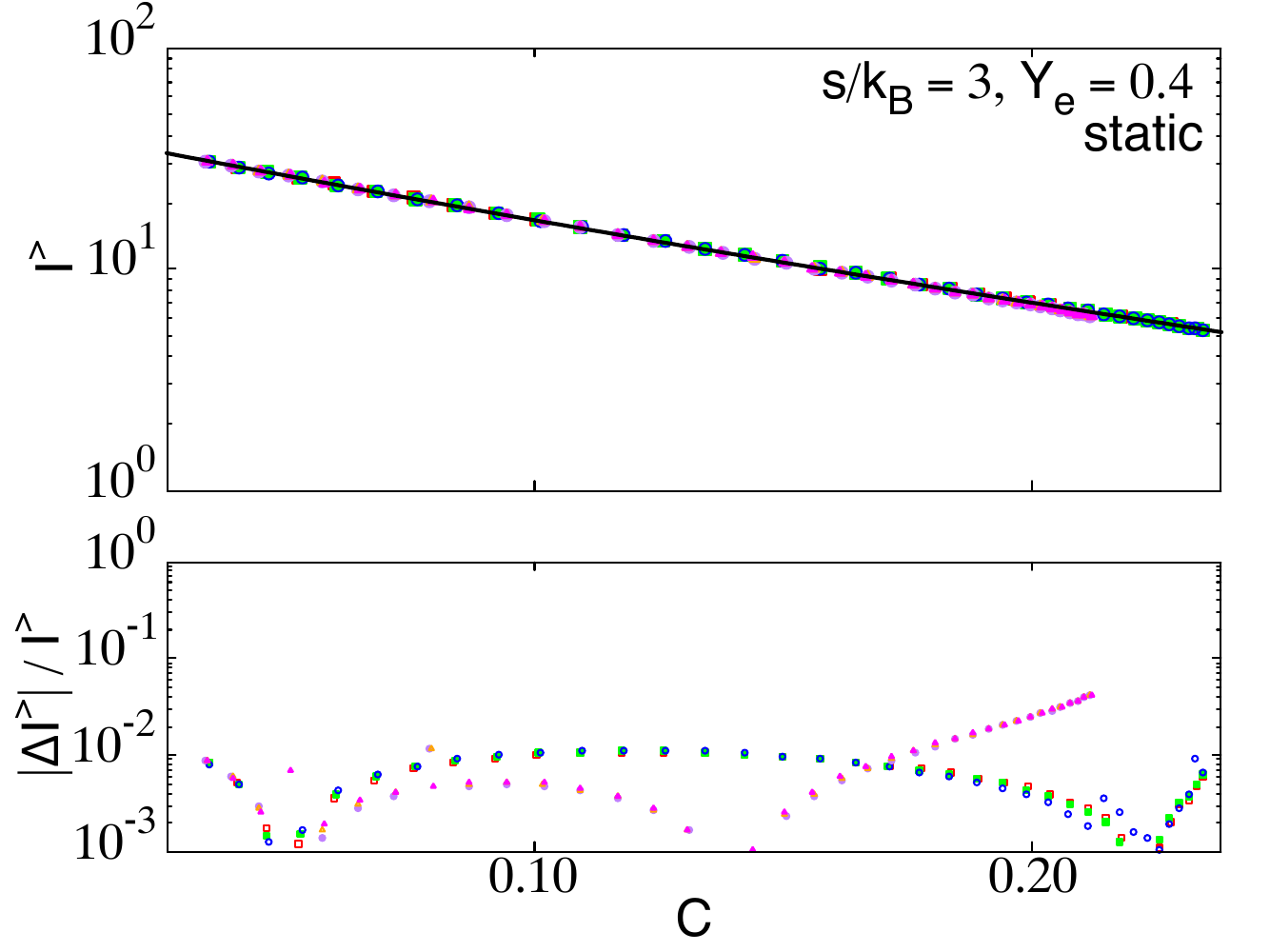}
    \caption{Same as in Fig.~\ref{fig:I_tilde_a_Fit_Error} but for the dimensionless moment of inertia $I^{>}(C)$ fitted using Eq.~\eqref{IbarC}. }
    \label{fig:I_bar_a_Fit_Error}
\end{figure*}

\begin{figure*}[!]
\includegraphics[width=0.37\linewidth]{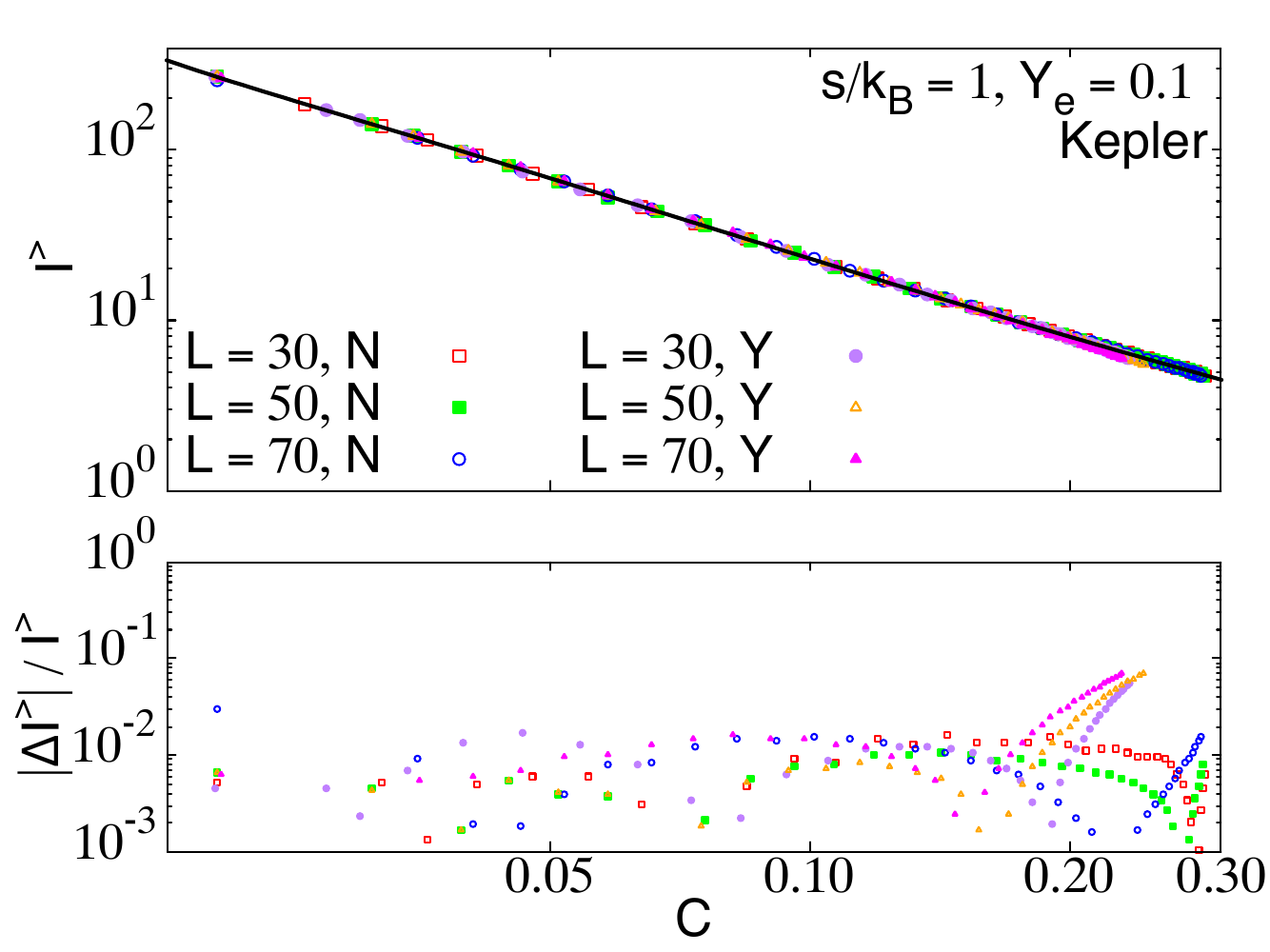}
\includegraphics[width=0.37\linewidth]{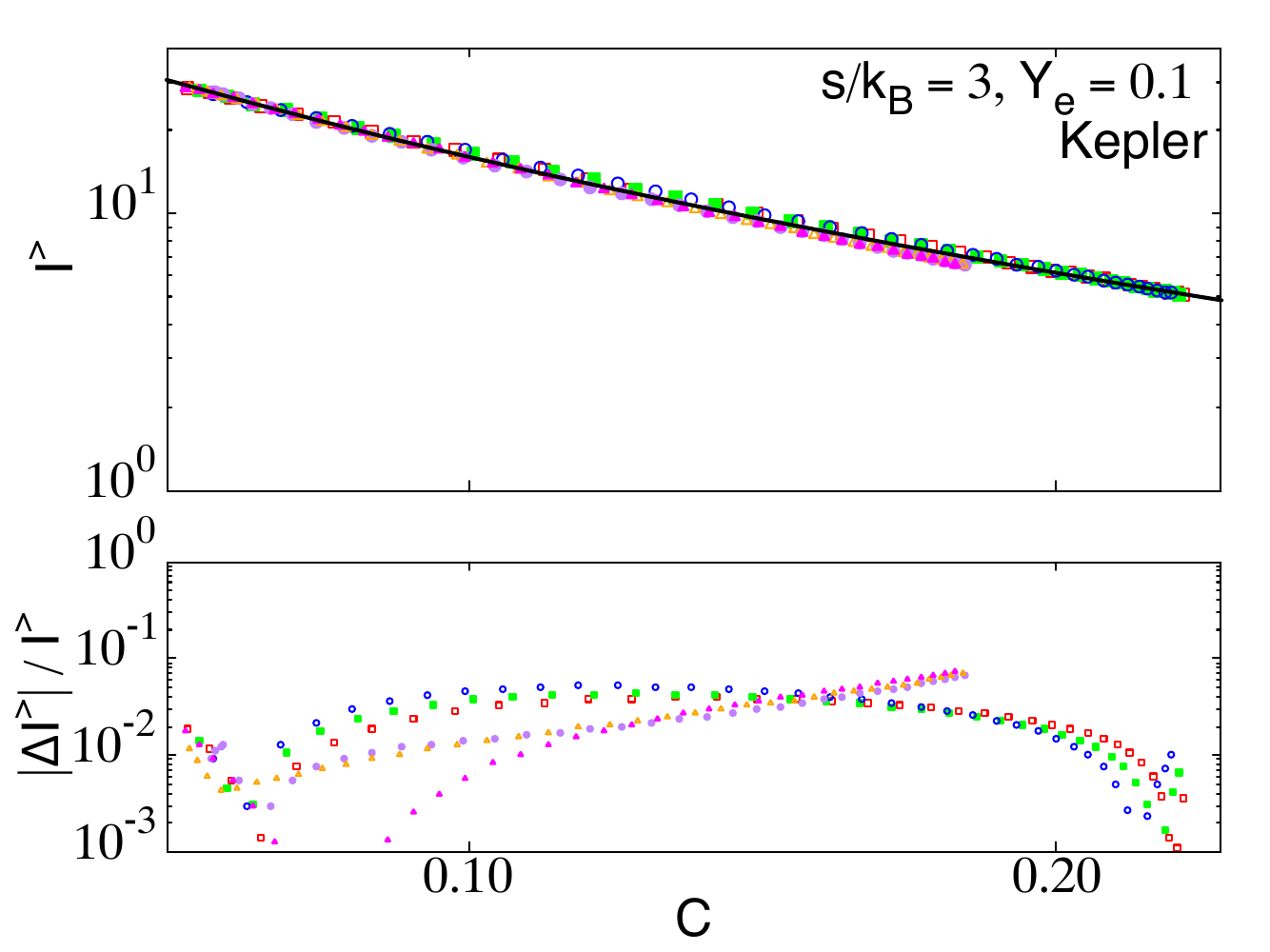}

\includegraphics[width=0.37\linewidth]{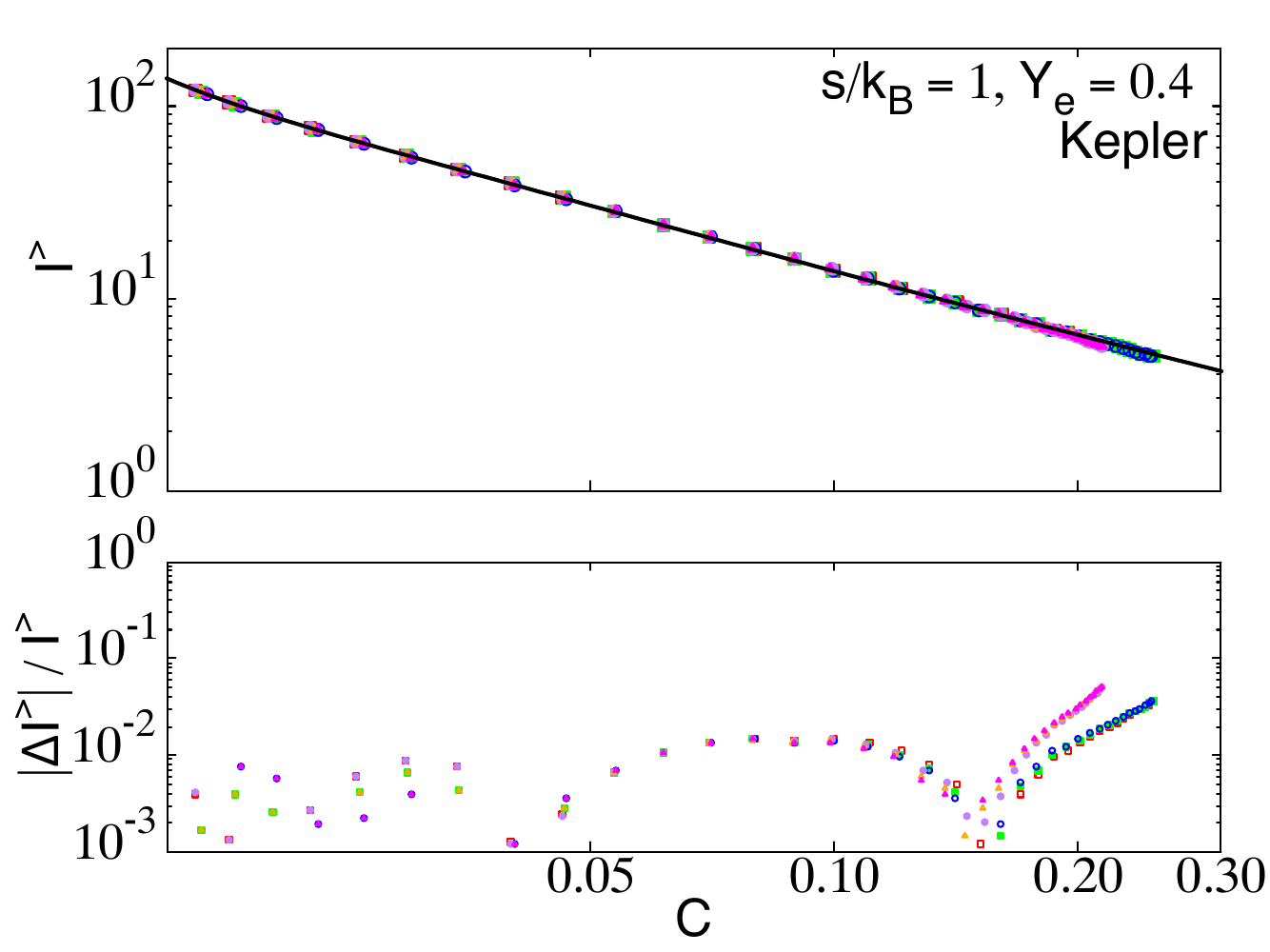}
\includegraphics[width=0.37\linewidth]{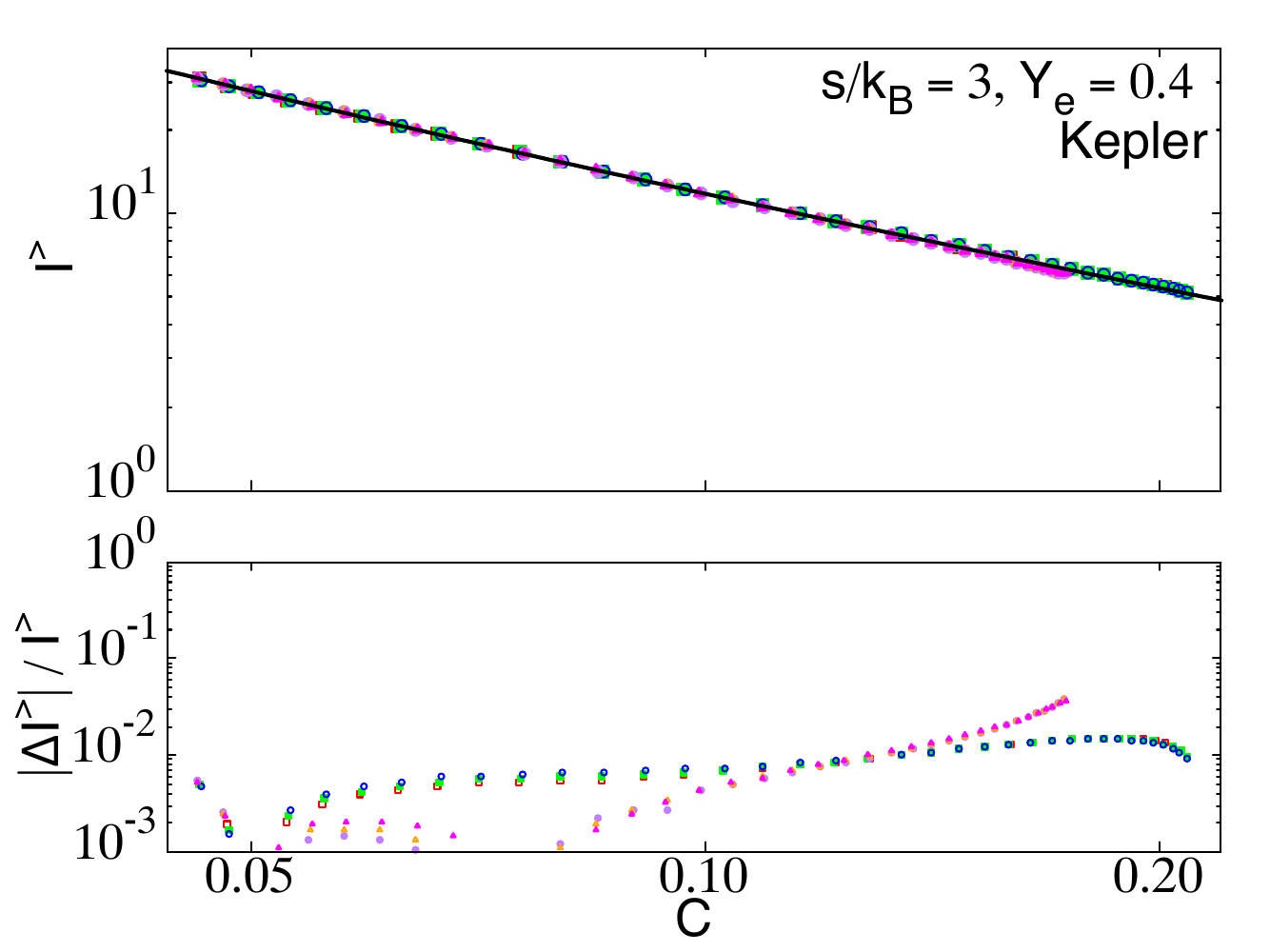}
    \caption{Same as in Fig.\ref{fig:I_bar_a_Fit_Error} but for the case of rotating stars at the mass-shedding limit.}
    \label{fig:Ibar_c_Fit_Error_kep}
\end{figure*}

A second universal relation was proposed in Ref.~\cite{Breu:2016}, who introduced an alternative normalization for the moment of inertia. These authors demonstrated that $I^{>} = I/M^3$ can be expressed as a series expansion in inverse powers of the compactness parameter
    \bea
    I^{>} (C) = \sum_{j=0}^{m} b_{j}\left[C^{j}\right]^{-1}.
    \label{IbarC}
    \eea
    As in the case of cold mature NS, the application of this fitting formula reveals that the universal relation achieves remarkable accuracy, with relative deviations of less than 10\% between the numerical fit and the calculated data for both static and maximally rotating (Keplerian) stellar configurations and given thermodynamic conditions. This level of agreement holds consistently, regardless of the chosen entropy per particle ($s/k_B$) and electron fraction ($Y_e$) combinations (with EoS sets including nucleonic and hyperonic compositions). 
The accuracy of this relation is illustrated through numerical results displayed in Fig.~\ref{fig:I_bar_a_Fit_Error} for 
static configurations, where we show the data points, numerical fits, and relative errors $|\Delta I^{>}|/I^{>}$ for $s/k_B = 1$ (left panels) and $s/k_B = 3$ (right panels), comprising cases with $Y_e = 0.1$ (top) and $Y_e = 0.4$ (bottom). The same for rotating configurations is shown in  Fig.~\ref{fig:Ibar_c_Fit_Error_kep}, organized using the same panel arrangement.
The polynomial fit parameters are contained in Table~\ref{tab:b_stat} for the static and in Table~\ref{tab:b_Kepler} for the Keplerian cases. As in the previous universal relation, two terms are used in the polynomial expansion of the high-entropy sequences, while four terms are needed for the low $s/k_B$ cases. Notably, in the latter case, these are the terms up to $m=3$ for the static configuration, with the respective set for the Keplerian ones, including the $1\leq m \leq 4$ terms.

\begin{figure*}[!]
\includegraphics[width=0.37\linewidth]{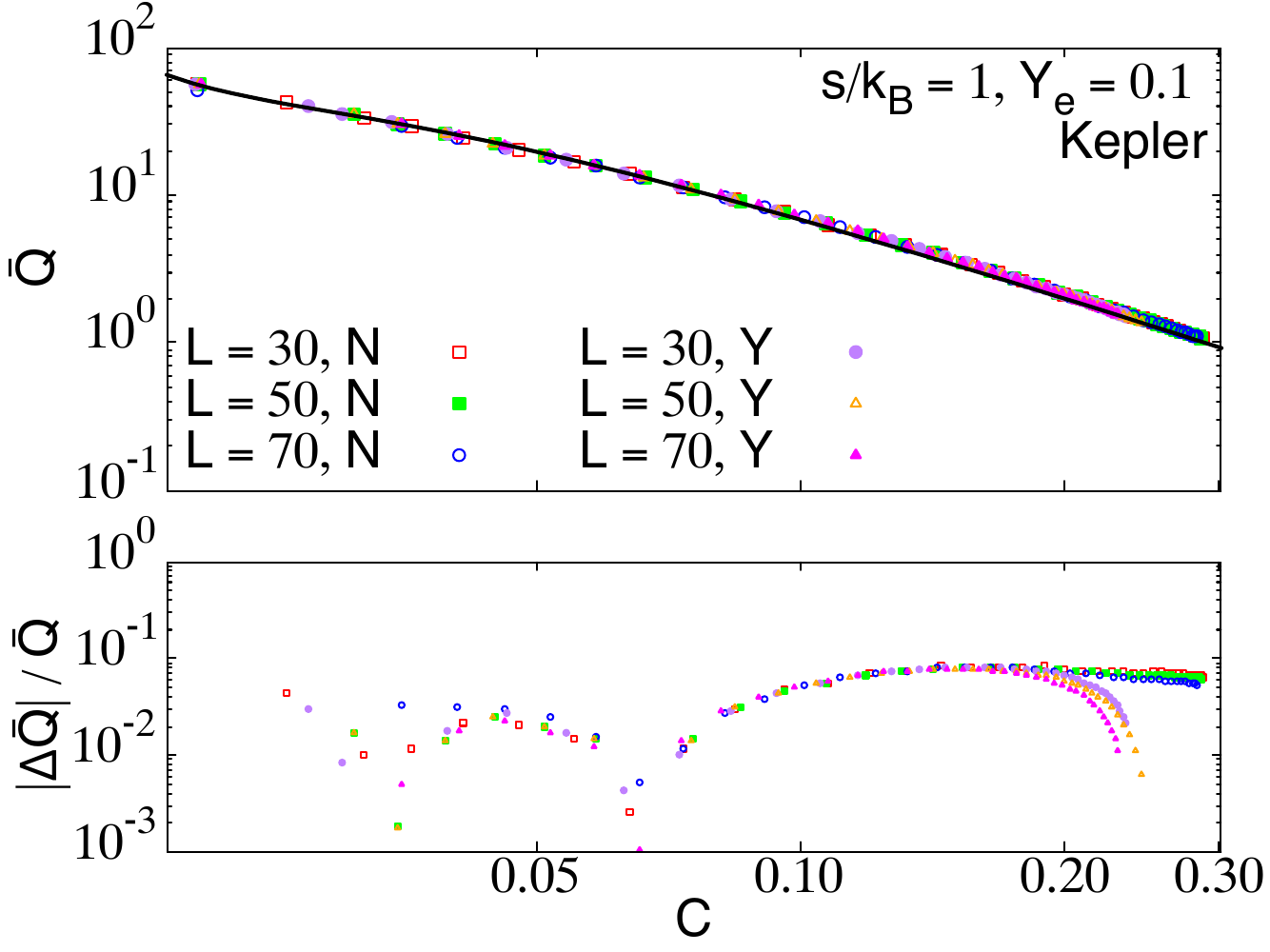}
\includegraphics[width=0.37\linewidth]{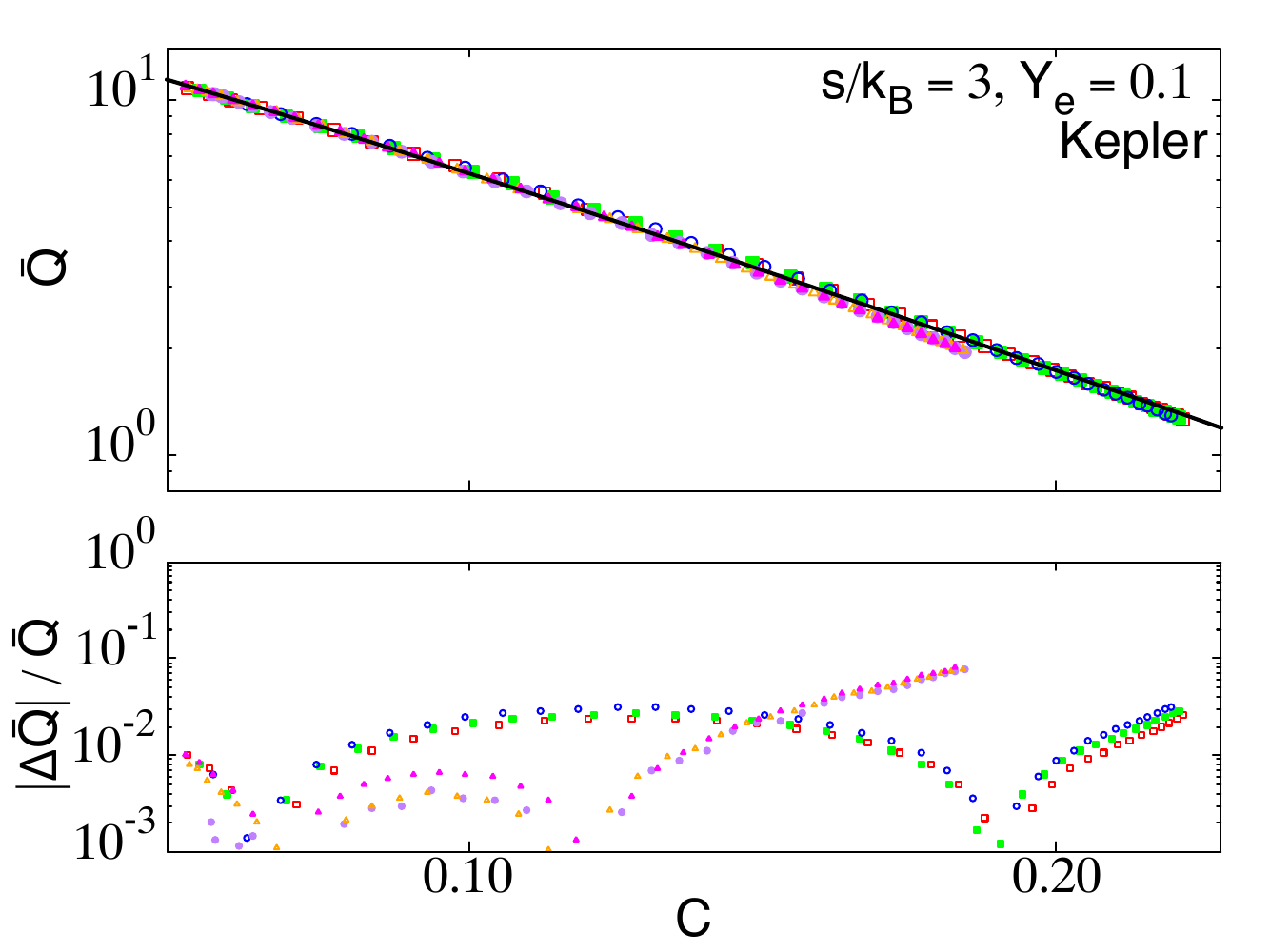}

\includegraphics[width=0.37\linewidth]{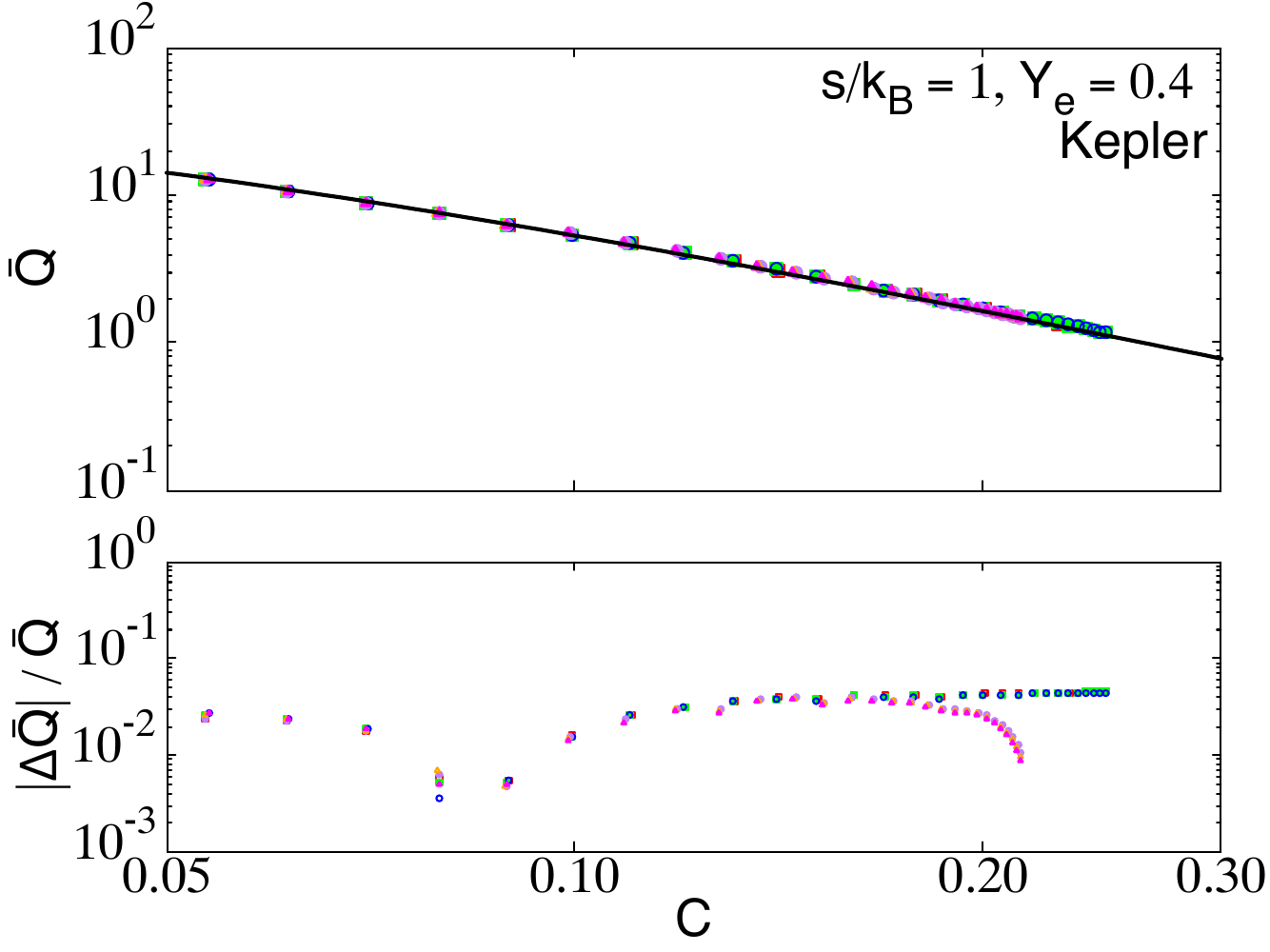}
\includegraphics[width=0.37\linewidth]{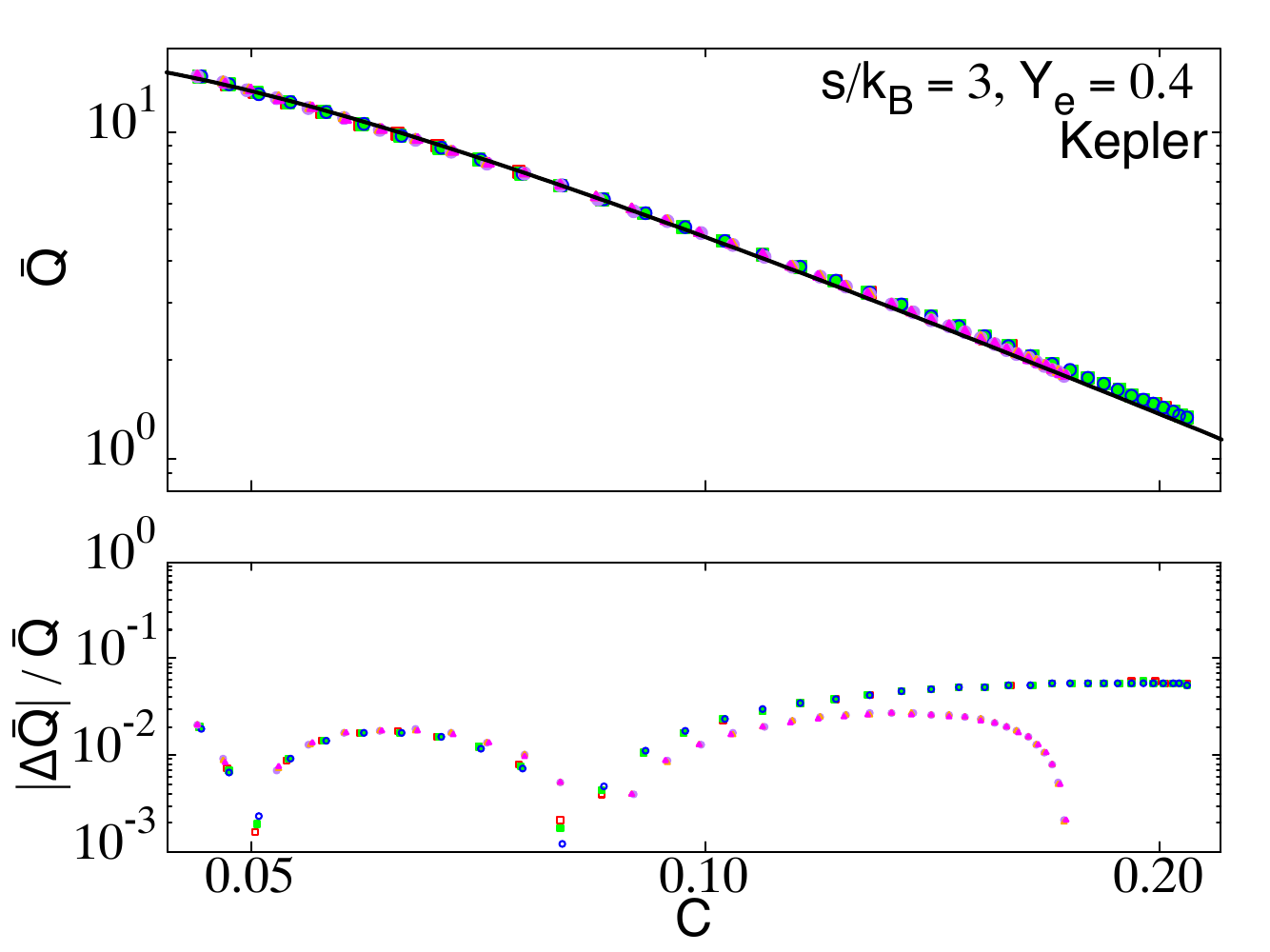}
    \caption{
    Same as in Fig.~\ref{fig:I_tilde_a_Fit_Error} but for the quadrupole moment $\bar{Q}(C)$ fitted using Eq.~\eqref{eq:QbarC}.
    }
    \label{fig:Q_bar_a_Fit_Error}
\end{figure*}

The third universal relation we are investigating involves the normalized quadrupole moment, defined as
    \bea
    \bar{Q} = -\frac{Q^{*}}{M^3\chi^2},
    \label{Qbardefinition}
    \eea
in which $Q^{*}$ is the dimensionful stellar mass quadrupole moment and $\chi=J/M^{2}$ is the dimensionless spin. As proposed in Ref.~\cite{Yagi:2013a}, this quantity can be expressed as a polynomial of powers of $C$, independent of the EoS. In the present work, however, we follow Ref.~\cite{Raduta_MNRAS_2020} and represent $\bar{Q}$ as a polynomial in negative powers of compactness, given by
    \bea\label{eq:QbarC}
    \bar{Q} = \sum_{0}^{j} c_{j} \left[C^{j}\right]^{-1}.
    \label{QbarC}
    \eea
Following the same data-fitting procedure as for the previous universal relations, we present in Fig.~\ref{fig:Q_bar_a_Fit_Error} the results for the same set of stellar sequences in the Keplerian case. For all parameter combinations, the quadratic term $(1/C)^{2}$ is found to provide the leading contribution. Furthermore, the high-entropy stellar sequences retain the property of requiring only two terms to accurately capture the universal relation between the normalized quadrupole moment and the compactness. The coefficients of the polynomial expansion in Eq.~\eqref{eq:QbarC} are provided in Table~\ref{tab:c_Kepler} of the Appendix.

\begin{figure*}[!]
\includegraphics[width=0.37\linewidth]{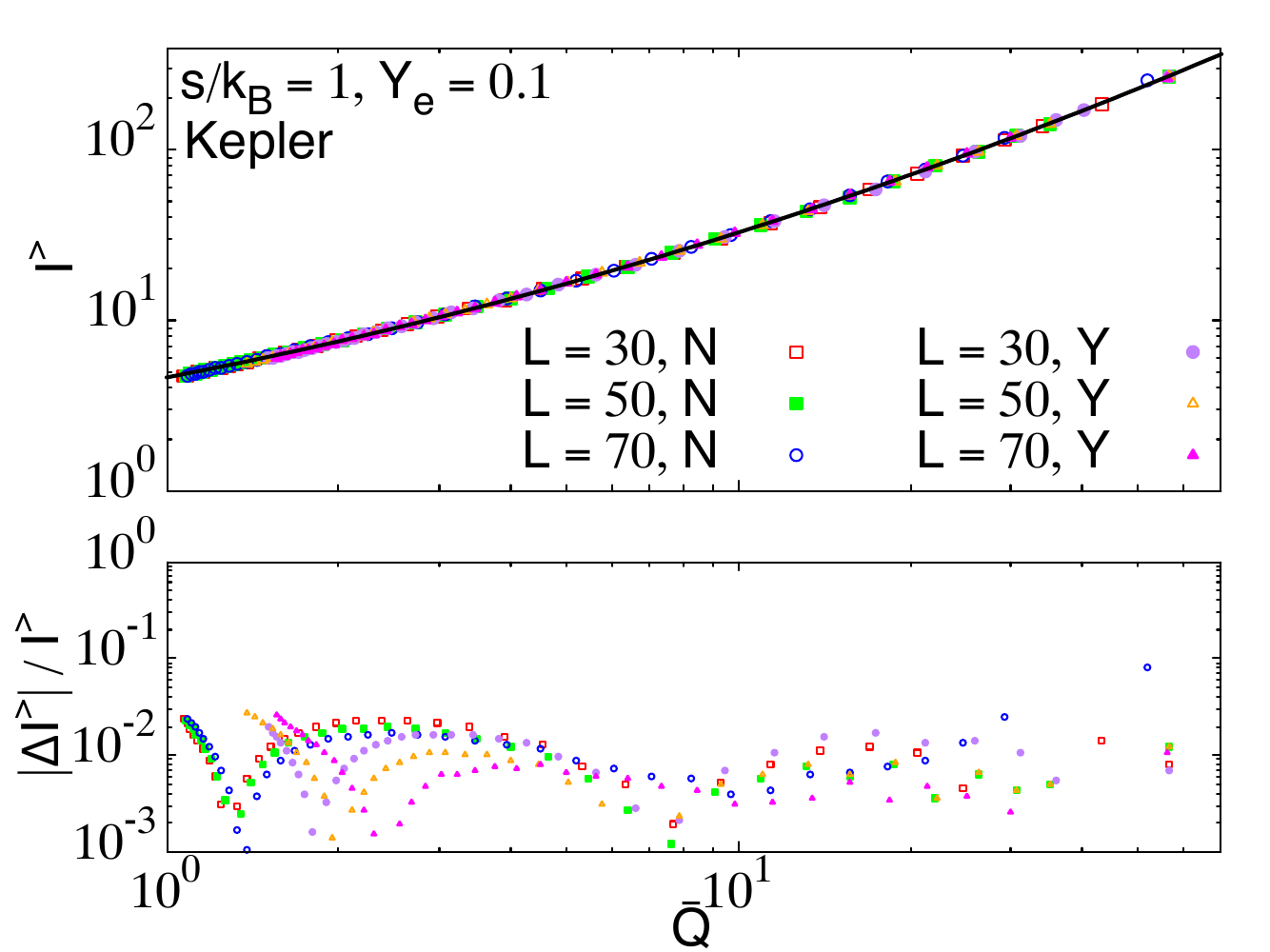}
\includegraphics[width=0.37\linewidth]{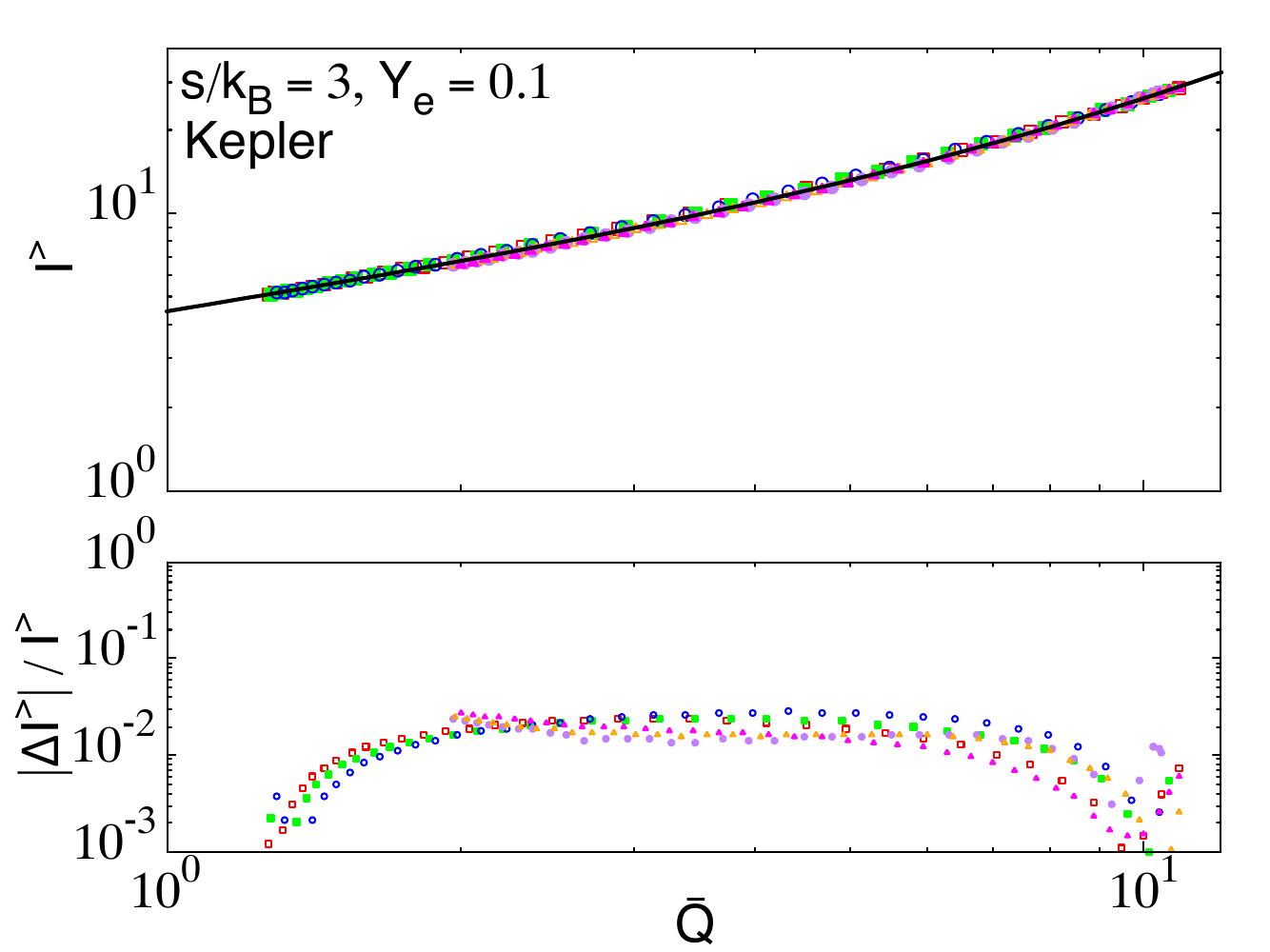}

\includegraphics[width=0.37\linewidth]{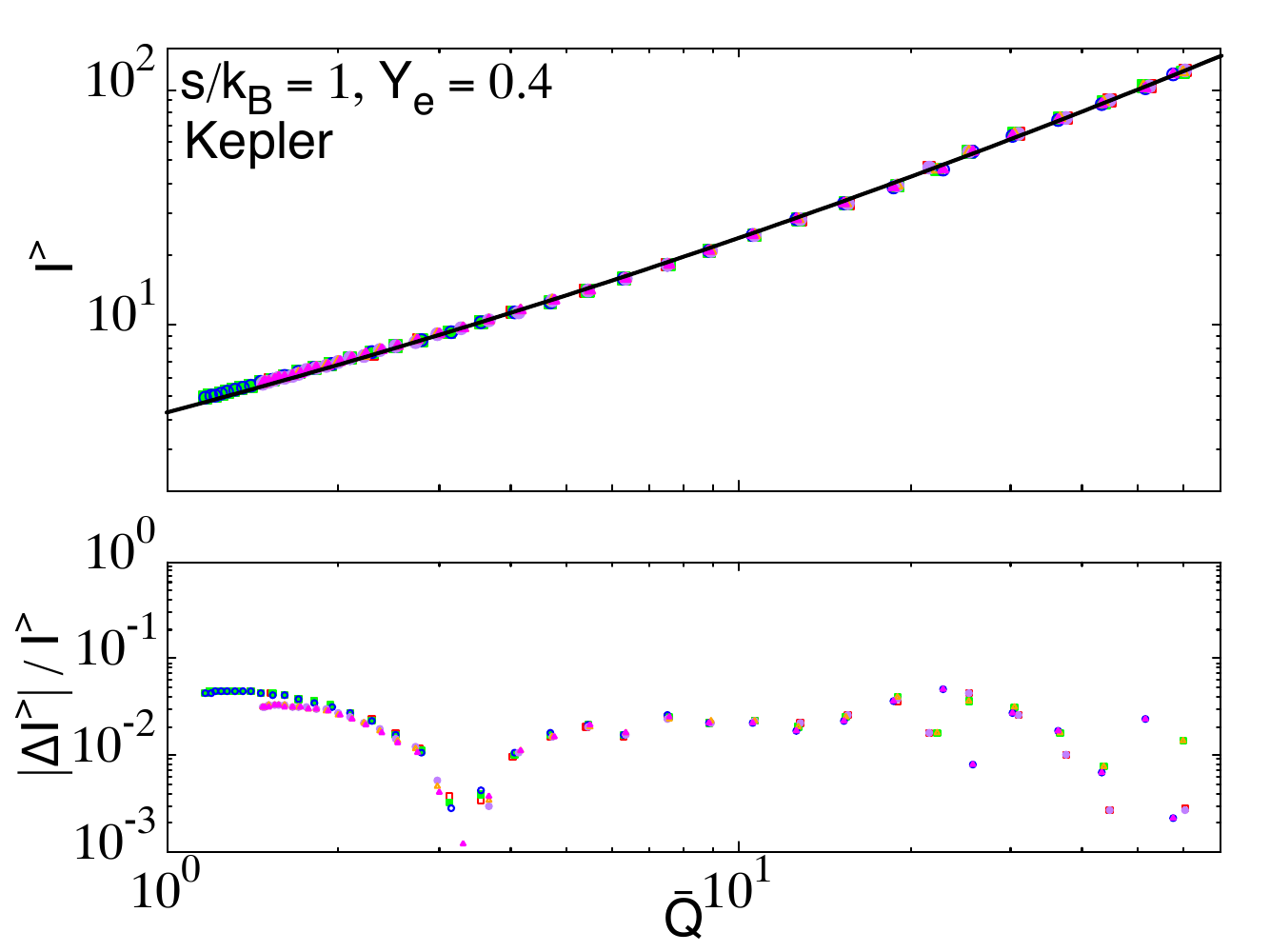}
\includegraphics[width=0.37\linewidth]{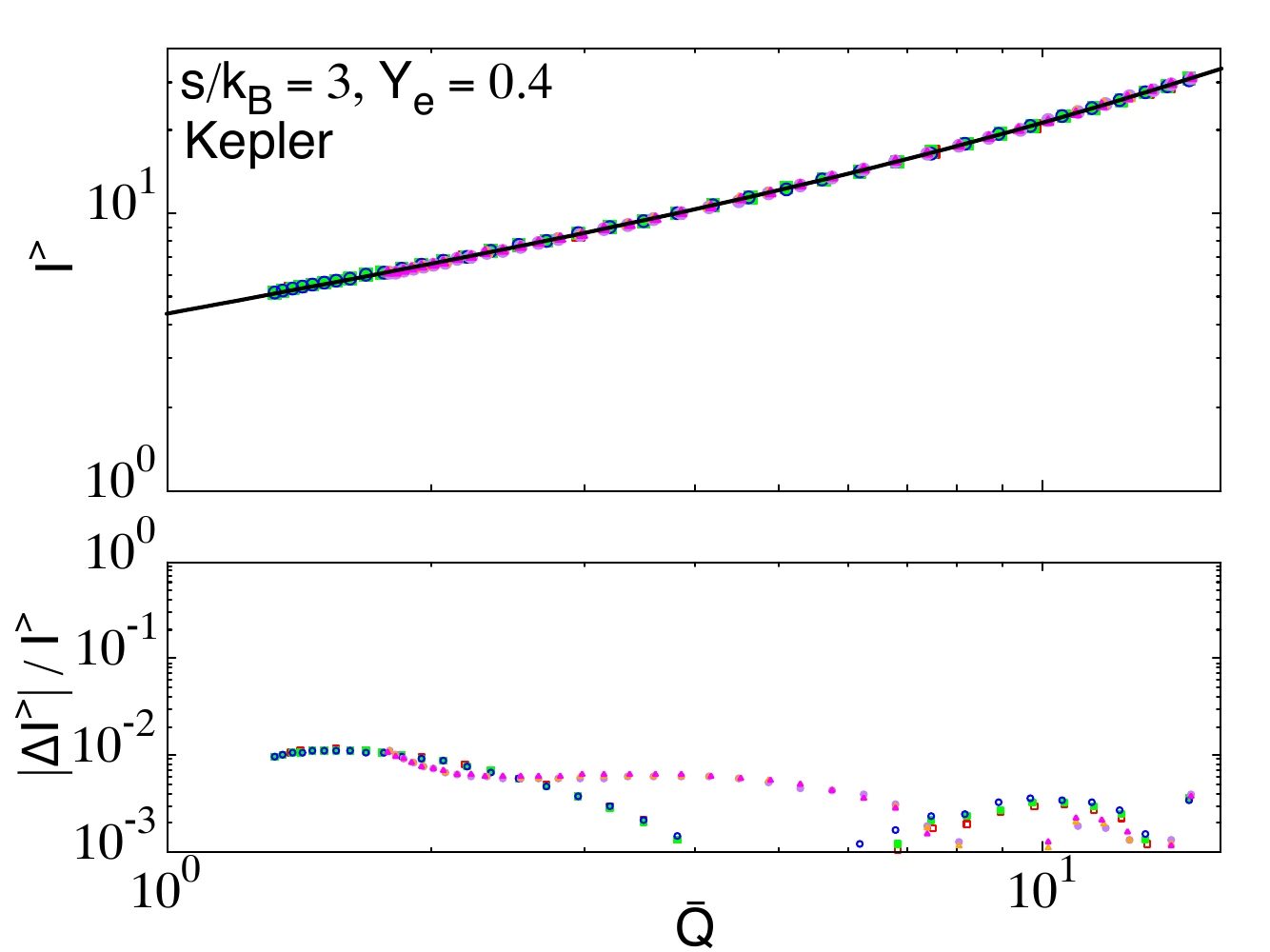}
    \caption{
    Same as in Fig.~\ref{fig:I_tilde_a_Fit_Error} but for dimensionless moment of inertia $I^{>}$ as a function of the dimensionless quadrupole moment $\bar{Q}$. }
    \label{fig:I_bar_Q_Fit_Error}
\end{figure*}

The next relation we examine is part of the $I$-Love-$Q$ universal relations, specifically, the one between the normalized moment of inertia $I^{>}$ and the normalized quadrupole moment $\bar{Q}$. Adopting the general form of those relations for slowly rotating stars, as introduced in Refs.~\cite{Yagi:2013a, Yagi:2013b}, we write 
    \bea\label{eq:IloveQ}
    \ln\,\bar I = \sum_{j=0}^{m}d_{j}\left(\ln\,\bar{Q}_{i}\right)^{j}.
    \eea 
    We then apply a numerical fitting procedure for the Keplerian stellar sequences that share a specific combination of $s/k_{B}, Y_{e}$ values. This choice is motivated by the proof~\cite{Doneva:2013rha} that those relations hold also for rapidly rotating stars, with the specific fit parameters modified accordingly to capture the effect of rotation. Fixing the thermodynamic conditions—the values of $s/k_{B}$ and $Y_{e}$—follows the requirement established in Ref.~\cite{Raduta_MNRAS_2020} for nonrotating stars, namely that universality is preserved only under specific thermodynamic conditions.
    The fit and the data are shown in Fig.~\ref{fig:I_bar_Q_Fit_Error} together with the relative deviation between the data points and the fitting curve derived from  Eq.~\eqref{eq:IloveQ} for all the thermodynamic combinations considered. The universal character of the relation is showcased by the remarkably low relative deviation of only a few percent for all cases. The details and exact values of the fitting parameters are included in Table~\ref{tab:d_Kepler}. Based on this conclusion, the validity of the $ I$–Love–$Q$ relations involving $\Lambda$ can be anticipated, especially in light of the strong correlations between $I^{>}$–$\Lambda$ and $\bar{Q}$–$\Lambda$ reported in previous studies~\cite{Yagi:2017}.

\begin{figure*}[!]
\includegraphics[width=0.37\linewidth]{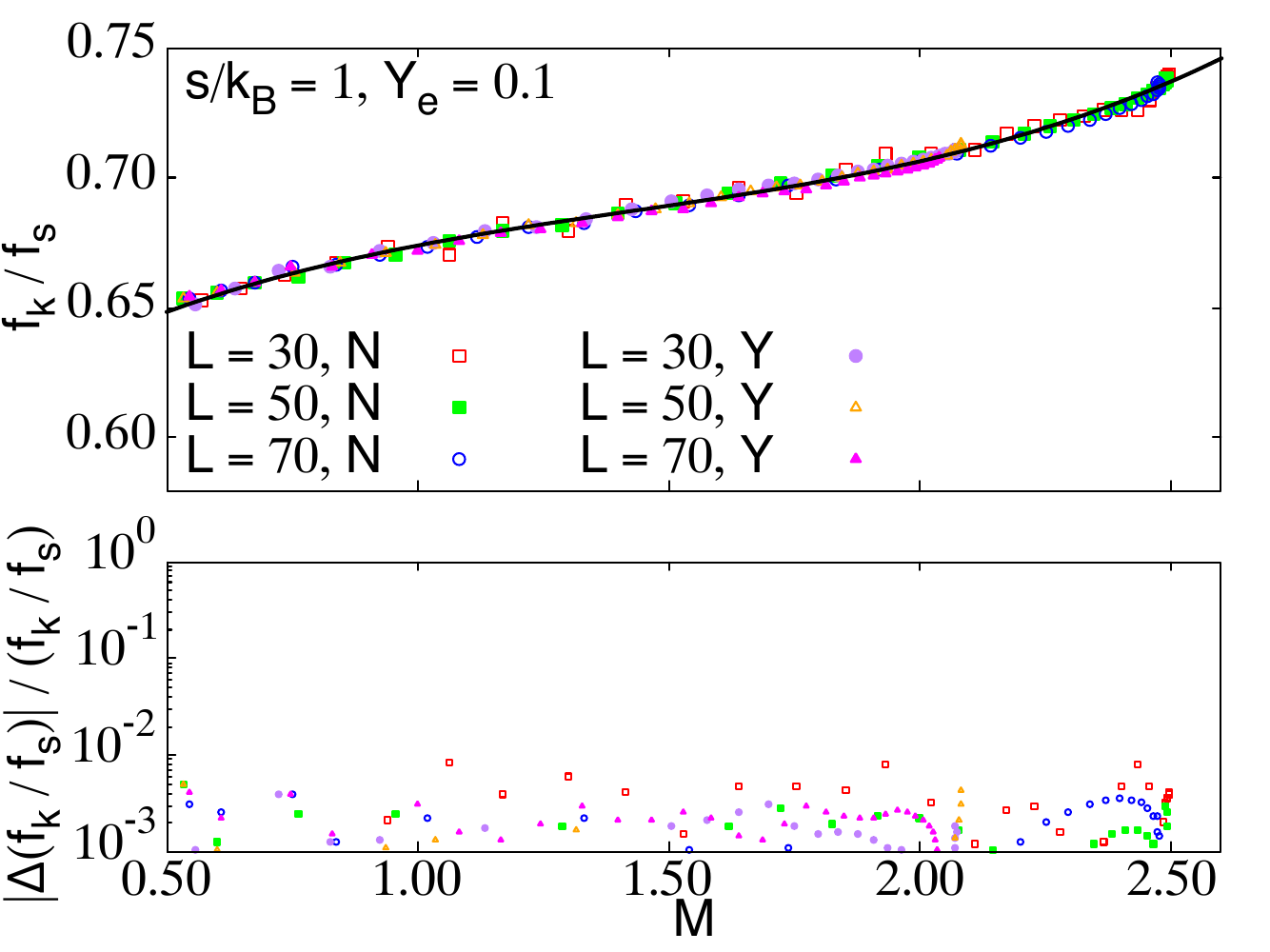}
\includegraphics[width=0.37\linewidth]{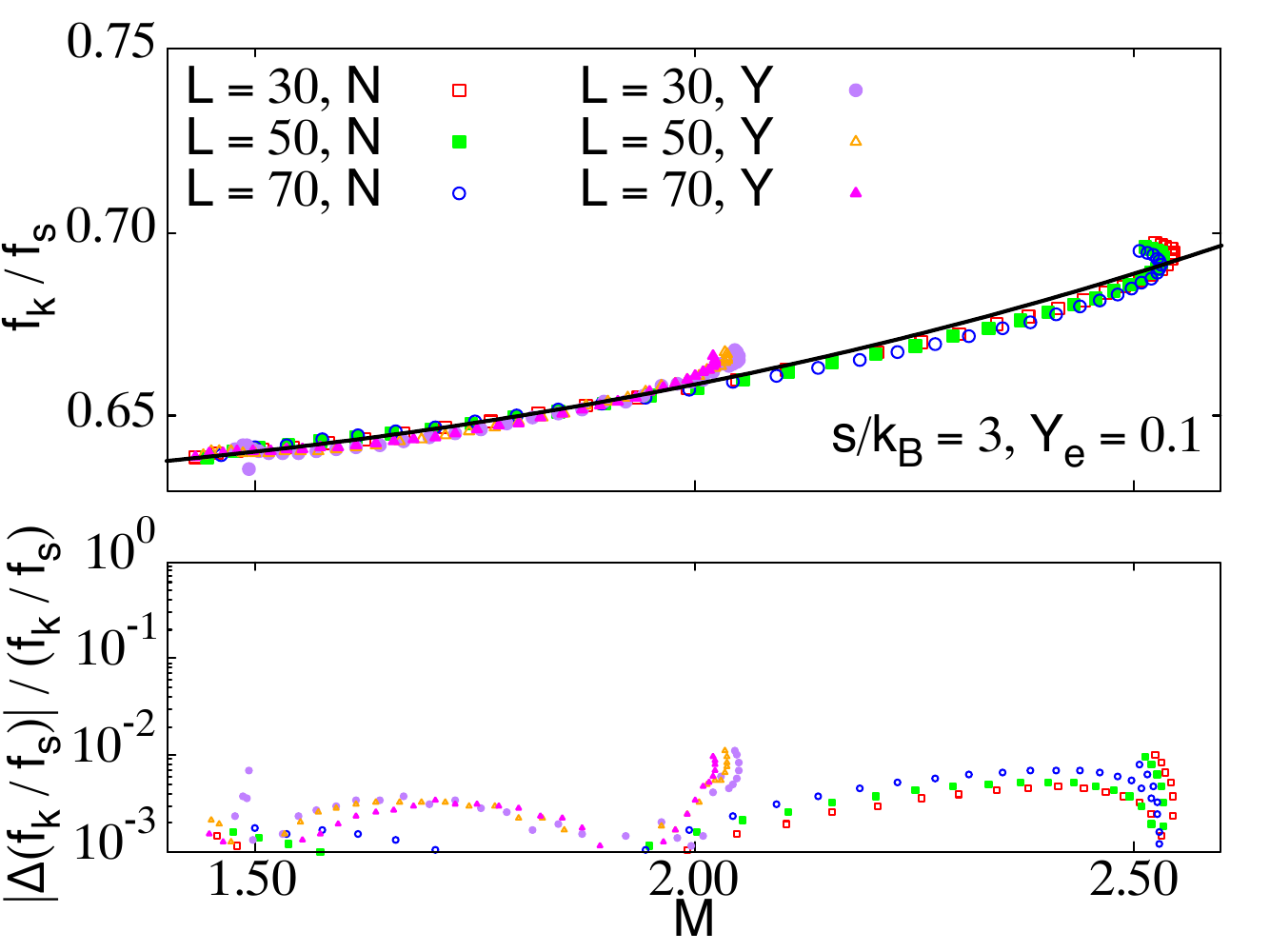}
\includegraphics[width=0.37\linewidth]{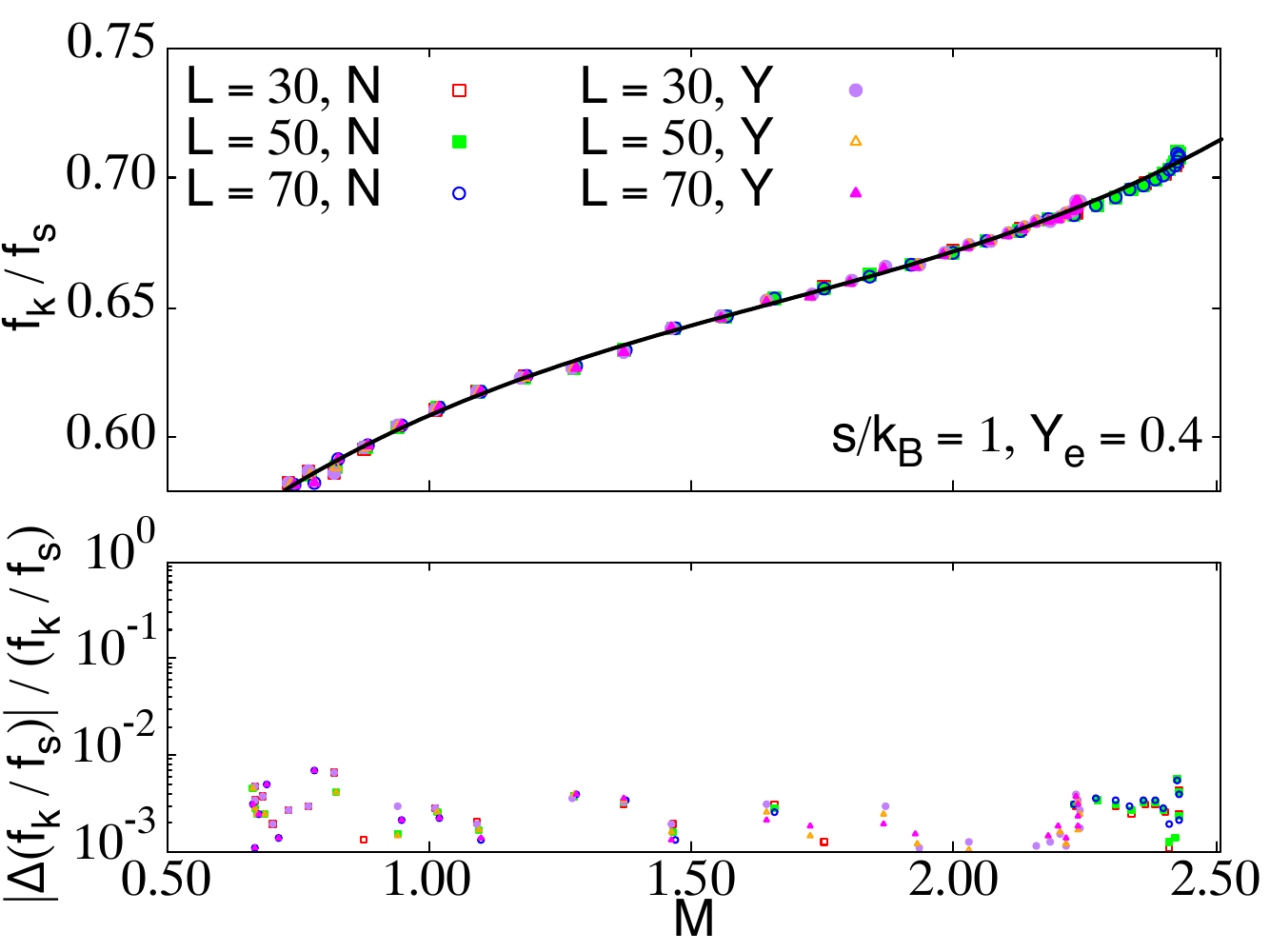}
\includegraphics[width=0.37\linewidth]{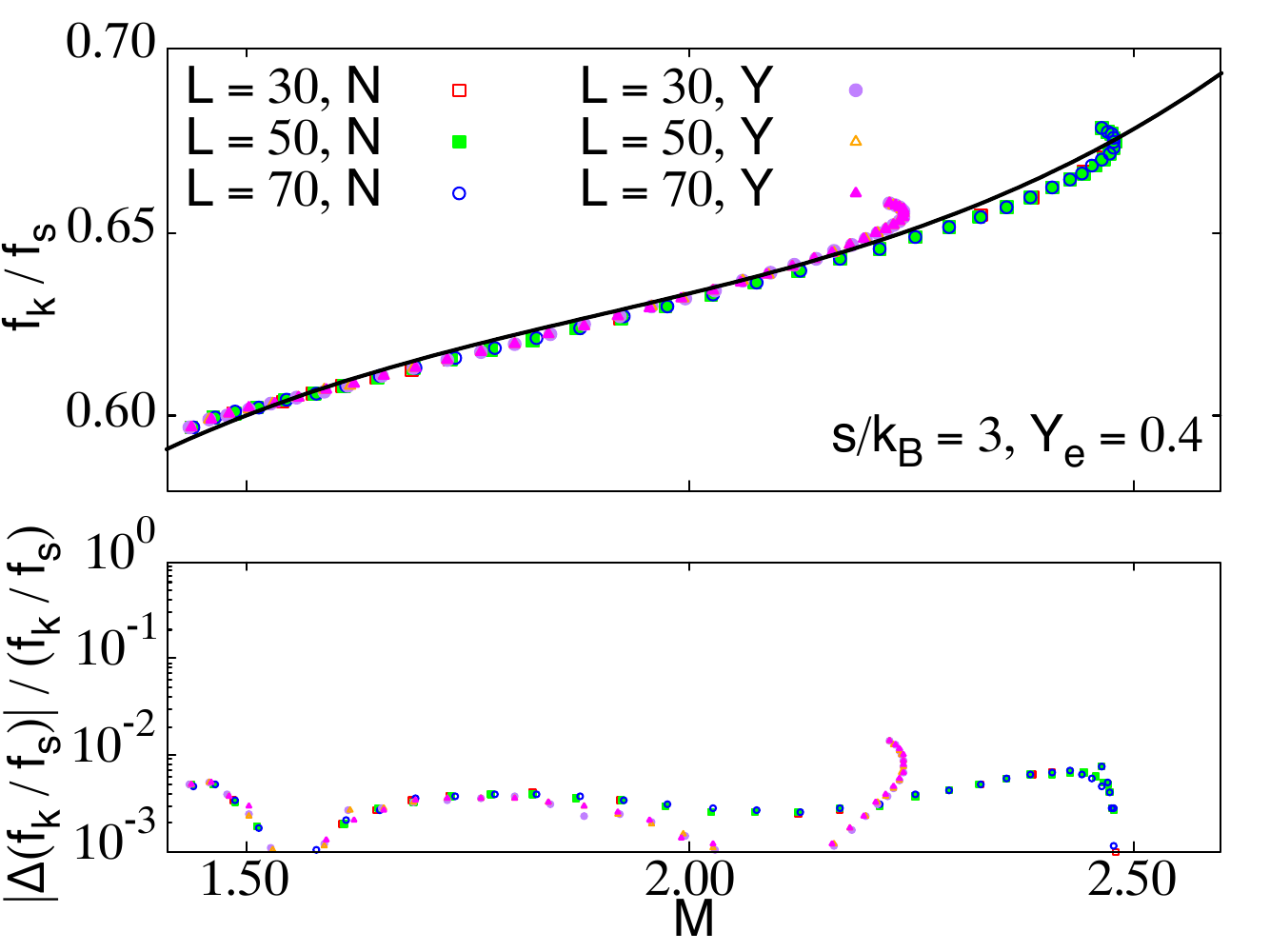}
    \caption{
The ratio of the Keplerian frequency of a maximally rotating NS, $f_{k}$, to the orbital frequency $f_{S}$ of a test particle in a circular equatorial orbit around a nonrotating star with the same central density is shown as a function of the mass $M$ of the static star. The lower panels display the relative residuals with respect to the fit obtained using Eqs.~\eqref{fratio_expand}. The remaining conventions are the same as in 
Fig.~\ref{fig:I_tilde_a_Fit_Error}.
}
\label{fig:freq_Ms_Fit_Error}
\end{figure*}

In their seminal work~\cite{Lattimer:2004}, Lattimer and Prakash proposed a scaling relation $f_K\approx 0.5701 f_{S}$ that was nearly EoS independent, where  $f_K$ is given by Eq.~\eqref{eq:Omega_K_scaling} and 
\bea\label{eq:f_S}
f_S = 1833\left(M/M_{\odot}\right)^{1/2}\left(10\,\text{km}/R\right)^{3/2} ~\text{Hz}
\eea
is the orbital frequency of a test particle in circular orbit 
around a spherical mass $M$ at orbital radius $R$. 
This  relation connects the maximum rotational 
frequency to the structural properties of the corresponding nonrotating reference star that shares the same central 
density as the rotating configuration; note that the frequency $f_S$  
has its origin in Newtonian orbital mechanics. 
To assess the accuracy of this
scaling relation, for our set of finite-temperature EoS, we have fitted the ratio 
\bea\label{fratio_expand}
\frac{f_{K}}{f_{S}} =  \sum_{j=0}^{m} u_j 
\left(\frac{M}{M_{\odot}}\right)^j.
\eea
with $m=3$.
Figure~\ref{fig:freq_Ms_Fit_Error} displays the  ratio $f_{\rm K}/f_S$ as a function of stellar mass $M$. It is seen  that the universality holds with quite high accuracy below 1\% in most  regimes, and deviations only arise close to the maximum mass  of the stellar sequence. Table~\ref{tab:u_Kepler} lists the fit  parameters for various combinations of the entropy per baryon and electron fraction, along with the error estimates.

\begin{figure}[H]
  \includegraphics[width=0.9\linewidth]{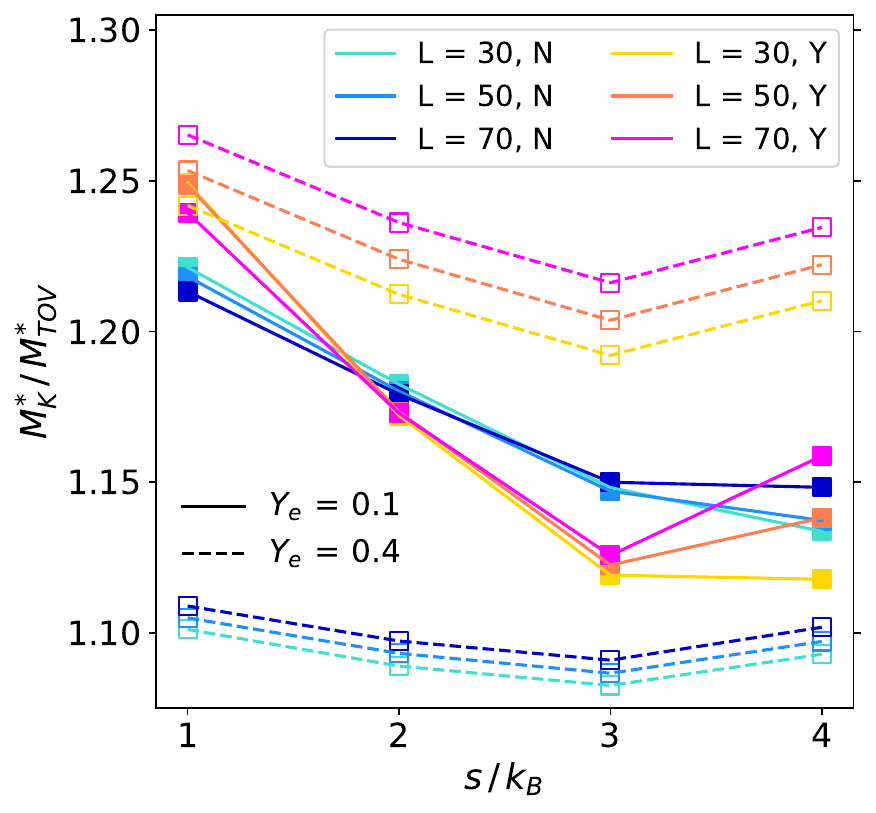}
    \caption{
      The ratio of $M_{\rm K}^{\star}$ to $M_{\rm TOV}^{\star}$   as a function of the entropy per baryon for two values of electron fraction
      $Y_e =0.1 $ and 0.4  for nucleonic and hyperonic compositions. 
}
    \label{fig:freq_Mmaxratios}
\end{figure} 
Next, we would like to address the problem of determining the maximum mass of cold NS from the gravitational wave events involving BNS mergers, using GW170817 as a prototype. Previously, various authors used the parameters extracted from this event to place an upper limit on the value of the maximum mass $M_{\rm TOV}^{\star}$ of static cold NS~\cite{Margalit2017,Rezzolla2018ApJ,Shibata2019PhRvD,Khadkikar:2021}.
The argument for an upper limit on $M_{\rm TOV}^{\star}$ in the GW170817 event proceeds as follows: the merger initially produces a HMNS with differential rotation. The HMNS spins down via gravitational and neutrino radiation and mass ejection, while internal dissipation removes differential rotation, leading to uniform rotation (magneto-dipole losses are negligible on ($\sim 10~{\rm ms}$) timescales).

At this stage, the star lies within the supramassive NS stability region, supported by uniform rotation. It subsequently crosses the stability line beyond which collapse occurs. While this crossing can, in principle, occur anywhere along the line connecting TOV mass $M_{\rm TOV}$ and the Keplerian mass $M_{\rm K}$, merger dynamics suggest it happens near $M_{\rm K}^{\star}$~see Ref.~\cite{Shibata2019PhRvD}, which questions this assumption and explores corrections (slower rotation allows a slightly higher maximum mass, so our estimate may be somewhat relaxed).  Reference~\cite{Khadkikar:2021} points out that the supramassive NS left by the BNS merger is hot; therefore, it is not identical to the Keplerian cold star. Therefore, its mass is not related to the $M_{\rm TOV}^{\star}$ mass via a universal relation; see their Fig.~7. Accounting for these effects relaxes the previously set limits on $M_{\rm TOV}^{\star}$, allowing for larger masses.

Given our set of EoS with systematic variations of $L_{\rm sym}$, we now reassess the relation between the cold NS maximum mass, $M_{\rm TOV}^{\star}$, and the maximum gravitational mass of the
{\it hot, isentropic} Keplerian configuration, $M_{\rm K}^{\star}$, for fixed $L_{\rm sym}$ and thermodynamic parameters. Figure~\ref{fig:freq_Mmaxratios} shows the ratio of these two masses as a function of entropy per baryon. It is evident that no universality holds for this ratio. For $Y_e=0.1$, the ratio varies within the range (1.15--1.22) for nucleonic stars and (1.12--1.25) for hyperonic stars, consistent with the ranges reported in Ref.~\cite{Khadkikar:2021}. Moreover, the larger range found for hyperonic stars aligns with their findings (see Fig.~10 in Ref.~\cite{Khadkikar:2021}).

The variations with $L_{\rm sym}$ are minor for nucleonic stars at fixed both $Y_e = 0.1$ and 0.4, but the ratio itself changes substantially when $Y_e$ is varied. A similar trend is observed for hyperonic configurations, although in this case the spread among different $L_{\rm sym}$ values is larger.

\section{Conclusions}
\label{sec:Conclusions} 

The objective of the present work was to investigate how variations in the nuclear symmetry energy slope parameter $L_{\rm sym}$ affect the global properties of compact stars at finite entropies and out of $\beta$-equilibrium, and to assess the universality of relations between these properties across different EoS, including both nuclear and hypernuclear matter EOS models. Our framework has important implications for the study of astrophysical transients such as supernovae and their associated PNS, as well as BNS  remnants. While we do not solve the full dynamical problem, taking “snapshots” of constant-entropy per baryon stages with a prescribed electron fraction provides valuable insight into the state of matter and the global properties of both static and Keplerian stellar configurations.

Our study of the mass-radius relation, which provides the key constraints on the static (or slowly rotating) compact objects, reveals significant structural differences between nucleonic and hyperonic stars at finite temperature. As expected, the inclusion of hyperons systematically softens the EoS, leading to reduced maximum masses for both static and rotating configurations. Let us remind that static $\beta$-equilibrated stars show consistency with NICER observations for canonical NSs ($1.4M_{\odot}$) and massive ones ($2M_{\odot}$). Rapidly rotating $\beta$-equilibrated nucleonic stars can achieve masses approaching three solar masses, making them viable candidates for ``mass-gap" compact objects observed in gravitational wave events like GW190814 and GW230529.  The key structural trends revealed by this study are: (a)~finite-temperature stars with lower $L_{\rm sym}$ values exhibit, as a rule, higher maximum masses, consistent with the zero-temperature case; (b)~isentropic stars generally have larger radii than cold counterparts due to envelope expansion, the radius being larger the larger is the value of $L_{\rm sym}$, except close to the maximum mass, where the trend reverses; (some exceptions still exist, such as the kink for $Y_e=0.1$ and $s/k_B =3$ for hypernuclear star in Fig.~\ref{fig:MR_fixed_sYe}--it can be traced to the matching point of the low- and high-density equations of state); (c)~increasing electron fraction from $Y_e=0.1$ to $Y_e=0.4$ causes pronounced envelope expansion and radius increase.

We have examined several universal relations that have been extensively studied at zero temperature across a variety of EoS models, both for nucleonic and hyperonic compact stars. Our work can be used in conjunction with  Ref.~\cite{Yeasin2025}, which employed zero-temperature nucleonic  EoS only, but the same CDF parametrization as ours, with systematic variations of the symmetry energy slope parameter $L_{\rm sym}$ and skewness. Our work extends this study to the case of finite-temperature isentropic stars (with the skewness parameter fixed in the present study).

We have additionally studied stellar configurations in the static and Keplerian limits, which bracket the sequences of rigidly rotating stars. The static limit provides a good approximation for most observed compact stars (e.g., pulsars), whereas the Keplerian limit is of interest not only as an extreme case, but also because of its relevance for constraining the maximum cold TOV mass from BNS postmerger remnants~\cite{Rezzolla2018ApJ,Shibata2019PhRvD,Khadkikar:2021}.
Our analysis addressed four types of universal relations: (a) the normalized moment of inertia expressed as a polynomial function of the stellar compactness; (b) an alternative relation employing a different normalization of the moment of inertia expanded as a polynomial in inverse powers of compactness;  (c) the logarithm of the normalized quadrupole moment expressed as a polynomial function of the logarithm of the moment of inertia, and finally (d) the ratio $f_K/f_S$ as function of static mass $M$ for 
same-central density static and Keplerian stars. We find that universality holds for these relations to better than 10\% accuracy across both static and maximally rotating (Keplerian) configurations, independent of entropy per particle or electron fraction for both nucleonic and hyperonic compositions. We find that for the polynomial expansions, sequences with higher entropy require fewer terms in the polynomial expansion for prescribed precision than those with lower entropy. Comprehensive tables with the fit coefficients for both static and rotating stars are provided in the Appendix to facilitate practical application of these universal relations in astrophysical modeling.

Finally, our analysis confirms that the inference of an upper bound on the cold TOV mass from GW170817 is more subtle than originally assumed~\cite{Khadkikar:2021}. The hot, supramassive remnant formed in a BNS merger cannot be mapped onto cold Keplerian sequences through a universal relation, as previously suggested. By systematically varying $L_{\rm sym}$ and considering both nucleonic and hyperonic compositions, we show that the ratio $M_{\rm K}^{\star}/M_{\rm TOV}^{\star}$ is not universal but depends on the thermodynamic conditions. For nucleonic stars, variations with $L_{\rm sym}$ are small, but the ratio changes significantly with $Y_e$. Hyperonic stars display a broader spread in this ratio, consistent with earlier findings~\cite{Khadkikar:2021}.

These findings have important implications for attempts to constrain the cold-star maximum mass using observations of hot postmerger remnants, such as those expected from future GW170817-like events. In particular, our results show that the maximum mass of a hot, rapidly rotating, isentropic star cannot be directly mapped onto the maximum mass of its cold, $\beta$-equilibrated counterpart through a universal relation. Consequently, the procedure employed in earlier studies--where a universal scaling between hot and cold maximum masses was assumed--requires corrections that take into account thermal and compositional effects.

\section*{Acknowledgments}
This work was  supported by the Polish National Science Center (NCN) Grant No. 2023/51/B/ST9/02798  and the Deutsche Forschungsgemeinschaft (DFG) Grant No. SE 1836/6-1 (A.~S and S.~T.). S.~T. is a member of the IMPRS for ``Quantum dynamics and Control'' at the Max Planck Institute for the Physics of Complex Systems, Dresden, Germany, and acknowledges its partial support. M.O. acknowledges support from the Agence Nationale de la Recherche (ANR) under Contract ANR-22-CE31-0001-01. 

\clearpage
\newpage
\appendix
\section{Tables of fit parameters}

\begin{table}[hbt]
\centering
\begin{tabular}{|c|c|c|c|c|}
\hline
\multicolumn{5}{|c|}{Fit parameters for $I^{<}(C)$ function for static stars} \\
\hline
$s/k_B$ & $Y_e$ & Parameter & Value & Asymptotic Standard Error \\
\hline
\multirow{10}{*}{1} & \multirow{6}{*}{0.1} 
  & $a_0$ & 0.131875 & $\pm 0.002136$ (1.62\%) \\
  &  & $a_1$ & 2.38623  & $\pm 0.04735$  (1.984\%) \\
  &  & $a_2$ & -6.65268 & $\pm 0.2385$   (3.585\%) \\
  &  & $a_4$ & 27.931   & $\pm 1.26$     (4.511\%) \\
  &  &  $\sigma$ & \multicolumn{2}{c|}{0.00621559} \\
  &  & $\chi^2$ & \multicolumn{2}{c|}{$3.86336 \times 10^{-5}$} \\
\cline{2-5}
  & \multirow{6}{*}{0.4} 
  & $a_0$ & 0.0189494 & $\pm 0.0005678$ (2.997\%) \\
  &  & $a_1$ & 1.82268   & $\pm 0.01046$  (0.5741\%) \\
  &  & $a_3$ & -15.0892  & $\pm 0.4789$   (3.174\%) \\
  &  & $a_4$ & 34.2486   & $\pm 1.389$    (4.055\%) \\
  &  &  $\sigma$ & \multicolumn{2}{c|}{0.00205755} \\
  &  & $\chi^2$ & \multicolumn{2}{c|}{$4.2335 \times 10^{-6}$} \\
\hline
\multirow{6}{*}{3} & \multirow{4}{*}{0.1} 
  & $a_0$ & 0.212531  & $\pm 0.0009303$ (0.4377\%) \\
  &  & $a_2$ & 2.39658   & $\pm 0.02544$  (1.061\%) \\
  &  &  $\sigma$ & \multicolumn{2}{c|}{0.0073166} \\
  &  & $\chi^2$ & \multicolumn{2}{c|}{$5.35326 \times 10^{-5}$} \\
\cline{2-5}
  & \multirow{4}{*}{0.4} 
  & $a_0$ & 0.0539777 & $\pm 0.0005996$ (1.111\%) \\
  &  & $a_1$ & 1.12242   & $\pm 0.003675$ (0.3275\%) \\
  &  &  $\sigma$ & \multicolumn{2}{c|}{0.00319011} \\
  &  & $\chi^2$ & \multicolumn{2}{c|}{$1.01768 \times 10^{-5}$} \\
\hline
\end{tabular}
\caption{Final fit parameters for the normalized moment of inertia $I^{<}(C)$ function for static stars with different entropy per particle ($s/k_B$) and electron fraction ($Y_e$). Each coefficient $a_j$ is listed alongside its asymptotic standard error, including the absolute $\pm$ value and the corresponding percentage error. The root mean square (RMS) of residuals $\sigma$ and reduced $\chi^2$ values are provided for each case. Fits converged after 4 to 6 iterations, depending on the case.}
\label{tab:a_stat}
\end{table}


\begin{table}[hbt]
\centering
\begin{tabular}{|c|c|c|c|c|}
\hline
\multicolumn{5}{|c|}{Fit parameters for $I^{<}(C)$ function for Keplerian stars} \\
\hline
$s/k_B$ & $Y_e$ & Parameter & Value & Asymptotic Standard Error \\
\hline
\multirow{7}{*}{1} & \multirow{5}{*}{0.1} 
  & $a_0$ & 0.0779513 & $\pm 0.002022$ (2.594\%) \\
  &  & $a_1$ & 2.08629 & $\pm 0.04755$ (2.279\%) \\
  &  & $a_2$ & -5.5578 & $\pm 0.2534$ (4.559\%) \\
  &  & $a_4$ & 25.739 & $\pm 1.502$ (5.836\%) \\
  &  &  $\sigma$ & \multicolumn{2}{c|}{0.00543607} \\
  &  & $\chi^2$ & \multicolumn{2}{c|}{$2.95509 \times 10^{-5}$} \\
\cline{2-5}
  & \multirow{5}{*}{0.4} 
  & $a_0$ & 0.00898396 & $\pm 0.0003964$ (4.412\%) \\
  &  & $a_1$ & 1.37539 & $\pm 0.007886$ (0.5734\%) \\
  &  & $a_2$ & -0.833162 & $\pm 0.03086$ (3.704\%) \\
  &  &  $\sigma$ & \multicolumn{2}{c|}{0.00192407} \\
  &  & $\chi^2$ & \multicolumn{2}{c|}{$3.70205 \times 10^{-6}$} \\
\hline
\multirow{5}{*}{3} & \multirow{3}{*}{0.1} 
  & $a_0$ & 0.135205 & $\pm 0.0007279$ (0.5384\%) \\
  &  & $a_2$ & 2.69406 & $\pm 0.02826$ (1.049\%) \\
  &  &  $\sigma$ & \multicolumn{2}{c|}{0.00599537} \\
  &  & $\chi^2$ & \multicolumn{2}{c|}{$3.59445 \times 10^{-5}$} \\
\cline{2-5}
  & \multirow{3}{*}{0.4} 
  & $a_0$ & 0.0208091 & $\pm 0.0004114$ (1.977\%) \\
  &  & $a_1$ & 0.974588 & $\pm 0.003172$ (0.3255\%) \\
  &  &  $\sigma$ & \multicolumn{2}{c|}{0.00234427} \\
  &  & $\chi^2$ & \multicolumn{2}{c|}{$5.49562 \times 10^{-6}$} \\
\hline
\end{tabular}
\caption{Final fit parameters for the normalized moment of inertia function $I^{<}(C)$ for Keplerian stars for different values $s/k_B$ and $Y_e$. Each coefficient $a_j$ is listed with its asymptotic standard error, showing both absolute $\pm$ values and percentage errors as well as $\sigma$ and reduced $\chi^2$. The fits converged after 3 to 5 iterations, depending on the case.}
\label{tab:a_Kepler}
\end{table}


\begin{table}[hbt]
\centering
\begin{tabular}{|c|c|c|c|c|}
\hline
\multicolumn{5}{|c|}{Fit parameters for $I^{>}(C)$ function for static stars} \\
\hline
$s/k_B$ & $Y_e$ & Parameter & Value & Asymptotic Standard Error \\
\hline
\multirow{8}{*}{1} & \multirow{5}{*}{0.1} 
  & $b_0$ & 0.724474 & $\pm 0.01161$ (1.603\%) \\
  &  & $b_1$ & 0.263631 & $\pm 0.001547$ (0.5869\%) \\
  &  & $b_2$ & -0.00352393 & $\pm 5.724 \times 10^{-5}$ (1.624\%) \\
  &  & $b_3$ & 2.6299 $\times 10^{-5}$ & $\pm 6.007 \times 10^{-7}$ (2.284\%) \\
  &  &  $\sigma$ & \multicolumn{2}{c|}{0.390008} \\
  &  & $\chi^2$ & \multicolumn{2}{c|}{0.152106} \\
\cline{2-5}
  & \multirow{8}{*}{0.4} 
  & $b_0$ & 1.2457 & $\pm 0.008549$ (0.6863\%) \\
  &  & $b_1$ & 0.0830036 & $\pm 0.001364$ (1.643\%) \\
  &  & $b_2$ & -0.00244892 & $\pm 6.145 \times 10^{-5}$ (2.509\%) \\
  &  & $b_3$ & 3.06685 $\times 10^{-5}$ & $\pm 8.197 \times 10^{-7}$ (2.673\%) \\
  &  &  $\sigma$ & \multicolumn{2}{c|}{0.229879} \\
  &  & $\chi^2$ & \multicolumn{2}{c|}{0.0528442} \\
\hline
\multirow{6}{*}{3} & \multirow{4}{*}{0.1} 
  & $b_1$ & 1.29701 & $\pm 0.006178$ (0.4764\%) \\
  &  & $b_3$ & 0.0106201 & $\pm 7.613 \times 10^{-5}$ (0.7169\%) \\
  &  &  $\sigma$ & \multicolumn{2}{c|}{0.284654} \\
  &  & $\chi^2$ & \multicolumn{2}{c|}{0.0810278} \\
\cline{2-5}
  & \multirow{5}{*}{0.4} 
  & $b_1$ & 1.14871 & $\pm 0.003049$ (0.2654\%) \\
  &  & $b_2$ & 0.0510812 & $\pm 0.0002636$ (0.516\%) \\
  &  &  $\sigma$ & \multicolumn{2}{c|}{0.123203} \\
  &  & $\chi^2$ & \multicolumn{2}{c|}{0.0151789} \\
\hline
\end{tabular}
\caption{Final fit parameters for the normalized moment of inertia function $I^{>}(C)$ for static stars for different values $s/k_B$ and $Y_e$. Each coefficient $b_j$ is listed with its asymptotic standard error, showing both absolute $\pm$ values and percentage errors as well as $\sigma$ and reduced $\chi^2$. The fits converged after 3 to 5 iterations, depending on the case. }
\label{tab:b_stat}
\end{table}
\begin{table}[hbt]
\centering
\begin{tabular}{|c|c|c|c|c|}
\hline
\multicolumn{5}{|c|}{Final fit parameters for $I^{>}(C)$ function for Keplerian stars} \\
\hline
$s/k_B$ & $Y_e$ & Parameter & Value & Asymptotic Standard Error \\
\hline
\multirow{6}{*}{1} & \multirow{5}{*}{0.1} 
  & $b_1$ & 0.773127 & $\pm 0.02213$ (2.863\%) \\
  &  & $b_2$ & 0.177592 & $\pm 0.003404$ (1.917\%) \\
  &  & $b_3$ & $-0.00270834$ & $\pm 0.0001442$ (5.325\%) \\
  &  & $b_4$ & $2.20504 \times 10^{-5}$ & $\pm 1.737 \times 10^{-6}$ (7.876\%) \\
  &  &  $\sigma$ & \multicolumn{2}{c|}{0.646099} \\
  &  & $\chi^2$ & \multicolumn{2}{c|}{0.417444} \\
\cline{2-5}
  & \multirow{6}{*}{0.4} 
  & $b_1$ & 1.16578 & $\pm 0.005418$ (0.4647\%) \\
  &  & $b_2$ & 0.0252728 & $\pm 0.0005984$ (2.368\%) \\
  &  & $b_3$ & $-0.000507427$ & $\pm 1.883 \times 10^{-5}$ (3.71\%) \\
  &  & $b_4$ & $4.99221 \times 10^{-6}$ & $\pm 1.755 \times 10^{-7}$ (3.515\%) \\
  &  &  $\sigma$ & \multicolumn{2}{c|}{0.207446} \\
  &  & $\chi^2$ & \multicolumn{2}{c|}{0.043034} \\
\hline
\multirow{4}{*}{3} & \multirow{3}{*}{0.1} 
  & $b_1$ & 1.09515 & $\pm 0.005682$ (0.5188\%) \\
  &  & $b_3$ & 0.00511678 & $\pm 4.506 \times 10^{-5}$ (0.8806\%) \\
  &  &  $\sigma$ & \multicolumn{2}{c|}{0.323431} \\
  &  & $\chi^2$ & \multicolumn{2}{c|}{0.104607} \\
\cline{2-5}
  & \multirow{3}{*}{0.4} 
  & $b_1$ & 0.97371 & $\pm 0.001805$ (0.1854\%) \\
  &  & $b_2$ & 0.0210514 & $\pm 0.0001133$ (0.538\%) \\
  &  &  $\sigma$ & \multicolumn{2}{c|}{0.099143} \\
  &  & $\chi^2$ & \multicolumn{2}{c|}{0.00982933} \\
\hline
\end{tabular}
\caption{Final fit parameters for the normalized moment of inertia function $I^{>}(C)$ for Keplerian stars for different values of $s/k_B$ and $Y_e$. Each coefficient $b_j$ is listed with its asymptotic standard error, showing both absolute $\pm$ values and percentage errors as well as $\sigma$ and reduced $\chi^2$. Fits converged after 
3 to 5 iterations, depending on the case.}
\label{tab:b_Kepler}
\end{table}


\begin{table}[hbt]
\centering
\begin{tabular}{|c|c|c|c|c|}
\hline
\multicolumn{5}{|c|}{Fit parameters for $\bar{Q}(C)$ function for Keplerian stars} \\
\hline
$s/k_B$ & $Y_e$ & Parameter & Value & Asymptotic Standard Error \\
\hline
\multirow{8}{*}{1} 
  & 0.1 & $c_2$ & 0.092219 & $\pm 0.0007614$ (0.8257\%) \\
  &     & $c_3$ & $-0.00265984$ & $\pm 4.568 \times 10^{-5}$ (1.717\%) \\
  &     & $c_4$ & $2.57633 \times 10^{-5}$ & $\pm 6.345 \times 10^{-7}$ (2.463\%) \\
  &     & $\sigma$ & \multicolumn{2}{c|}{0.409734} \\
  &     & $\chi^2$ & \multicolumn{2}{c|}{0.167882} \\
\cline{2-5}
  & 0.4 & $c_2$ & 0.0802341 & $\pm 0.0006126$ (0.7635\%) \\
  &     & $c_3$ & $-0.0032548$ & $\pm 4.817 \times 10^{-5}$ (1.48\%) \\
  &     & $c_4$ & $5.83428 \times 10^{-5}$ & $\pm 1.205 \times 10^{-6}$ (2.065\%) \\
  &     & $c_5$ & $-3.64188 \times 10^{-7}$ & $\pm 9.576 \times 10^{-9}$ (2.63\%) \\
  &     & $\sigma$ & \multicolumn{2}{c|}{0.26174} \\
  &     & $\chi^2$ & \multicolumn{2}{c|}{0.0685078} \\
\hline
\multirow{5}{*}{3} 
  & 0.1 & $c_2$ & 0.0766376 & $\pm 0.0003013$ (0.3932\%) \\
  &     & $c_3$ & $-0.0014566$ & $\pm 2.544 \times 10^{-5}$ (1.747\%) \\
  &     & $\sigma$ & \multicolumn{2}{c|}{0.0776467} \\
  &     & $\chi^2$ & \multicolumn{2}{c|}{0.006029} \\
\cline{2-5}
  & 0.4 & $c_2$ & 0.0618689 & $\pm 0.0001961$ (0.317\%) \\
  &     & $c_3$ & $-0.00143371$ & $\pm 1.069 \times 10^{-5}$ (0.745\%) \\
  &     & $\sigma$ & \multicolumn{2}{c|}{0.106413} \\
  &     & $\chi^2$ & \multicolumn{2}{c|}{0.0113238} \\
\hline
\end{tabular}
\caption{Final fit parameters for the normalized quadrupole moment function $\bar{Q}(C)$ for Keplerian stars for different values of $s/k_B$ and $Y_e$. Each coefficient $c_j$ is listed with its asymptotic standard error,  showing both absolute $\pm$ values and percentage errors as well as $\sigma$ and reduced $\chi^2$. Fits converged after 8 to 10 iterations for $s/k_B = 1$ and 4 to 5 iterations for  $s/k_B = 3$.}
\label{tab:c_Kepler}
\end{table}

\begin{table}[hbt]
\centering
\begin{tabular}{|c|c|c|c|c|}
\hline
\multicolumn{5}{|c|}{Fit parameters for $I^{>}(\bar{Q})$ function for Keplerian stars} \\
\hline
$s/k_B$ & $Y_e$ & Parameter & Value & Asymptotic Standard Error \\
\hline
\multirow{6}{*}{1} 
  & 0.1 & $d_0$ & 1.52565 & $\pm 0.02288$ (1.5\%) \\
  &     & $d_1$ & 0.641557 & $\pm 0.01494$ (2.329\%) \\
  &     & $d_2$ & 0.0915705 & $\pm 0.002432$ (2.656\%) \\
  &     & $\sigma$ & \multicolumn{2}{c|}{1.39406} \\
  &     & $\chi^2$ & \multicolumn{2}{c|}{1.9434} \\
\cline{2-5}
  & 0.4 & $d_0$ & 1.45064 & $\pm 0.01411$ (0.9723\%) \\
  &     & $d_1$ & 0.643363 & $\pm 0.00988$ (1.536\%) \\
  &     & $d_2$ & 0.0414671 & $\pm 0.00170$ (4.099\%) \\
  &     & $\sigma$ & \multicolumn{2}{c|}{0.696258} \\
  &     & $\chi^2$ & \multicolumn{2}{c|}{0.484775} \\
\hline
\multirow{6}{*}{3} 
  & 0.1 & $d_0$ & 1.48988 & $\pm 0.005991$ (0.4021\%) \\
  &     & $d_1$ & 0.589783 & $\pm 0.005797$ (0.9829\%) \\
  &     & $d_3$ & 0.0343887 & $\pm 0.0006794$ (1.976\%) \\
  &     & $\sigma$ & \multicolumn{2}{c|}{0.181995} \\
  &     & $\chi^2$ & \multicolumn{2}{c|}{0.0331221} \\
\cline{2-5}
  & 0.4 & $d_0$ & 1.46976 & $\pm 0.001549$ (0.1054\%) \\
  &     & $d_1$ & 0.591108 & $\pm 0.001356$ (0.2294\%) \\
  &     & $d_3$ & 0.0190608 & $\pm 0.0001278$ (0.6707\%) \\
  &     & $\sigma$ & \multicolumn{2}{c|}{0.0520311} \\
  &     & $\chi^2$ & \multicolumn{2}{c|}{0.00270724} \\
\hline
\end{tabular}
\caption{Final fit parameters for the normalized moment of inertia function $I^{>}(\bar{Q})$ for Keplerian stars for various values of $s/k_B$ and $Y_e$. Parameters $d_j$ are listed with their asymptotic standard errors showing both absolute $\pm$ values and percentage errors as well as $\sigma$ and reduced $\chi^2$. Fits converged after 5 to 7 iterations for $s/k_B=1$ and 4 to 6 iterations for $s/k_B=3$. }
\label{tab:d_Kepler}
\end{table}


\begin{table}[hbt]
\centering
\begin{tabular}{|c|c|l|l|l|}
\hline
\multicolumn{5}{|c|}{Fit parameters for the $f_k / f_s(M)$ relation for stellar sequences} \\
\hline
$s/k_B$ & $Y_e$ & Parameter & Value & Asymptotic Standard Error \\
\hline
\multirow{12}{*}{1} 
  & 0.1 & $u_0$ & 0.600765 & $\pm 0.0008552$ (0.1423\%) \\
  &     & $u_1$ & 0.125411 & $\pm 0.002426$ (1.935\%) \\
  &     & $u_2$ & -0.0681173 & $\pm 0.001882$ (2.764\%) \\
  &     & $u_3$ & 0.0159017 & $\pm 0.0004325$ (2.72\%) \\
  &     &  $\sigma$ & \multicolumn{2}{c|}{0.00160819} \\
  &     & $\chi^2$ & \multicolumn{2}{c|}{$2.58629 \times 10^{-6}$} \\
\cline{2-5}
  & 0.4 & $u_0$ & 0.443066 & $\pm 0.002645$ (0.5969\%) \\
  &     & $u_1$ & 0.268923 & $\pm 0.006012$ (2.236\%) \\
  &     & $u_2$ & -0.129426 & $\pm 0.004127$ (3.188\%) \\
  &     & $u_3$ & 0.0260622 & $\pm 0.0008811$ (3.381\%) \\
  &     &  $\sigma$ & \multicolumn{2}{c|}{0.00157744} \\
  &     & $\chi^2$ & \multicolumn{2}{c|}{$2.48831 \times 10^{-6}$} \\
\hline
\multirow{8}{*}{3} 
  & 0.1 & $u_0$ & 0.62709 & $\pm 0.0003721$ (0.05934\%) \\
  &     & $u_3$ & 0.00394009 & $\pm 3.952 \times 10^{-5}$ (1.003\%) \\
  &     &  $\sigma$ & \multicolumn{2}{c|}{0.00261356} \\
  &     & $\chi^2$ & \multicolumn{2}{c|}{$6.83072 \times 10^{-6}$} \\
\cline{2-5}
  & 0.4 & $u_1$ & 0.878086 & $\pm 0.004564$ (0.5198\%) \\
  &     & $u_2$ & -0.432365 & $\pm 0.004614$ (1.067\%) \\
  &     & $u_3$ & 0.0758315 & $\pm 0.001144$ (1.509\%) \\
  &     &  $\sigma$ & \multicolumn{2}{c|}{0.00309912} \\
  &     & $\chi^2$ & \multicolumn{2}{c|}{$9.60454 \times 10^{-6}$} \\
\hline
\end{tabular}
\caption{Final fit parameters of the $f_k / f_s(M_s)$ relation for stellar sequences at various values of  $s/k_B$ and $Y_e$. Parameters $u_j$ are given with their asymptotic standard errors (absolute and relative) along with $\sigma$ and $\chi^2$ 
Fits converged after 3 to 6 iterations, depending on the case.
}
\label{tab:u_Kepler}
\end{table}

\clearpage
\newpage

\begin{thebibliography}{86}%
\makeatletter
\providecommand \@ifxundefined [1]{%
 \@ifx{#1\undefined}
}%
\providecommand \@ifnum [1]{%
 \ifnum #1\expandafter \@firstoftwo
 \else \expandafter \@secondoftwo
 \fi
}%
\providecommand \@ifx [1]{%
 \ifx #1\expandafter \@firstoftwo
 \else \expandafter \@secondoftwo
 \fi
}%
\providecommand \natexlab [1]{#1}%
\providecommand \enquote  [1]{``#1''}%
\providecommand \bibnamefont  [1]{#1}%
\providecommand \bibfnamefont [1]{#1}%
\providecommand \citenamefont [1]{#1}%
\providecommand \href@noop [0]{\@secondoftwo}%
\providecommand \href [0]{\begingroup \@sanitize@url \@href}%
\providecommand \@href[1]{\@@startlink{#1}\@@href}%
\providecommand \@@href[1]{\endgroup#1\@@endlink}%
\providecommand \@sanitize@url [0]{\catcode `\\12\catcode `\$12\catcode
  `\&12\catcode `\#12\catcode `\^12\catcode `\_12\catcode `\%12\relax}%
\providecommand \@@startlink[1]{}%
\providecommand \@@endlink[0]{}%
\providecommand \url  [0]{\begingroup\@sanitize@url \@url }%
\providecommand \@url [1]{\endgroup\@href {#1}{\urlprefix }}%
\providecommand \urlprefix  [0]{URL }%
\providecommand \Eprint [0]{\href }%
\providecommand \doibase [0]{http://dx.doi.org/}%
\providecommand \selectlanguage [0]{\@gobble}%
\providecommand \bibinfo  [0]{\@secondoftwo}%
\providecommand \bibfield  [0]{\@secondoftwo}%
\providecommand \translation [1]{[#1]}%
\providecommand \BibitemOpen [0]{}%
\providecommand \bibitemStop [0]{}%
\providecommand \bibitemNoStop [0]{.\EOS\space}%
\providecommand \EOS [0]{\spacefactor3000\relax}%
\providecommand \BibitemShut  [1]{\csname bibitem#1\endcsname}%
\let\auto@bib@innerbib\@empty
\bibitem [{\citenamefont {Lattimer}\ and\ \citenamefont
  {Prakash}(2016)}]{Lattimer:2015nhk}%
  \BibitemOpen
  \bibfield  {author} {\bibinfo {author} {\bibfnamefont {J.~M.}\ \bibnamefont
  {Lattimer}}\ and\ \bibinfo {author} {\bibfnamefont {M.}~\bibnamefont
  {Prakash}},\ }\href {\doibase 10.1016/j.physrep.2015.12.005} {\bibfield
  {journal} {\bibinfo  {journal} {Phys. Rept.}\ }\textbf {\bibinfo {volume}
  {621}},\ \bibinfo {pages} {127} (\bibinfo {year} {2016})},\ \Eprint
  {http://arxiv.org/abs/1512.07820} {arXiv:1512.07820 [astro-ph.SR]}
  \BibitemShut {NoStop}%
\bibitem [{\citenamefont {Oertel}\ \emph {et~al.}(2017)\citenamefont {Oertel},
  \citenamefont {Hempel}, \citenamefont {Klahn},\ and\ \citenamefont
  {Typel}}]{Oertel_RMP_2017}%
  \BibitemOpen
  \bibfield  {author} {\bibinfo {author} {\bibfnamefont {M.}~\bibnamefont
  {Oertel}}, \bibinfo {author} {\bibfnamefont {M.}~\bibnamefont {Hempel}},
  \bibinfo {author} {\bibfnamefont {T.}~\bibnamefont {Klahn}}, \ and\ \bibinfo
  {author} {\bibfnamefont {S.}~\bibnamefont {Typel}},\ }\href {\doibase
  10.1103/RevModPhys.89.015007} {\bibfield  {journal} {\bibinfo  {journal}
  {Rev. Mod. Phys.}\ }\textbf {\bibinfo {volume} {89}},\ \bibinfo {pages}
  {015007} (\bibinfo {year} {2017})}\BibitemShut {NoStop}%
\bibitem [{\citenamefont {{Sedrakian}}\ \emph {et~al.}(2023)\citenamefont
  {{Sedrakian}}, \citenamefont {{Li}},\ and\ \citenamefont
  {{Weber}}}]{Sedrakian2023}%
  \BibitemOpen
  \bibfield  {author} {\bibinfo {author} {\bibfnamefont {A.}~\bibnamefont
  {{Sedrakian}}}, \bibinfo {author} {\bibfnamefont {J.~J.}\ \bibnamefont
  {{Li}}}, \ and\ \bibinfo {author} {\bibfnamefont {F.}~\bibnamefont
  {{Weber}}},\ }\href {\doibase 10.1016/j.ppnp.2023.104041} {\bibfield
  {journal} {\bibinfo  {journal} {Progress in Particle and Nuclear Physics}\
  }\textbf {\bibinfo {volume} {131}},\ \bibinfo {eid} {104041} (\bibinfo {year}
  {2023})},\ \Eprint {http://arxiv.org/abs/2212.01086} {arXiv:2212.01086
  [nucl-th]} \BibitemShut {NoStop}%
\bibitem [{\citenamefont {Cromartie}\ \emph {et~al.}(2020)\citenamefont
  {Cromartie}, \citenamefont {Fonseca}, \citenamefont {Ransom} \emph
  {et~al.}}]{NANOGrav:2019}%
  \BibitemOpen
  \bibfield  {author} {\bibinfo {author} {\bibfnamefont {H.~T.}\ \bibnamefont
  {Cromartie}}, \bibinfo {author} {\bibfnamefont {E.}~\bibnamefont {Fonseca}},
  \bibinfo {author} {\bibfnamefont {S.~M.}\ \bibnamefont {Ransom}},  \emph
  {et~al.} (\bibinfo {collaboration} {NANOGrav Collaboration}),\ }\href {\doibase
  10.1038/s41550-019-0880-2} {\bibfield  {journal} {\bibinfo  {journal} {Nat.
  Astron.}\ }\textbf {\bibinfo {volume} {4}},\ \bibinfo {pages} {72} (\bibinfo
  {year} {2020})},\ \Eprint {http://arxiv.org/abs/1904.06759} {arXiv:1904.06759
  [astro-ph.HE]} \BibitemShut {NoStop}%
\bibitem [{\citenamefont {Fonseca}\ \emph {et~al.}(2021)\citenamefont {Fonseca}
  \emph {et~al.}}]{Fonseca:2021}%
  \BibitemOpen
  \bibfield  {author} {\bibinfo {author} {\bibfnamefont {E.}~\bibnamefont
  {Fonseca}} \emph {et~al.},\ }\href {\doibase 10.3847/2041-8213/ac03b8}
  {\bibfield  {journal} {\bibinfo  {journal} {Astrophys. J. Lett.}\ }\textbf
  {\bibinfo {volume} {915}},\ \bibinfo {pages} {L12} (\bibinfo {year}
  {2021})},\ \Eprint {http://arxiv.org/abs/2104.00880} {arXiv:2104.00880
  [astro-ph.HE]} \BibitemShut {NoStop}%
\bibitem [{\citenamefont {Riley}\ \emph {et~al.}(2021)\citenamefont {Riley},
  \citenamefont {Watts}, \citenamefont {Ray} \emph {et~al.}}]{NICER:2021a}%
  \BibitemOpen
  \bibfield  {author} {\bibinfo {author} {\bibfnamefont {T.~E.}\ \bibnamefont
  {Riley}}, \bibinfo {author} {\bibfnamefont {A.~L.}\ \bibnamefont {Watts}},
  \bibinfo {author} {\bibfnamefont {P.~S.}\ \bibnamefont {Ray}},  \emph
  {et~al.},\ }\href {\doibase 10.3847/2041-8213/ac0a81} {\bibfield  {journal}
  {\bibinfo  {journal} {ApJL}\ }\textbf {\bibinfo {volume} {918}},\ \bibinfo
  {pages} {L27} (\bibinfo {year} {2021})},\ \Eprint
  {http://arxiv.org/abs/2105.06980} {arXiv:2105.06980 [astro-ph.HE]}
  \BibitemShut {NoStop}%
\bibitem [{\citenamefont {Miller}\ \emph {et~al.}(2021)\citenamefont {Miller},
  \citenamefont {Lamb}, \citenamefont {Dittmann} \emph {et~al.}}]{NICER:2021b}%
  \BibitemOpen
  \bibfield  {author} {\bibinfo {author} {\bibfnamefont {M.~C.}\ \bibnamefont
  {Miller}}, \bibinfo {author} {\bibfnamefont {F.~K.}\ \bibnamefont {Lamb}},
  \bibinfo {author} {\bibfnamefont {A.~J.}\ \bibnamefont {Dittmann}},  \emph
  {et~al.},\ }\href {\doibase 10.3847/2041-8213/ac089b} {\bibfield  {journal}
  {\bibinfo  {journal} {ApJL}\ }\textbf {\bibinfo {volume} {918}},\ \bibinfo
  {pages} {L28} (\bibinfo {year} {2021})},\ \Eprint
  {http://arxiv.org/abs/2105.06979} {arXiv:2105.06979 [astro-ph.HE]}
  \BibitemShut {NoStop}%
\bibitem [{\citenamefont {Abbott}\ \emph {et~al.}(2017)\citenamefont {Abbott}
  \emph {et~al.}}]{Abbott_2017}%
  \BibitemOpen
  \bibfield  {author} {\bibinfo {author} {\bibfnamefont {B.~P.}\ \bibnamefont
  {Abbott}} \emph {et~al.} (\bibinfo {collaboration} {LIGO Scientific
  Collaboration and Virgo Collaboration}),\ }\href {\doibase
  10.1103/PhysRevLett.119.161101} {\bibfield  {journal} {\bibinfo  {journal}
  {Phys. Rev. Lett.}\ }\textbf {\bibinfo {volume} {119}},\ \bibinfo {pages}
  {161101} (\bibinfo {year} {2017})}\BibitemShut {NoStop}%
\bibitem [{\citenamefont {Adhikari}\ \emph {et~al.}(2021)\citenamefont
  {Adhikari}, \citenamefont {Albataineh}, \citenamefont {Androic},
  \citenamefont {Aniol}, \citenamefont {Armstrong} \emph
  {et~al.}}]{PREX-II:2021}%
  \BibitemOpen
  \bibfield  {author} {\bibinfo {author} {\bibfnamefont {D.}~\bibnamefont
  {Adhikari}}, \bibinfo {author} {\bibfnamefont {H.}~\bibnamefont
  {Albataineh}}, \bibinfo {author} {\bibfnamefont {D.}~\bibnamefont {Androic}},
  \bibinfo {author} {\bibfnamefont {K.}~\bibnamefont {Aniol}}, \bibinfo
  {author} {\bibfnamefont {D.~S.}\ \bibnamefont {Armstrong}},  \emph {et~al.}
  (\bibinfo {collaboration} {PREX Collaboration}),\ }\href {\doibase
  10.1103/PhysRevLett.126.172502} {\bibfield  {journal} {\bibinfo  {journal}
  {PhRvL}\ }\textbf {\bibinfo {volume} {126}},\ \bibinfo {pages} {172502}
  (\bibinfo {year} {2021})},\ \Eprint {http://arxiv.org/abs/2102.10767}
  {arXiv:2102.10767 [nucl-ex]} \BibitemShut {NoStop}%
\bibitem [{\citenamefont {{Reinhard}}\ \emph {et~al.}(2021)\citenamefont
  {{Reinhard}}, \citenamefont {{Roca-Maza}},\ and\ \citenamefont
  {{Nazarewicz}}}]{Reinhard:2021}%
  \BibitemOpen
  \bibfield  {author} {\bibinfo {author} {\bibfnamefont {P.-G.}\ \bibnamefont
  {{Reinhard}}}, \bibinfo {author} {\bibfnamefont {X.}~\bibnamefont
  {{Roca-Maza}}}, \ and\ \bibinfo {author} {\bibfnamefont {W.}~\bibnamefont
  {{Nazarewicz}}},\ }\href {\doibase 10.1103/PhysRevLett.127.232501} {\bibfield
   {journal} {\bibinfo  {journal} {Phys. Rev. Lett.}\ }\textbf {\bibinfo
  {volume} {127}},\ \bibinfo {pages} {232501} (\bibinfo {year} {2021})},\
  \Eprint {http://arxiv.org/abs/2105.15050} {arXiv:2105.15050 [nucl-th]}
  \BibitemShut {NoStop}%
\bibitem [{\citenamefont {Adhikari}\ \emph {et~al.}(2022)\citenamefont
  {Adhikari}, \citenamefont {Albataineh}, \citenamefont {Androic} \emph
  {et~al.}}]{CREX:2022}%
  \BibitemOpen
  \bibfield  {author} {\bibinfo {author} {\bibfnamefont {D.}~\bibnamefont
  {Adhikari}}, \bibinfo {author} {\bibfnamefont {H.}~\bibnamefont
  {Albataineh}}, \bibinfo {author} {\bibfnamefont {D.}~\bibnamefont {Androic}},
   \emph {et~al.} (\bibinfo {collaboration} {CREX Collaboration}),\ }\href {\doibase
  10.1103/PhysRevLett.129.042501} {\bibfield  {journal} {\bibinfo  {journal}
  {PhRvL}\ }\textbf {\bibinfo {volume} {129}},\ \bibinfo {pages} {042501}
  (\bibinfo {year} {2022})},\ \Eprint {http://arxiv.org/abs/2205.11593}
  {arXiv:2205.11593 [nucl-ex]} \BibitemShut {NoStop}%
\bibitem [{\citenamefont {Lattimer}(2023)}]{Lattimer:2023}%
  \BibitemOpen
  \bibfield  {author} {\bibinfo {author} {\bibfnamefont {J.~M.}\ \bibnamefont
  {Lattimer}},\ }\href {\doibase 10.3390/particles6010003} {\bibfield
  {journal} {\bibinfo  {journal} {Particles}\ }\textbf {\bibinfo {volume}
  {6}},\ \bibinfo {pages} {30} (\bibinfo {year} {2023})},\ \Eprint
  {http://arxiv.org/abs/2301.03666} {arXiv:2301.03666 [nucl-th]} \BibitemShut
  {NoStop}%
\bibitem [{\citenamefont {Estee}\ \emph {et~al.}(2021)\citenamefont {Estee},
  \citenamefont {Lynch}, \citenamefont {Tsang} \emph {et~al.}}]{SpiRIT:2021}%
  \BibitemOpen
  \bibfield  {author} {\bibinfo {author} {\bibfnamefont {J.}~\bibnamefont
  {Estee}}, \bibinfo {author} {\bibfnamefont {W.~G.}\ \bibnamefont {Lynch}},
  \bibinfo {author} {\bibfnamefont {C.~Y.}\ \bibnamefont {Tsang}},  \emph
  {et~al.} (\bibinfo {collaboration} {SpiRIT Collaboration}),\ }\href {\doibase
  10.1103/PhysRevLett.126.162701} {\bibfield  {journal} {\bibinfo  {journal}
  {Phys. Rev. Lett.}\ }\textbf {\bibinfo {volume} {126}},\ \bibinfo {pages}
  {162701} (\bibinfo {year} {2021})},\ \Eprint
  {http://arxiv.org/abs/2103.06861} {arXiv:2103.06861 [nucl-ex]} \BibitemShut
  {NoStop}%
\bibitem [{\citenamefont {Giacalone}\ \emph {et~al.}(2023)\citenamefont
  {Giacalone}, \citenamefont {Nijs},\ and\ \citenamefont {van~der
  Schee}}]{Giacalone:2023}%
  \BibitemOpen
  \bibfield  {author} {\bibinfo {author} {\bibfnamefont {G.}~\bibnamefont
  {Giacalone}}, \bibinfo {author} {\bibfnamefont {G.}~\bibnamefont {Nijs}}, \
  and\ \bibinfo {author} {\bibfnamefont {W.}~\bibnamefont {van~der Schee}},\
  }\href {\doibase 10.1103/PhysRevLett.131.202302} {\bibfield  {journal}
  {\bibinfo  {journal} {Phys. Rev. Lett.}\ }\textbf {\bibinfo {volume} {131}},\
  \bibinfo {pages} {202302} (\bibinfo {year} {2023})},\ \Eprint
  {http://arxiv.org/abs/2305.00015} {arXiv:2305.00015 [nucl-th]} \BibitemShut
  {NoStop}%
\bibitem [{\citenamefont {{Shen}}\ \emph {et~al.}(1998)\citenamefont {{Shen}},
  \citenamefont {{Toki}}, \citenamefont {{Oyamatsu}},\ and\ \citenamefont
  {{Sumiyoshi}}}]{Shen1998}%
  \BibitemOpen
  \bibfield  {author} {\bibinfo {author} {\bibfnamefont {H.}~\bibnamefont
  {{Shen}}}, \bibinfo {author} {\bibfnamefont {H.}~\bibnamefont {{Toki}}},
  \bibinfo {author} {\bibfnamefont {K.}~\bibnamefont {{Oyamatsu}}}, \ and\
  \bibinfo {author} {\bibfnamefont {K.}~\bibnamefont {{Sumiyoshi}}},\ }\href
  {\doibase 10.1016/S0375-9474(98)00236-X} {\bibfield  {journal} {\bibinfo
  {journal} {\nphysa}\ }\textbf {\bibinfo {volume} {637}},\ \bibinfo {pages}
  {435} (\bibinfo {year} {1998})},\ \Eprint
  {http://arxiv.org/abs/nucl-th/9805035} {arXiv:nucl-th/9805035 [nucl-th]}
  \BibitemShut {NoStop}%
\bibitem [{\citenamefont {Hempel}\ and\ \citenamefont
  {Schaffner-Bielich}(2010)}]{Hempel:2009mc}%
  \BibitemOpen
  \bibfield  {author} {\bibinfo {author} {\bibfnamefont {M.}~\bibnamefont
  {Hempel}}\ and\ \bibinfo {author} {\bibfnamefont {J.}~\bibnamefont
  {Schaffner-Bielich}},\ }\href {\doibase 10.1016/j.nuclphysa.2010.02.010}
  {\bibfield  {journal} {\bibinfo  {journal} {Nucl. Phys. A}\ }\textbf
  {\bibinfo {volume} {837}},\ \bibinfo {pages} {210} (\bibinfo {year}
  {2010})},\ \Eprint {http://arxiv.org/abs/0911.4073} {arXiv:0911.4073
  [nucl-th]} \BibitemShut {NoStop}%
\bibitem [{\citenamefont {Hempel}\ \emph {et~al.}(2012)\citenamefont {Hempel},
  \citenamefont {Fischer}, \citenamefont {Schaffner-Bielich},\ and\
  \citenamefont {Liebendorfer}}]{Hempel:2011mk}%
  \BibitemOpen
  \bibfield  {author} {\bibinfo {author} {\bibfnamefont {M.}~\bibnamefont
  {Hempel}}, \bibinfo {author} {\bibfnamefont {T.}~\bibnamefont {Fischer}},
  \bibinfo {author} {\bibfnamefont {J.}~\bibnamefont {Schaffner-Bielich}}, \
  and\ \bibinfo {author} {\bibfnamefont {M.}~\bibnamefont {Liebendorfer}},\
  }\href {\doibase 10.1088/0004-637X/748/1/70} {\bibfield  {journal} {\bibinfo
  {journal} {Astrophys. J.}\ }\textbf {\bibinfo {volume} {748}},\ \bibinfo
  {pages} {70} (\bibinfo {year} {2012})},\ \Eprint
  {http://arxiv.org/abs/1108.0848} {arXiv:1108.0848 [astro-ph.HE]} \BibitemShut
  {NoStop}%
\bibitem [{\citenamefont {Steiner}\ \emph {et~al.}(2013)\citenamefont
  {Steiner}, \citenamefont {Hempel},\ and\ \citenamefont
  {Fischer}}]{Steiner:2012rk}%
  \BibitemOpen
  \bibfield  {author} {\bibinfo {author} {\bibfnamefont {A.~W.}\ \bibnamefont
  {Steiner}}, \bibinfo {author} {\bibfnamefont {M.}~\bibnamefont {Hempel}}, \
  and\ \bibinfo {author} {\bibfnamefont {T.}~\bibnamefont {Fischer}},\ }\href
  {\doibase 10.1088/0004-637X/774/1/17} {\bibfield  {journal} {\bibinfo
  {journal} {Astrophys. J.}\ }\textbf {\bibinfo {volume} {774}},\ \bibinfo
  {pages} {17} (\bibinfo {year} {2013})},\ \Eprint
  {http://arxiv.org/abs/1207.2184} {arXiv:1207.2184 [astro-ph.SR]} \BibitemShut
  {NoStop}%
\bibitem [{\citenamefont {Furusawa}\ \emph {et~al.}(2017)\citenamefont
  {Furusawa}, \citenamefont {Sumiyoshi}, \citenamefont {Yamada},\ and\
  \citenamefont {Suzuki}}]{FURUSAWA2017}%
  \BibitemOpen
  \bibfield  {author} {\bibinfo {author} {\bibfnamefont {S.}~\bibnamefont
  {Furusawa}}, \bibinfo {author} {\bibfnamefont {K.}~\bibnamefont {Sumiyoshi}},
  \bibinfo {author} {\bibfnamefont {S.}~\bibnamefont {Yamada}}, \ and\ \bibinfo
  {author} {\bibfnamefont {H.}~\bibnamefont {Suzuki}},\ }\href {\doibase
  https://doi.org/10.1016/j.nuclphysa.2016.09.002} {\bibfield  {journal}
  {\bibinfo  {journal} {Nuclear Physics A}\ }\textbf {\bibinfo {volume}
  {957}},\ \bibinfo {pages} {188} (\bibinfo {year} {2017})}\BibitemShut
  {NoStop}%
\bibitem [{\citenamefont {Ishizuka}\ \emph {et~al.}(2008)\citenamefont
  {Ishizuka}, \citenamefont {Ohnishi}, \citenamefont {Tsubakihara},
  \citenamefont {Sumiyoshi},\ and\ \citenamefont {Yamada}}]{Ishizuka:2008gr}%
  \BibitemOpen
  \bibfield  {author} {\bibinfo {author} {\bibfnamefont {C.}~\bibnamefont
  {Ishizuka}}, \bibinfo {author} {\bibfnamefont {A.}~\bibnamefont {Ohnishi}},
  \bibinfo {author} {\bibfnamefont {K.}~\bibnamefont {Tsubakihara}}, \bibinfo
  {author} {\bibfnamefont {K.}~\bibnamefont {Sumiyoshi}}, \ and\ \bibinfo
  {author} {\bibfnamefont {S.}~\bibnamefont {Yamada}},\ }\href {\doibase
  10.1088/0954-3899/35/8/085201} {\bibfield  {journal} {\bibinfo  {journal} {J.
  Phys. G}\ }\textbf {\bibinfo {volume} {35}},\ \bibinfo {pages} {085201}
  (\bibinfo {year} {2008})},\ \Eprint {http://arxiv.org/abs/0802.2318}
  {arXiv:0802.2318 [nucl-th]} \BibitemShut {NoStop}%
\bibitem [{\citenamefont {Colucci}\ and\ \citenamefont
  {Sedrakian}(2013)}]{Colucci:2013pya}%
  \BibitemOpen
  \bibfield  {author} {\bibinfo {author} {\bibfnamefont {G.}~\bibnamefont
  {Colucci}}\ and\ \bibinfo {author} {\bibfnamefont {A.}~\bibnamefont
  {Sedrakian}},\ }\href {\doibase 10.1103/PhysRevC.87.055806} {\bibfield
  {journal} {\bibinfo  {journal} {Phys. Rev. C}\ }\textbf {\bibinfo {volume}
  {87}},\ \bibinfo {pages} {055806} (\bibinfo {year} {2013})},\ \Eprint
  {http://arxiv.org/abs/1302.6925} {arXiv:1302.6925 [nucl-th]} \BibitemShut
  {NoStop}%
\bibitem [{\citenamefont {Marques}\ \emph
  {et~al.}(2017{\natexlab{a}})\citenamefont {Marques}, \citenamefont {Oertel},
  \citenamefont {Hempel},\ and\ \citenamefont {Novak}}]{Marques_2017}%
  \BibitemOpen
  \bibfield  {author} {\bibinfo {author} {\bibfnamefont {M.}~\bibnamefont
  {Marques}}, \bibinfo {author} {\bibfnamefont {M.}~\bibnamefont {Oertel}},
  \bibinfo {author} {\bibfnamefont {M.}~\bibnamefont {Hempel}}, \ and\ \bibinfo
  {author} {\bibfnamefont {J.}~\bibnamefont {Novak}},\ }\href {\doibase
  10.1103/PhysRevC.96.045806} {\bibfield  {journal} {\bibinfo  {journal} {Phys.
  Rev. C}\ }\textbf {\bibinfo {volume} {96}},\ \bibinfo {pages} {045806}
  (\bibinfo {year} {2017}{\natexlab{a}})},\ \Eprint
  {http://arxiv.org/abs/1706.02913} {arXiv:1706.02913 [nucl-th]} \BibitemShut
  {NoStop}%
\bibitem [{\citenamefont {Fortin}\ \emph {et~al.}(2018)\citenamefont {Fortin},
  \citenamefont {Oertel},\ and\ \citenamefont
  {Provid{\^e}ncia}}]{Fortin_PASA_2018}%
  \BibitemOpen
  \bibfield  {author} {\bibinfo {author} {\bibfnamefont {M.}~\bibnamefont
  {Fortin}}, \bibinfo {author} {\bibfnamefont {M.}~\bibnamefont {Oertel}}, \
  and\ \bibinfo {author} {\bibfnamefont {C.}~\bibnamefont {Provid{\^e}ncia}},\
  }\href {\doibase 10.1017/pasa.2018.32} {\bibfield  {journal} {\bibinfo
  {journal} {Publ. Astron. Soc. Aust.}\ }\textbf {\bibinfo {volume} {35}},\
  \bibinfo {pages} {44} (\bibinfo {year} {2018})},\ \Eprint
  {http://arxiv.org/abs/1711.09427} {arXiv:1711.09427 [astro-ph.HE]}
  \BibitemShut {NoStop}%
\bibitem [{\citenamefont {Raduta}\ \emph {et~al.}(2020)\citenamefont {Raduta},
  \citenamefont {Oertel},\ and\ \citenamefont {Sedrakian}}]{Raduta_MNRAS_2020}%
  \BibitemOpen
  \bibfield  {author} {\bibinfo {author} {\bibfnamefont {A.~R.}\ \bibnamefont
  {Raduta}}, \bibinfo {author} {\bibfnamefont {M.}~\bibnamefont {Oertel}}, \
  and\ \bibinfo {author} {\bibfnamefont {A.}~\bibnamefont {Sedrakian}},\ }\href
  {\doibase 10.1093/mnras/staa2491} {\bibfield  {journal} {\bibinfo  {journal}
  {Mon. Not. Roy. Astron. Soc.}\ }\textbf {\bibinfo {volume} {499}},\ \bibinfo
  {pages} {914} (\bibinfo {year} {2020})},\ \Eprint
  {http://arxiv.org/abs/2008.00213} {arXiv:2008.00213 [nucl-th]} \BibitemShut
  {NoStop}%
\bibitem [{\citenamefont {Stone}\ \emph
  {et~al.}(2021{\natexlab{a}})\citenamefont {Stone}, \citenamefont {Dexheimer},
  \citenamefont {Guichon}, \citenamefont {Thomas},\ and\ \citenamefont
  {Typel}}]{Stone:2019blq}%
  \BibitemOpen
  \bibfield  {author} {\bibinfo {author} {\bibfnamefont {J.~R.}\ \bibnamefont
  {Stone}}, \bibinfo {author} {\bibfnamefont {V.}~\bibnamefont {Dexheimer}},
  \bibinfo {author} {\bibfnamefont {P.~A.~M.}\ \bibnamefont {Guichon}},
  \bibinfo {author} {\bibfnamefont {A.~W.}\ \bibnamefont {Thomas}}, \ and\
  \bibinfo {author} {\bibfnamefont {S.}~\bibnamefont {Typel}},\ }\href
  {\doibase 10.1093/mnras/staa4006} {\bibfield  {journal} {\bibinfo  {journal}
  {Mon. Not. Roy. Astron. Soc.}\ }\textbf {\bibinfo {volume} {502}},\ \bibinfo
  {pages} {3476} (\bibinfo {year} {2021}{\natexlab{a}})},\ \Eprint
  {http://arxiv.org/abs/1906.11100} {arXiv:1906.11100 [nucl-th]} \BibitemShut
  {NoStop}%
\bibitem [{\citenamefont {Sedrakian}\ and\ \citenamefont
  {Harutyunyan}(2022)}]{Sedrakian:2022kgj}%
  \BibitemOpen
  \bibfield  {author} {\bibinfo {author} {\bibfnamefont {A.}~\bibnamefont
  {Sedrakian}}\ and\ \bibinfo {author} {\bibfnamefont {A.}~\bibnamefont
  {Harutyunyan}},\ }\href {\doibase 10.1140/epja/s10050-022-00792-w} {\bibfield
   {journal} {\bibinfo  {journal} {Eur. Phys. J. A}\ }\textbf {\bibinfo
  {volume} {58}},\ \bibinfo {pages} {137} (\bibinfo {year} {2022})},\ \Eprint
  {http://arxiv.org/abs/2202.12083} {arXiv:2202.12083 [nucl-th]} \BibitemShut
  {NoStop}%
\bibitem [{\citenamefont {Sedrakian}\ and\ \citenamefont
  {Harutyunyan}(2021)}]{Sedrakian:2021qjw}%
  \BibitemOpen
  \bibfield  {author} {\bibinfo {author} {\bibfnamefont {A.}~\bibnamefont
  {Sedrakian}}\ and\ \bibinfo {author} {\bibfnamefont {A.}~\bibnamefont
  {Harutyunyan}},\ }\href {\doibase 10.3390/universe7100382} {\bibfield
  {journal} {\bibinfo  {journal} {Universe}\ }\textbf {\bibinfo {volume} {7}},\
  \bibinfo {pages} {382} (\bibinfo {year} {2021})},\ \Eprint
  {http://arxiv.org/abs/2109.01919} {arXiv:2109.01919 [nucl-th]} \BibitemShut
  {NoStop}%
\bibitem [{\citenamefont {Kochankovski}\ \emph {et~al.}(2022)\citenamefont
  {Kochankovski}, \citenamefont {Ramos},\ and\ \citenamefont
  {Tolos}}]{Kochankovski:2022rid}%
  \BibitemOpen
  \bibfield  {author} {\bibinfo {author} {\bibfnamefont {H.}~\bibnamefont
  {Kochankovski}}, \bibinfo {author} {\bibfnamefont {A.}~\bibnamefont {Ramos}},
  \ and\ \bibinfo {author} {\bibfnamefont {L.}~\bibnamefont {Tolos}},\ }\href
  {\doibase 10.1093/mnras/stac2671} {\bibfield  {journal} {\bibinfo  {journal}
  {Mon. Not. Roy. Astron. Soc.}\ }\textbf {\bibinfo {volume} {517}},\ \bibinfo
  {pages} {507} (\bibinfo {year} {2022})},\ \bibinfo {note} {[Erratum:
  Mon.Not.Roy.Astron.Soc. 518, 6376]},\ \Eprint
  {http://arxiv.org/abs/2206.11266} {arXiv:2206.11266 [astro-ph.HE]}
  \BibitemShut {NoStop}%
\bibitem [{\citenamefont {{Tsiopelas}}\ \emph {et~al.}(2024)\citenamefont
  {{Tsiopelas}}, \citenamefont {{Sedrakian}},\ and\ \citenamefont
  {{Oertel}}}]{Tsiopelas2024EPJA}%
  \BibitemOpen
  \bibfield  {author} {\bibinfo {author} {\bibfnamefont {S.}~\bibnamefont
  {{Tsiopelas}}}, \bibinfo {author} {\bibfnamefont {A.}~\bibnamefont
  {{Sedrakian}}}, \ and\ \bibinfo {author} {\bibfnamefont {M.}~\bibnamefont
  {{Oertel}}},\ }\href {\doibase 10.1140/epja/s10050-024-01351-1} {\bibfield
  {journal} {\bibinfo  {journal} {European Physical Journal A}\ }\textbf
  {\bibinfo {volume} {60}},\ \bibinfo {eid} {127} (\bibinfo {year} {2024})},\
  \Eprint {http://arxiv.org/abs/2406.00484} {arXiv:2406.00484 [nucl-th]}
  \BibitemShut {NoStop}%
\bibitem [{\citenamefont {Barman}\ and\ \citenamefont
  {Chatterjee}(2025)}]{Barman:2025qrm}%
  \BibitemOpen
  \bibfield  {author} {\bibinfo {author} {\bibfnamefont {N.}~\bibnamefont
  {Barman}}\ and\ \bibinfo {author} {\bibfnamefont {D.}~\bibnamefont
  {Chatterjee}},\ }\href@noop {} {\  (\bibinfo {year} {2025})},\ \Eprint
  {http://arxiv.org/abs/2506.03288} {arXiv:2506.03288 [astro-ph.HE]}
  \BibitemShut {NoStop}%
\bibitem [{\citenamefont {Typel}\ \emph {et~al.}(2015)\citenamefont {Typel},
  \citenamefont {Oertel},\ and\ \citenamefont {Kl{\"a}hn}}]{Typel:2013rza}%
  \BibitemOpen
  \bibfield  {author} {\bibinfo {author} {\bibfnamefont {S.}~\bibnamefont
  {Typel}}, \bibinfo {author} {\bibfnamefont {M.}~\bibnamefont {Oertel}}, \
  and\ \bibinfo {author} {\bibfnamefont {T.}~\bibnamefont {Kl{\"a}hn}},\ }\href
  {\doibase 10.1134/S1063779615040061} {\bibfield  {journal} {\bibinfo
  {journal} {Phys. Part. Nucl.}\ }\textbf {\bibinfo {volume} {46}},\ \bibinfo
  {pages} {633} (\bibinfo {year} {2015})},\ \Eprint
  {http://arxiv.org/abs/1307.5715} {arXiv:1307.5715 [astro-ph.SR]} \BibitemShut
  {NoStop}%
\bibitem [{\citenamefont {{Dexheimer}}\ \emph {et~al.}(2022)\citenamefont
  {{Dexheimer}}, \citenamefont {{Mancini}}, \citenamefont {{Oertel}},
  \citenamefont {{Provid{\^e}ncia}}, \citenamefont {{Tolos}},\ and\
  \citenamefont {{Typel}}}]{Dexheimer2022Parti}%
  \BibitemOpen
  \bibfield  {author} {\bibinfo {author} {\bibfnamefont {V.}~\bibnamefont
  {{Dexheimer}}}, \bibinfo {author} {\bibfnamefont {M.}~\bibnamefont
  {{Mancini}}}, \bibinfo {author} {\bibfnamefont {M.}~\bibnamefont {{Oertel}}},
  \bibinfo {author} {\bibfnamefont {C.}~\bibnamefont {{Provid{\^e}ncia}}},
  \bibinfo {author} {\bibfnamefont {L.}~\bibnamefont {{Tolos}}}, \ and\
  \bibinfo {author} {\bibfnamefont {S.}~\bibnamefont {{Typel}}},\ }\href
  {\doibase 10.3390/particles5030028} {\bibfield  {journal} {\bibinfo
  {journal} {Particles}\ }\textbf {\bibinfo {volume} {5}},\ \bibinfo {pages}
  {346} (\bibinfo {year} {2022})},\ \Eprint {http://arxiv.org/abs/2311.04715}
  {arXiv:2311.04715 [nucl-th]} \BibitemShut {NoStop}%
\bibitem [{\citenamefont {Typel}\ \emph {et~al.}(2022)\citenamefont {Typel}
  \emph {et~al.}}]{ComposeCoreTeam:2022ddl}%
  \BibitemOpen
  \bibfield  {author} {\bibinfo {author} {\bibfnamefont {S.}~\bibnamefont
  {Typel}} \emph {et~al.} (\bibinfo {collaboration} {CompOSE Core Team}),\
  }\href {\doibase 10.1140/epja/s10050-022-00847-y} {\bibfield  {journal}
  {\bibinfo  {journal} {Eur. Phys. J. A}\ }\textbf {\bibinfo {volume} {58}},\
  \bibinfo {pages} {221} (\bibinfo {year} {2022})},\ \Eprint
  {http://arxiv.org/abs/2203.03209} {arXiv:2203.03209 [astro-ph.HE]}
  \BibitemShut {NoStop}%
\bibitem [{\citenamefont {{Li}}\ and\ \citenamefont
  {{Sedrakian}}(2023)}]{Li2023ApJ}%
  \BibitemOpen
  \bibfield  {author} {\bibinfo {author} {\bibfnamefont {J.-J.}\ \bibnamefont
  {{Li}}}\ and\ \bibinfo {author} {\bibfnamefont {A.}~\bibnamefont
  {{Sedrakian}}},\ }\href {\doibase 10.3847/1538-4357/acfa73} {\bibfield
  {journal} {\bibinfo  {journal} {\apj}\ }\textbf {\bibinfo {volume} {957}},\
  \bibinfo {eid} {41} (\bibinfo {year} {2023})},\ \Eprint
  {http://arxiv.org/abs/2308.14457} {arXiv:2308.14457 [nucl-th]} \BibitemShut
  {NoStop}%
\bibitem [{\citenamefont {{Margalit}}\ and\ \citenamefont
  {{Metzger}}(2017)}]{Margalit2017}%
  \BibitemOpen
  \bibfield  {author} {\bibinfo {author} {\bibfnamefont {B.}~\bibnamefont
  {{Margalit}}}\ and\ \bibinfo {author} {\bibfnamefont {B.~D.}\ \bibnamefont
  {{Metzger}}},\ }\href {\doibase 10.3847/2041-8213/aa991c} {\bibfield
  {journal} {\bibinfo  {journal} {\apjl}\ }\textbf {\bibinfo {volume} {850}},\
  \bibinfo {eid} {L19} (\bibinfo {year} {2017})},\ \Eprint
  {http://arxiv.org/abs/1710.05938} {arXiv:1710.05938 [astro-ph.HE]}
  \BibitemShut {NoStop}%
\bibitem [{\citenamefont {{Rezzolla}}\ \emph {et~al.}(2018)\citenamefont
  {{Rezzolla}}, \citenamefont {{Most}},\ and\ \citenamefont
  {{Weih}}}]{Rezzolla2018ApJ}%
  \BibitemOpen
  \bibfield  {author} {\bibinfo {author} {\bibfnamefont {L.}~\bibnamefont
  {{Rezzolla}}}, \bibinfo {author} {\bibfnamefont {E.~R.}\ \bibnamefont
  {{Most}}}, \ and\ \bibinfo {author} {\bibfnamefont {L.~R.}\ \bibnamefont
  {{Weih}}},\ }\href {\doibase 10.3847/2041-8213/aaa401} {\bibfield  {journal}
  {\bibinfo  {journal} {\apjl}\ }\textbf {\bibinfo {volume} {852}},\ \bibinfo
  {eid} {L25} (\bibinfo {year} {2018})},\ \Eprint
  {http://arxiv.org/abs/1711.00314} {arXiv:1711.00314 [astro-ph.HE]}
  \BibitemShut {NoStop}%
\bibitem [{\citenamefont {{Shibata}}\ \emph {et~al.}(2019)\citenamefont
  {{Shibata}}, \citenamefont {{Zhou}}, \citenamefont {{Kiuchi}},\ and\
  \citenamefont {{Fujibayashi}}}]{Shibata2019PhRvD}%
  \BibitemOpen
  \bibfield  {author} {\bibinfo {author} {\bibfnamefont {M.}~\bibnamefont
  {{Shibata}}}, \bibinfo {author} {\bibfnamefont {E.}~\bibnamefont {{Zhou}}},
  \bibinfo {author} {\bibfnamefont {K.}~\bibnamefont {{Kiuchi}}}, \ and\
  \bibinfo {author} {\bibfnamefont {S.}~\bibnamefont {{Fujibayashi}}},\ }\href
  {\doibase 10.1103/PhysRevD.100.023015} {\bibfield  {journal} {\bibinfo
  {journal} {\prd}\ }\textbf {\bibinfo {volume} {100}},\ \bibinfo {eid}
  {023015} (\bibinfo {year} {2019})},\ \Eprint
  {http://arxiv.org/abs/1905.03656} {arXiv:1905.03656 [astro-ph.HE]}
  \BibitemShut {NoStop}%
\bibitem [{\citenamefont {{Khadkikar}}\ \emph {et~al.}(2021)\citenamefont
  {{Khadkikar}}, \citenamefont {{Raduta}}, \citenamefont {{Oertel}},\ and\
  \citenamefont {{Sedrakian}}}]{Khadkikar:2021}%
  \BibitemOpen
  \bibfield  {author} {\bibinfo {author} {\bibfnamefont {S.}~\bibnamefont
  {{Khadkikar}}}, \bibinfo {author} {\bibfnamefont {A.~R.}\ \bibnamefont
  {{Raduta}}}, \bibinfo {author} {\bibfnamefont {M.}~\bibnamefont {{Oertel}}},
  \ and\ \bibinfo {author} {\bibfnamefont {A.~e.}\ \bibnamefont
  {{Sedrakian}}},\ }\href {\doibase 10.1103/PhysRevC.103.055811} {\bibfield
  {journal} {\bibinfo  {journal} {PhRvC}\ }\textbf {\bibinfo {volume} {103}},\
  \bibinfo {eid} {055811} (\bibinfo {year} {2021})},\ \Eprint
  {http://arxiv.org/abs/2102.00988} {arXiv:2102.00988 [astro-ph.HE]}
  \BibitemShut {NoStop}%
\bibitem [{\citenamefont {Yagi}\ and\ \citenamefont
  {Yunes}(2013{\natexlab{a}})}]{Yagi:2013a}%
  \BibitemOpen
  \bibfield  {author} {\bibinfo {author} {\bibfnamefont {K.}~\bibnamefont
  {Yagi}}\ and\ \bibinfo {author} {\bibfnamefont {N.}~\bibnamefont {Yunes}},\
  }\href {\doibase 10.1126/science.1236462} {\bibfield  {journal} {\bibinfo
  {journal} {Science}\ }\textbf {\bibinfo {volume} {341}},\ \bibinfo {pages}
  {365} (\bibinfo {year} {2013}{\natexlab{a}})}\BibitemShut {NoStop}%
\bibitem [{\citenamefont {Yagi}\ and\ \citenamefont
  {Yunes}(2013{\natexlab{b}})}]{Yagi:2013b}%
  \BibitemOpen
  \bibfield  {author} {\bibinfo {author} {\bibfnamefont {K.}~\bibnamefont
  {Yagi}}\ and\ \bibinfo {author} {\bibfnamefont {N.}~\bibnamefont {Yunes}},\
  }\href {\doibase 10.1103/PhysRevD.88.023009} {\bibfield  {journal} {\bibinfo
  {journal} {Phys. Rev. D}\ }\textbf {\bibinfo {volume} {88}},\ \bibinfo
  {pages} {023009} (\bibinfo {year} {2013}{\natexlab{b}})},\ \Eprint
  {http://arxiv.org/abs/1303.1528} {arXiv:1303.1528 [gr-qc]} \BibitemShut
  {NoStop}%
\bibitem [{\citenamefont {Yagi}\ and\ \citenamefont {Yunes}(2017)}]{Yagi:2017}%
  \BibitemOpen
  \bibfield  {author} {\bibinfo {author} {\bibfnamefont {K.}~\bibnamefont
  {Yagi}}\ and\ \bibinfo {author} {\bibfnamefont {N.}~\bibnamefont {Yunes}},\
  }\href {\doibase 10.1016/j.physrep.2017.03.002} {\bibfield  {journal}
  {\bibinfo  {journal} {Phys. Rep.}\ }\textbf {\bibinfo {volume} {681}},\
  \bibinfo {pages} {1} (\bibinfo {year} {2017})},\ \Eprint
  {http://arxiv.org/abs/1608.02582} {arXiv:1608.02582 [gr-qc]} \BibitemShut
  {NoStop}%
\bibitem [{\citenamefont {Doneva}\ \emph
  {et~al.}(2013{\natexlab{a}})\citenamefont {Doneva}, \citenamefont
  {Yazadjiev}, \citenamefont {Stergioulas},\ and\ \citenamefont
  {Kokkotas}}]{Doneva:2013}%
  \BibitemOpen
  \bibfield  {author} {\bibinfo {author} {\bibfnamefont {D.~D.}\ \bibnamefont
  {Doneva}}, \bibinfo {author} {\bibfnamefont {S.~S.}\ \bibnamefont
  {Yazadjiev}}, \bibinfo {author} {\bibfnamefont {N.}~\bibnamefont
  {Stergioulas}}, \ and\ \bibinfo {author} {\bibfnamefont {K.~D.}\ \bibnamefont
  {Kokkotas}},\ }\href {\doibase 10.1088/2041-8205/781/1/L6} {\bibfield
  {journal} {\bibinfo  {journal} {Astrophys. J. Lett.}\ }\textbf {\bibinfo
  {volume} {781}},\ \bibinfo {pages} {L6} (\bibinfo {year}
  {2013}{\natexlab{a}})},\ \Eprint {http://arxiv.org/abs/1310.7436}
  {arXiv:1310.7436 [gr-qc]} \BibitemShut {NoStop}%
\bibitem [{\citenamefont {Pappas}\ and\ \citenamefont
  {Apostolatos}(2014)}]{Pappas:2014}%
  \BibitemOpen
  \bibfield  {author} {\bibinfo {author} {\bibfnamefont {G.}~\bibnamefont
  {Pappas}}\ and\ \bibinfo {author} {\bibfnamefont {T.~A.}\ \bibnamefont
  {Apostolatos}},\ }\href {\doibase 10.1103/PhysRevLett.112.121101} {\bibfield
  {journal} {\bibinfo  {journal} {Phys. Rev. Lett.}\ }\textbf {\bibinfo
  {volume} {112}},\ \bibinfo {pages} {121101} (\bibinfo {year} {2014})},\
  \Eprint {http://arxiv.org/abs/1311.5508} {arXiv:1311.5508 [gr-qc]}
  \BibitemShut {NoStop}%
\bibitem [{\citenamefont {Chakrabarti}\ \emph {et~al.}(2014)\citenamefont
  {Chakrabarti}, \citenamefont {Delsate}, \citenamefont {G\"urlebeck},\ and\
  \citenamefont {Steinhoff}}]{Chakrabarti:2014}%
  \BibitemOpen
  \bibfield  {author} {\bibinfo {author} {\bibfnamefont {S.}~\bibnamefont
  {Chakrabarti}}, \bibinfo {author} {\bibfnamefont {T.}~\bibnamefont
  {Delsate}}, \bibinfo {author} {\bibfnamefont {N.}~\bibnamefont
  {G\"urlebeck}}, \ and\ \bibinfo {author} {\bibfnamefont {J.}~\bibnamefont
  {Steinhoff}},\ }\href {\doibase 10.1103/PhysRevLett.112.201102} {\bibfield
  {journal} {\bibinfo  {journal} {Phys. Rev. Lett.}\ }\textbf {\bibinfo
  {volume} {112}},\ \bibinfo {pages} {201102} (\bibinfo {year} {2014})},\
  \Eprint {http://arxiv.org/abs/1311.6509} {arXiv:1311.6509 [gr-qc]}
  \BibitemShut {NoStop}%
\bibitem [{\citenamefont {Cipolletta}\ \emph {et~al.}(2015)\citenamefont
  {Cipolletta}, \citenamefont {Cherubini}, \citenamefont {Filippi},
  \citenamefont {Rueda},\ and\ \citenamefont {Ruffini}}]{Cipolletta:2015}%
  \BibitemOpen
  \bibfield  {author} {\bibinfo {author} {\bibfnamefont {F.}~\bibnamefont
  {Cipolletta}}, \bibinfo {author} {\bibfnamefont {C.}~\bibnamefont
  {Cherubini}}, \bibinfo {author} {\bibfnamefont {S.}~\bibnamefont {Filippi}},
  \bibinfo {author} {\bibfnamefont {J.~A.}\ \bibnamefont {Rueda}}, \ and\
  \bibinfo {author} {\bibfnamefont {R.}~\bibnamefont {Ruffini}},\ }\href
  {\doibase 10.1103/PhysRevD.92.023007} {\bibfield  {journal} {\bibinfo
  {journal} {Phys. Rev. D}\ }\textbf {\bibinfo {volume} {92}},\ \bibinfo
  {pages} {023007} (\bibinfo {year} {2015})},\ \Eprint
  {http://arxiv.org/abs/1506.05926} {arXiv:1506.05926 [astro-ph.SR]}
  \BibitemShut {NoStop}%
\bibitem [{\citenamefont {Breu}\ and\ \citenamefont
  {Rezzolla}(2016)}]{Breu:2016}%
  \BibitemOpen
  \bibfield  {author} {\bibinfo {author} {\bibfnamefont {C.}~\bibnamefont
  {Breu}}\ and\ \bibinfo {author} {\bibfnamefont {L.}~\bibnamefont
  {Rezzolla}},\ }\href {\doibase 10.1093/mnras/stw575} {\bibfield  {journal}
  {\bibinfo  {journal} {Mon. Not. Roy. Astron. Soc.}\ }\textbf {\bibinfo
  {volume} {459}},\ \bibinfo {pages} {646} (\bibinfo {year} {2016})},\ \Eprint
  {http://arxiv.org/abs/1601.06083} {arXiv:1601.06083 [gr-qc]} \BibitemShut
  {NoStop}%
\bibitem [{\citenamefont {Riahi}\ \emph {et~al.}(2019)\citenamefont {Riahi},
  \citenamefont {Kalantari},\ and\ \citenamefont
  {Rueda~Hernandez}}]{Riahi:2019}%
  \BibitemOpen
  \bibfield  {author} {\bibinfo {author} {\bibfnamefont {R.}~\bibnamefont
  {Riahi}}, \bibinfo {author} {\bibfnamefont {S.~Z.}\ \bibnamefont
  {Kalantari}}, \ and\ \bibinfo {author} {\bibfnamefont {J.~A.}\ \bibnamefont
  {Rueda~Hernandez}},\ }\href {\doibase 10.1103/PhysRevD.99.043004} {\bibfield
  {journal} {\bibinfo  {journal} {Phys. Rev. D}\ }\textbf {\bibinfo {volume}
  {99}},\ \bibinfo {pages} {043004} (\bibinfo {year} {2019})},\ \Eprint
  {http://arxiv.org/abs/1902.00349} {arXiv:1902.00349 [astro-ph.HE]}
  \BibitemShut {NoStop}%
\bibitem [{\citenamefont {{Riahi}}\ \emph {et~al.}(2019)\citenamefont
  {{Riahi}}, \citenamefont {{Kalantari}},\ and\ \citenamefont
  {{Rueda}}}]{Riahi2019}%
  \BibitemOpen
  \bibfield  {author} {\bibinfo {author} {\bibfnamefont {R.}~\bibnamefont
  {{Riahi}}}, \bibinfo {author} {\bibfnamefont {S.~Z.}\ \bibnamefont
  {{Kalantari}}}, \ and\ \bibinfo {author} {\bibfnamefont {J.~A.}\ \bibnamefont
  {{Rueda}}},\ }\href {\doibase 10.1103/PhysRevD.99.043004} {\bibfield
  {journal} {\bibinfo  {journal} {\prd}\ }\textbf {\bibinfo {volume} {99}},\
  \bibinfo {eid} {043004} (\bibinfo {year} {2019})},\ \Eprint
  {http://arxiv.org/abs/1902.00349} {arXiv:1902.00349 [astro-ph.HE]}
  \BibitemShut {NoStop}%
\bibitem [{\citenamefont {Koliogiannis}\ and\ \citenamefont
  {Moustakidis}(2020)}]{Koliogiannis:2020}%
  \BibitemOpen
  \bibfield  {author} {\bibinfo {author} {\bibfnamefont {P.~S.}\ \bibnamefont
  {Koliogiannis}}\ and\ \bibinfo {author} {\bibfnamefont {C.~C.}\ \bibnamefont
  {Moustakidis}},\ }\href {\doibase 10.1103/PhysRevC.101.015805} {\bibfield
  {journal} {\bibinfo  {journal} {Phys. Rev. C}\ }\textbf {\bibinfo {volume}
  {101}},\ \bibinfo {pages} {015805} (\bibinfo {year} {2020})},\ \Eprint
  {http://arxiv.org/abs/1907.13375} {arXiv:1907.13375 [nucl-th]} \BibitemShut
  {NoStop}%
\bibitem [{\citenamefont {{Konstantinou}}\ and\ \citenamefont
  {{Morsink}}(2022)}]{Konstantinou:2022}%
  \BibitemOpen
  \bibfield  {author} {\bibinfo {author} {\bibfnamefont {A.}~\bibnamefont
  {{Konstantinou}}}\ and\ \bibinfo {author} {\bibfnamefont {S.~M.}\
  \bibnamefont {{Morsink}}},\ }\href {\doibase 10.3847/1538-4357/ac7b86}
  {\bibfield  {journal} {\bibinfo  {journal} {\apj}\ }\textbf {\bibinfo
  {volume} {934}},\ \bibinfo {eid} {139} (\bibinfo {year} {2022})},\ \Eprint
  {http://arxiv.org/abs/2206.12515} {arXiv:2206.12515 [astro-ph.HE]}
  \BibitemShut {NoStop}%
\bibitem [{\citenamefont {Largani}\ \emph {et~al.}(2022)\citenamefont
  {Largani}, \citenamefont {Fischer}, \citenamefont {Sedrakian}, \citenamefont
  {Cierniak}, \citenamefont {Alvarez-Castillo},\ and\ \citenamefont
  {Blaschke}}]{Largani:2022}%
  \BibitemOpen
  \bibfield  {author} {\bibinfo {author} {\bibfnamefont {N.~K.}\ \bibnamefont
  {Largani}}, \bibinfo {author} {\bibfnamefont {T.}~\bibnamefont {Fischer}},
  \bibinfo {author} {\bibfnamefont {A.}~\bibnamefont {Sedrakian}}, \bibinfo
  {author} {\bibfnamefont {M.}~\bibnamefont {Cierniak}}, \bibinfo {author}
  {\bibfnamefont {D.~E.}\ \bibnamefont {Alvarez-Castillo}}, \ and\ \bibinfo
  {author} {\bibfnamefont {D.~B.}\ \bibnamefont {Blaschke}},\ }\href {\doibase
  10.1093/mnras/stac1916} {\bibfield  {journal} {\bibinfo  {journal} {Mon. Not.
  Roy. Astron. Soc.}\ }\textbf {\bibinfo {volume} {515}},\ \bibinfo {pages}
  {3539} (\bibinfo {year} {2022})},\ \Eprint {http://arxiv.org/abs/2112.10439}
  {arXiv:2112.10439 [astro-ph.HE]} \BibitemShut {NoStop}%
\bibitem [{\citenamefont {{Li}}\ \emph {et~al.}(2023)\citenamefont {{Li}},
  \citenamefont {{Sedrakian}},\ and\ \citenamefont {{Weber}}}]{Li2023PhRvC}%
  \BibitemOpen
  \bibfield  {author} {\bibinfo {author} {\bibfnamefont {J.~J.}\ \bibnamefont
  {{Li}}}, \bibinfo {author} {\bibfnamefont {A.}~\bibnamefont {{Sedrakian}}}, \
  and\ \bibinfo {author} {\bibfnamefont {F.}~\bibnamefont {{Weber}}},\ }\href
  {\doibase 10.1103/PhysRevC.108.025810} {\bibfield  {journal} {\bibinfo
  {journal} {\prc}\ }\textbf {\bibinfo {volume} {108}},\ \bibinfo {eid}
  {025810} (\bibinfo {year} {2023})},\ \Eprint
  {http://arxiv.org/abs/2306.14190} {arXiv:2306.14190 [nucl-th]} \BibitemShut
  {NoStop}%
\bibitem [{\citenamefont {Martinon}\ \emph {et~al.}(2014)\citenamefont
  {Martinon}, \citenamefont {Maselli}, \citenamefont {Gualtieri},\ and\
  \citenamefont {Ferrari}}]{Martinon:2014}%
  \BibitemOpen
  \bibfield  {author} {\bibinfo {author} {\bibfnamefont {G.}~\bibnamefont
  {Martinon}}, \bibinfo {author} {\bibfnamefont {A.}~\bibnamefont {Maselli}},
  \bibinfo {author} {\bibfnamefont {L.}~\bibnamefont {Gualtieri}}, \ and\
  \bibinfo {author} {\bibfnamefont {V.}~\bibnamefont {Ferrari}},\ }\href
  {\doibase 10.1103/PhysRevD.90.064026} {\bibfield  {journal} {\bibinfo
  {journal} {Phys. Rev. D}\ }\textbf {\bibinfo {volume} {90}},\ \bibinfo
  {pages} {064026} (\bibinfo {year} {2014})},\ \Eprint
  {http://arxiv.org/abs/1406.7661} {arXiv:1406.7661 [gr-qc]} \BibitemShut
  {NoStop}%
\bibitem [{\citenamefont {Marques}\ \emph
  {et~al.}(2017{\natexlab{b}})\citenamefont {Marques}, \citenamefont {Oertel},
  \citenamefont {Hempel},\ and\ \citenamefont {Novak}}]{Marques:2017}%
  \BibitemOpen
  \bibfield  {author} {\bibinfo {author} {\bibfnamefont {M.}~\bibnamefont
  {Marques}}, \bibinfo {author} {\bibfnamefont {M.}~\bibnamefont {Oertel}},
  \bibinfo {author} {\bibfnamefont {M.}~\bibnamefont {Hempel}}, \ and\ \bibinfo
  {author} {\bibfnamefont {J.}~\bibnamefont {Novak}},\ }\href {\doibase
  10.1103/PhysRevC.96.045806} {\bibfield  {journal} {\bibinfo  {journal} {Phys.
  Rev. C}\ }\textbf {\bibinfo {volume} {96}},\ \bibinfo {pages} {045806}
  (\bibinfo {year} {2017}{\natexlab{b}})},\ \Eprint
  {http://arxiv.org/abs/1706.02913} {arXiv:1706.02913 [nucl-th]} \BibitemShut
  {NoStop}%
\bibitem [{\citenamefont {Lenka}\ \emph {et~al.}(2019)\citenamefont {Lenka},
  \citenamefont {Char},\ and\ \citenamefont {Banik}}]{Lenka:2018}%
  \BibitemOpen
  \bibfield  {author} {\bibinfo {author} {\bibfnamefont {S.~S.}\ \bibnamefont
  {Lenka}}, \bibinfo {author} {\bibfnamefont {P.}~\bibnamefont {Char}}, \ and\
  \bibinfo {author} {\bibfnamefont {S.}~\bibnamefont {Banik}},\ }\href
  {\doibase 10.1088/1361-6471/ab36a2} {\bibfield  {journal} {\bibinfo
  {journal} {J. Phys. G}\ }\textbf {\bibinfo {volume} {46}},\ \bibinfo {pages}
  {105201} (\bibinfo {year} {2019})},\ \Eprint
  {http://arxiv.org/abs/1805.09492} {arXiv:1805.09492 [astro-ph.HE]}
  \BibitemShut {NoStop}%
\bibitem [{\citenamefont {Stone}\ \emph
  {et~al.}(2021{\natexlab{b}})\citenamefont {Stone}, \citenamefont {Dexheimer},
  \citenamefont {Guichon}, \citenamefont {Thomas},\ and\ \citenamefont
  {Typel}}]{Stone:2019}%
  \BibitemOpen
  \bibfield  {author} {\bibinfo {author} {\bibfnamefont {J.~R.}\ \bibnamefont
  {Stone}}, \bibinfo {author} {\bibfnamefont {V.}~\bibnamefont {Dexheimer}},
  \bibinfo {author} {\bibfnamefont {P.~A.~M.}\ \bibnamefont {Guichon}},
  \bibinfo {author} {\bibfnamefont {A.~W.}\ \bibnamefont {Thomas}}, \ and\
  \bibinfo {author} {\bibfnamefont {S.}~\bibnamefont {Typel}},\ }\href
  {\doibase 10.1093/mnras/staa4006} {\bibfield  {journal} {\bibinfo  {journal}
  {Mon. Not. Roy. Astron. Soc.}\ }\textbf {\bibinfo {volume} {502}},\ \bibinfo
  {pages} {3476} (\bibinfo {year} {2021}{\natexlab{b}})},\ \Eprint
  {http://arxiv.org/abs/1906.11100} {arXiv:1906.11100 [nucl-th]} \BibitemShut
  {NoStop}%
\bibitem [{\citenamefont {Agathos}\ \emph {et~al.}(2015)\citenamefont
  {Agathos}, \citenamefont {Meidam}, \citenamefont {Pozzo}, \citenamefont {Li},
  \citenamefont {Urban}, \citenamefont {Broeck}, \citenamefont {Nagar},\ and\
  \citenamefont {Bernuzzi}}]{Agathos2015}%
  \BibitemOpen
  \bibfield  {author} {\bibinfo {author} {\bibfnamefont {M.}~\bibnamefont
  {Agathos}}, \bibinfo {author} {\bibfnamefont {J.}~\bibnamefont {Meidam}},
  \bibinfo {author} {\bibfnamefont {W.~D.}\ \bibnamefont {Pozzo}}, \bibinfo
  {author} {\bibfnamefont {T.~G.~F.}\ \bibnamefont {Li}}, \bibinfo {author}
  {\bibfnamefont {M.}~\bibnamefont {Urban}}, \bibinfo {author} {\bibfnamefont
  {C.~V.~D.}\ \bibnamefont {Broeck}}, \bibinfo {author} {\bibfnamefont
  {A.}~\bibnamefont {Nagar}}, \ and\ \bibinfo {author} {\bibfnamefont
  {A.}~\bibnamefont {Bernuzzi}},\ }\href {\doibase 10.1103/PhysRevD.92.023012}
  {\bibfield  {journal} {\bibinfo  {journal} {Phys. Rev. D}\ }\textbf {\bibinfo
  {volume} {92}},\ \bibinfo {pages} {023012} (\bibinfo {year} {2015})},\
  \Eprint {http://arxiv.org/abs/1503.05405} {arXiv:1503.05405 [gr-qc]}
  \BibitemShut {NoStop}%
\bibitem [{\citenamefont {Chatziioannou}\ \emph {et~al.}(2018)\citenamefont
  {Chatziioannou}, \citenamefont {Haster},\ and\ \citenamefont
  {Zimmerman}}]{Chatziioannou2018}%
  \BibitemOpen
  \bibfield  {author} {\bibinfo {author} {\bibfnamefont {K.}~\bibnamefont
  {Chatziioannou}}, \bibinfo {author} {\bibfnamefont {C.-J.}\ \bibnamefont
  {Haster}}, \ and\ \bibinfo {author} {\bibfnamefont {A.}~\bibnamefont
  {Zimmerman}},\ }\href {\doibase 10.1103/PhysRevD.97.104036} {\bibfield
  {journal} {\bibinfo  {journal} {Phys. Rev. D}\ }\textbf {\bibinfo {volume}
  {97}},\ \bibinfo {pages} {104036} (\bibinfo {year} {2018})},\ \Eprint
  {http://arxiv.org/abs/1804.03221} {arXiv:1804.03221 [gr-qc]} \BibitemShut
  {NoStop}%
\bibitem [{\citenamefont {Carson}\ \emph {et~al.}(2019)\citenamefont {Carson},
  \citenamefont {Steiner},\ and\ \citenamefont {Yagi}}]{Carson2019}%
  \BibitemOpen
  \bibfield  {author} {\bibinfo {author} {\bibfnamefont {Z.}~\bibnamefont
  {Carson}}, \bibinfo {author} {\bibfnamefont {A.~W.}\ \bibnamefont {Steiner}},
  \ and\ \bibinfo {author} {\bibfnamefont {K.}~\bibnamefont {Yagi}},\ }\href
  {\doibase 10.1103/PhysRevD.99.043010} {\bibfield  {journal} {\bibinfo
  {journal} {Phys. Rev. D}\ }\textbf {\bibinfo {volume} {99}},\ \bibinfo
  {pages} {043010} (\bibinfo {year} {2019})},\ \Eprint
  {http://arxiv.org/abs/1812.08910} {arXiv:1812.08910 [astro-ph.HE]}
  \BibitemShut {NoStop}%
\bibitem [{\citenamefont {Suleiman}\ and\ \citenamefont
  {Read}(2024)}]{Suleiman:2024ztn}%
  \BibitemOpen
  \bibfield  {author} {\bibinfo {author} {\bibfnamefont {L.}~\bibnamefont
  {Suleiman}}\ and\ \bibinfo {author} {\bibfnamefont {J.}~\bibnamefont
  {Read}},\ }\href {\doibase 10.1103/PhysRevD.109.103029} {\bibfield  {journal}
  {\bibinfo  {journal} {Phys. Rev. D}\ }\textbf {\bibinfo {volume} {109}},\
  \bibinfo {pages} {103029} (\bibinfo {year} {2024})},\ \Eprint
  {http://arxiv.org/abs/2402.01948} {arXiv:2402.01948 [astro-ph.HE]}
  \BibitemShut {NoStop}%
\bibitem [{\citenamefont {{Yeasin}}\ \emph {et~al.}(2025)\citenamefont
  {{Yeasin}}, \citenamefont {{Tsiopelas}}, \citenamefont {{Sedrakian}},\ and\
  \citenamefont {{Li}}}]{Yeasin2025}%
  \BibitemOpen
  \bibfield  {author} {\bibinfo {author} {\bibfnamefont {P.~S.}\ \bibnamefont
  {{Yeasin}}}, \bibinfo {author} {\bibfnamefont {S.}~\bibnamefont
  {{Tsiopelas}}}, \bibinfo {author} {\bibfnamefont {A.}~\bibnamefont
  {{Sedrakian}}}, \ and\ \bibinfo {author} {\bibfnamefont {J.-J.}\ \bibnamefont
  {{Li}}},\ }\href {\doibase 10.1103/mz9y-hqlw} {\bibfield  {journal} {\bibinfo
   {journal} {\prd}\ }\textbf {\bibinfo {volume} {112}},\ \bibinfo {eid}
  {063028} (\bibinfo {year} {2025})},\ \Eprint
  {http://arxiv.org/abs/2507.01093} {arXiv:2507.01093 [astro-ph.HE]}
  \BibitemShut {NoStop}%
\bibitem [{\citenamefont {Roberts}\ \emph {et~al.}(2012)\citenamefont
  {Roberts}, \citenamefont {Shen}, \citenamefont {Cirigliano}, \citenamefont
  {Pons}, \citenamefont {Reddy},\ and\ \citenamefont
  {Woosley}}]{Roberts:2011yw}%
  \BibitemOpen
  \bibfield  {author} {\bibinfo {author} {\bibfnamefont {L.~F.}\ \bibnamefont
  {Roberts}}, \bibinfo {author} {\bibfnamefont {G.}~\bibnamefont {Shen}},
  \bibinfo {author} {\bibfnamefont {V.}~\bibnamefont {Cirigliano}}, \bibinfo
  {author} {\bibfnamefont {J.~A.}\ \bibnamefont {Pons}}, \bibinfo {author}
  {\bibfnamefont {S.}~\bibnamefont {Reddy}}, \ and\ \bibinfo {author}
  {\bibfnamefont {S.~E.}\ \bibnamefont {Woosley}},\ }\href {\doibase
  10.1103/PhysRevLett.108.061103} {\bibfield  {journal} {\bibinfo  {journal}
  {Phys. Rev. Lett.}\ }\textbf {\bibinfo {volume} {108}},\ \bibinfo {pages}
  {061103} (\bibinfo {year} {2012})},\ \Eprint {http://arxiv.org/abs/1112.0335}
  {arXiv:1112.0335 [astro-ph.HE]} \BibitemShut {NoStop}%
\bibitem [{\citenamefont {Pascal}\ \emph {et~al.}(2022)\citenamefont {Pascal},
  \citenamefont {Novak},\ and\ \citenamefont {Oertel}}]{Pascal:2022qeg}%
  \BibitemOpen
  \bibfield  {author} {\bibinfo {author} {\bibfnamefont {A.}~\bibnamefont
  {Pascal}}, \bibinfo {author} {\bibfnamefont {J.}~\bibnamefont {Novak}}, \
  and\ \bibinfo {author} {\bibfnamefont {M.}~\bibnamefont {Oertel}},\ }\href
  {\doibase 10.1093/mnras/stac016} {\bibfield  {journal} {\bibinfo  {journal}
  {Mon. Not. Roy. Astron. Soc.}\ }\textbf {\bibinfo {volume} {511}},\ \bibinfo
  {pages} {356} (\bibinfo {year} {2022})},\ \Eprint
  {http://arxiv.org/abs/2201.01955} {arXiv:2201.01955 [nucl-th]} \BibitemShut
  {NoStop}%
\bibitem [{\citenamefont {{Radice}}\ \emph {et~al.}(2016)\citenamefont
  {{Radice}}, \citenamefont {{Galeazzi}}, \citenamefont {{Lippuner}},
  \citenamefont {{Roberts}}, \citenamefont {{Ott}},\ and\ \citenamefont
  {{Rezzolla}}}]{Radice2016}%
  \BibitemOpen
  \bibfield  {author} {\bibinfo {author} {\bibfnamefont {D.}~\bibnamefont
  {{Radice}}}, \bibinfo {author} {\bibfnamefont {F.}~\bibnamefont
  {{Galeazzi}}}, \bibinfo {author} {\bibfnamefont {J.}~\bibnamefont
  {{Lippuner}}}, \bibinfo {author} {\bibfnamefont {L.~F.}\ \bibnamefont
  {{Roberts}}}, \bibinfo {author} {\bibfnamefont {C.~D.}\ \bibnamefont
  {{Ott}}}, \ and\ \bibinfo {author} {\bibfnamefont {L.}~\bibnamefont
  {{Rezzolla}}},\ }\href {\doibase 10.1093/mnras/stw1227} {\bibfield  {journal}
  {\bibinfo  {journal} {\mnras}\ }\textbf {\bibinfo {volume} {460}},\ \bibinfo
  {pages} {3255} (\bibinfo {year} {2016})},\ \Eprint
  {http://arxiv.org/abs/1601.02426} {arXiv:1601.02426 [astro-ph.HE]}
  \BibitemShut {NoStop}%
\bibitem [{\citenamefont {{Sekiguchi}}\ \emph {et~al.}(2016)\citenamefont
  {{Sekiguchi}}, \citenamefont {{Kiuchi}}, \citenamefont {{Kyutoku}},
  \citenamefont {{Shibata}},\ and\ \citenamefont
  {{Taniguchi}}}]{Sekiguchi2016}%
  \BibitemOpen
  \bibfield  {author} {\bibinfo {author} {\bibfnamefont {Y.}~\bibnamefont
  {{Sekiguchi}}}, \bibinfo {author} {\bibfnamefont {K.}~\bibnamefont
  {{Kiuchi}}}, \bibinfo {author} {\bibfnamefont {K.}~\bibnamefont {{Kyutoku}}},
  \bibinfo {author} {\bibfnamefont {M.}~\bibnamefont {{Shibata}}}, \ and\
  \bibinfo {author} {\bibfnamefont {K.}~\bibnamefont {{Taniguchi}}},\ }\href
  {\doibase 10.1103/PhysRevD.93.124046} {\bibfield  {journal} {\bibinfo
  {journal} {\prd}\ }\textbf {\bibinfo {volume} {93}},\ \bibinfo {eid} {124046}
  (\bibinfo {year} {2016})},\ \Eprint {http://arxiv.org/abs/1603.01918}
  {arXiv:1603.01918 [astro-ph.HE]} \BibitemShut {NoStop}%
\bibitem [{\citenamefont {{Radice}}\ \emph {et~al.}(2020)\citenamefont
  {{Radice}}, \citenamefont {{Bernuzzi}},\ and\ \citenamefont
  {{Perego}}}]{Radice2020}%
  \BibitemOpen
  \bibfield  {author} {\bibinfo {author} {\bibfnamefont {D.}~\bibnamefont
  {{Radice}}}, \bibinfo {author} {\bibfnamefont {S.}~\bibnamefont
  {{Bernuzzi}}}, \ and\ \bibinfo {author} {\bibfnamefont {A.}~\bibnamefont
  {{Perego}}},\ }\href {\doibase 10.1146/annurev-nucl-013120-114541} {\bibfield
   {journal} {\bibinfo  {journal} {Annual Review of Nuclear and Particle
  Science}\ }\textbf {\bibinfo {volume} {70}},\ \bibinfo {pages} {95} (\bibinfo
  {year} {2020})},\ \Eprint {http://arxiv.org/abs/2002.03863} {arXiv:2002.03863
  [astro-ph.HE]} \BibitemShut {NoStop}%
\bibitem [{\citenamefont {{Sumiyoshi}}\ \emph {et~al.}(2021)\citenamefont
  {{Sumiyoshi}}, \citenamefont {{Fujibayashi}}, \citenamefont {{Sekiguchi}},\
  and\ \citenamefont {{Shibata}}}]{Sumiyoshi2021}%
  \BibitemOpen
  \bibfield  {author} {\bibinfo {author} {\bibfnamefont {K.}~\bibnamefont
  {{Sumiyoshi}}}, \bibinfo {author} {\bibfnamefont {S.}~\bibnamefont
  {{Fujibayashi}}}, \bibinfo {author} {\bibfnamefont {Y.}~\bibnamefont
  {{Sekiguchi}}}, \ and\ \bibinfo {author} {\bibfnamefont {M.}~\bibnamefont
  {{Shibata}}},\ }\href {\doibase 10.3847/1538-4357/abce63} {\bibfield
  {journal} {\bibinfo  {journal} {\apj}\ }\textbf {\bibinfo {volume} {907}},\
  \bibinfo {eid} {92} (\bibinfo {year} {2021})},\ \Eprint
  {http://arxiv.org/abs/2010.10865} {arXiv:2010.10865 [astro-ph.HE]}
  \BibitemShut {NoStop}%
\bibitem [{\citenamefont {{Espino}}\ \emph {et~al.}(2024)\citenamefont
  {{Espino}}, \citenamefont {{Hammond}}, \citenamefont {{Radice}},
  \citenamefont {{Bernuzzi}}, \citenamefont {{Gamba}}, \citenamefont {{Zappa}},
  \citenamefont {{Micchi}},\ and\ \citenamefont {{Perego}}}]{Espino2024}%
  \BibitemOpen
  \bibfield  {author} {\bibinfo {author} {\bibfnamefont {P.~L.}\ \bibnamefont
  {{Espino}}}, \bibinfo {author} {\bibfnamefont {P.}~\bibnamefont {{Hammond}}},
  \bibinfo {author} {\bibfnamefont {D.}~\bibnamefont {{Radice}}}, \bibinfo
  {author} {\bibfnamefont {S.}~\bibnamefont {{Bernuzzi}}}, \bibinfo {author}
  {\bibfnamefont {R.}~\bibnamefont {{Gamba}}}, \bibinfo {author} {\bibfnamefont
  {F.}~\bibnamefont {{Zappa}}}, \bibinfo {author} {\bibfnamefont {L.~F.~L.}\
  \bibnamefont {{Micchi}}}, \ and\ \bibinfo {author} {\bibfnamefont
  {A.}~\bibnamefont {{Perego}}},\ }\href {\doibase
  10.1103/PhysRevLett.132.211001} {\bibfield  {journal} {\bibinfo  {journal}
  {\prl}\ }\textbf {\bibinfo {volume} {132}},\ \bibinfo {eid} {211001}
  (\bibinfo {year} {2024})},\ \Eprint {http://arxiv.org/abs/2311.00031}
  {arXiv:2311.00031 [astro-ph.HE]} \BibitemShut {NoStop}%
\bibitem [{\citenamefont {{Foucart}}(2023)}]{Foucart2023}%
  \BibitemOpen
  \bibfield  {author} {\bibinfo {author} {\bibfnamefont {F.}~\bibnamefont
  {{Foucart}}},\ }\href {\doibase 10.1007/s41115-023-00016-y} {\bibfield
  {journal} {\bibinfo  {journal} {Living Reviews in Computational
  Astrophysics}\ }\textbf {\bibinfo {volume} {9}},\ \bibinfo {eid} {1}
  (\bibinfo {year} {2023})},\ \Eprint {http://arxiv.org/abs/2209.02538}
  {arXiv:2209.02538 [astro-ph.HE]} \BibitemShut {NoStop}%
\bibitem [{\citenamefont {Oertel}\ \emph {et~al.}(2012)\citenamefont {Oertel},
  \citenamefont {Fantina},\ and\ \citenamefont {Novak}}]{Oertel:2012qd}%
  \BibitemOpen
  \bibfield  {author} {\bibinfo {author} {\bibfnamefont {M.}~\bibnamefont
  {Oertel}}, \bibinfo {author} {\bibfnamefont {A.~F.}\ \bibnamefont {Fantina}},
  \ and\ \bibinfo {author} {\bibfnamefont {J.}~\bibnamefont {Novak}},\ }\href
  {\doibase 10.1103/PhysRevC.85.055806} {\bibfield  {journal} {\bibinfo
  {journal} {Phys. Rev. C}\ }\textbf {\bibinfo {volume} {85}},\ \bibinfo
  {pages} {055806} (\bibinfo {year} {2012})},\ \Eprint
  {http://arxiv.org/abs/1202.2679} {arXiv:1202.2679 [nucl-th]} \BibitemShut
  {NoStop}%
\bibitem [{\citenamefont {Oertel}\ \emph {et~al.}(2016)\citenamefont {Oertel},
  \citenamefont {Gulminelli}, \citenamefont {Provid{\^e}ncia},\ and\
  \citenamefont {Raduta}}]{Oertel:2016xsn}%
  \BibitemOpen
  \bibfield  {author} {\bibinfo {author} {\bibfnamefont {M.}~\bibnamefont
  {Oertel}}, \bibinfo {author} {\bibfnamefont {F.}~\bibnamefont {Gulminelli}},
  \bibinfo {author} {\bibfnamefont {C.}~\bibnamefont {Provid{\^e}ncia}}, \ and\
  \bibinfo {author} {\bibfnamefont {A.~R.}\ \bibnamefont {Raduta}},\ }\href
  {\doibase 10.1140/epja/i2016-16050-1} {\bibfield  {journal} {\bibinfo
  {journal} {Eur. Phys. J. A}\ }\textbf {\bibinfo {volume} {52}},\ \bibinfo
  {pages} {50} (\bibinfo {year} {2016})},\ \Eprint
  {http://arxiv.org/abs/1601.00435} {arXiv:1601.00435 [nucl-th]} \BibitemShut
  {NoStop}%
\bibitem [{\citenamefont {{Stergioulas}}(2003)}]{Stergioulas2003}%
  \BibitemOpen
  \bibfield  {author} {\bibinfo {author} {\bibfnamefont {N.}~\bibnamefont
  {{Stergioulas}}},\ }\href {\doibase 10.12942/lrr-2003-3} {\bibfield
  {journal} {\bibinfo  {journal} {Living Reviews in Relativity}\ }\textbf
  {\bibinfo {volume} {6}},\ \bibinfo {eid} {3} (\bibinfo {year} {2003})},\
  \Eprint {http://arxiv.org/abs/gr-qc/0302034} {arXiv:gr-qc/0302034 [gr-qc]}
  \BibitemShut {NoStop}%
\bibitem [{\citenamefont {{Goussard}}\ \emph {et~al.}(1997)\citenamefont
  {{Goussard}}, \citenamefont {{Haensel}},\ and\ \citenamefont
  {{Zdunik}}}]{Goussard1997}%
  \BibitemOpen
  \bibfield  {author} {\bibinfo {author} {\bibfnamefont {J.~O.}\ \bibnamefont
  {{Goussard}}}, \bibinfo {author} {\bibfnamefont {P.}~\bibnamefont
  {{Haensel}}}, \ and\ \bibinfo {author} {\bibfnamefont {J.~L.}\ \bibnamefont
  {{Zdunik}}},\ }\href {\doibase 10.48550/arXiv.astro-ph/9610265} {\bibfield
  {journal} {\bibinfo  {journal} {\aap}\ }\textbf {\bibinfo {volume} {321}},\
  \bibinfo {pages} {822} (\bibinfo {year} {1997})},\ \Eprint
  {http://arxiv.org/abs/astro-ph/9610265} {arXiv:astro-ph/9610265 [astro-ph]}
  \BibitemShut {NoStop}%
\bibitem [{\citenamefont {Cook}\ \emph {et~al.}(1994)\citenamefont {Cook},
  \citenamefont {Shapiro},\ and\ \citenamefont {Teukolsky}}]{Cook:1994}%
  \BibitemOpen
  \bibfield  {author} {\bibinfo {author} {\bibfnamefont {G.~B.}\ \bibnamefont
  {Cook}}, \bibinfo {author} {\bibfnamefont {S.~L.}\ \bibnamefont {Shapiro}}, \
  and\ \bibinfo {author} {\bibfnamefont {S.~A.}\ \bibnamefont {Teukolsky}},\
  }\href@noop {} {\bibfield  {journal} {\bibinfo  {journal} {Astrophys. J.}\
  }\textbf {\bibinfo {volume} {422}},\ \bibinfo {pages} {227} (\bibinfo {year}
  {1994})}\BibitemShut {NoStop}%
\bibitem [{\citenamefont {{Haensel}}\ \emph {et~al.}(2009)\citenamefont
  {{Haensel}}, \citenamefont {{Zdunik}}, \citenamefont {{Bejger}},\ and\
  \citenamefont {{Lattimer}}}]{Haensel2009}%
  \BibitemOpen
  \bibfield  {author} {\bibinfo {author} {\bibfnamefont {P.}~\bibnamefont
  {{Haensel}}}, \bibinfo {author} {\bibfnamefont {J.~L.}\ \bibnamefont
  {{Zdunik}}}, \bibinfo {author} {\bibfnamefont {M.}~\bibnamefont {{Bejger}}},
  \ and\ \bibinfo {author} {\bibfnamefont {J.~M.}\ \bibnamefont {{Lattimer}}},\
  }\href {\doibase 10.1051/0004-6361/200811605} {\bibfield  {journal} {\bibinfo
   {journal} {\aap}\ }\textbf {\bibinfo {volume} {502}},\ \bibinfo {pages}
  {605} (\bibinfo {year} {2009})},\ \Eprint {http://arxiv.org/abs/0901.1268}
  {arXiv:0901.1268 [astro-ph.SR]} \BibitemShut {NoStop}%
\bibitem [{\citenamefont {Chandrasekhar}(1969)}]{Chandrasekhar1969}%
  \BibitemOpen
  \bibfield  {author} {\bibinfo {author} {\bibfnamefont {S.}~\bibnamefont
  {Chandrasekhar}},\ }\href@noop {} {\emph {\bibinfo {title} {Ellipsoidal
  Figures of Equilibrium}}}\ (\bibinfo  {publisher} {Yale University Press},\
  \bibinfo {address} {New Haven and London},\ \bibinfo {year}
  {1969})\BibitemShut {NoStop}%
\bibitem [{\citenamefont {{Shibata}}\ \emph {et~al.}(2000)\citenamefont
  {{Shibata}}, \citenamefont {{Baumgarte}},\ and\ \citenamefont
  {{Shapiro}}}]{Shibata2000ApJ}%
  \BibitemOpen
  \bibfield  {author} {\bibinfo {author} {\bibfnamefont {M.}~\bibnamefont
  {{Shibata}}}, \bibinfo {author} {\bibfnamefont {T.~W.}\ \bibnamefont
  {{Baumgarte}}}, \ and\ \bibinfo {author} {\bibfnamefont {S.~L.}\ \bibnamefont
  {{Shapiro}}},\ }\href {\doibase 10.1086/309525} {\bibfield  {journal}
  {\bibinfo  {journal} {\apj}\ }\textbf {\bibinfo {volume} {542}},\ \bibinfo
  {pages} {453} (\bibinfo {year} {2000})},\ \Eprint
  {http://arxiv.org/abs/astro-ph/0005378} {arXiv:astro-ph/0005378 [astro-ph]}
  \BibitemShut {NoStop}%
\bibitem [{\citenamefont {Baiotti}\ \emph {et~al.}(2007)\citenamefont
  {Baiotti}, \citenamefont {Pietri}, \citenamefont {Manca},\ and\ \citenamefont
  {Rezzolla}}]{BaiottiPhysRevD}%
  \BibitemOpen
  \bibfield  {author} {\bibinfo {author} {\bibfnamefont {L.}~\bibnamefont
  {Baiotti}}, \bibinfo {author} {\bibfnamefont {R.~D.}\ \bibnamefont {Pietri}},
  \bibinfo {author} {\bibfnamefont {G.~M.}\ \bibnamefont {Manca}}, \ and\
  \bibinfo {author} {\bibfnamefont {L.}~\bibnamefont {Rezzolla}},\ }\href
  {\doibase 10.1103/PhysRevD.75.044023} {\bibfield  {journal} {\bibinfo
  {journal} {Phys. Rev. D}\ }\textbf {\bibinfo {volume} {75}},\ \bibinfo
  {pages} {044023} (\bibinfo {year} {2007})}\BibitemShut {NoStop}%
\bibitem [{\citenamefont {{Gondek-Rosi{\'n}ska}}\ and\ \citenamefont
  {{Gourgoulhon}}(2002)}]{Gondek2002}%
  \BibitemOpen
  \bibfield  {author} {\bibinfo {author} {\bibfnamefont {D.}~\bibnamefont
  {{Gondek-Rosi{\'n}ska}}}\ and\ \bibinfo {author} {\bibfnamefont
  {E.}~\bibnamefont {{Gourgoulhon}}},\ }\href {\doibase
  10.1103/PhysRevD.66.044021} {\bibfield  {journal} {\bibinfo  {journal}
  {\prd}\ }\textbf {\bibinfo {volume} {66}},\ \bibinfo {eid} {044021} (\bibinfo
  {year} {2002})},\ \Eprint {http://arxiv.org/abs/gr-qc/0205102}
  {arXiv:gr-qc/0205102 [gr-qc]} \BibitemShut {NoStop}%
\bibitem [{\citenamefont {Shibata}\ and\ \citenamefont
  {Karino}(2004)}]{Shibata2004}%
  \BibitemOpen
  \bibfield  {author} {\bibinfo {author} {\bibfnamefont {M.}~\bibnamefont
  {Shibata}}\ and\ \bibinfo {author} {\bibfnamefont {S.}~\bibnamefont
  {Karino}},\ }\href {\doibase 10.1103/PhysRevD.70.084022} {\bibfield
  {journal} {\bibinfo  {journal} {Phys. Rev. D}\ }\textbf {\bibinfo {volume}
  {70}},\ \bibinfo {pages} {084022} (\bibinfo {year} {2004})}\BibitemShut
  {NoStop}%
\bibitem [{\citenamefont {{Saijo}}\ and\ \citenamefont
  {{Gourgoulhon}}(2006)}]{Saijo2006}%
  \BibitemOpen
  \bibfield  {author} {\bibinfo {author} {\bibfnamefont {M.}~\bibnamefont
  {{Saijo}}}\ and\ \bibinfo {author} {\bibfnamefont {E.}~\bibnamefont
  {{Gourgoulhon}}},\ }\href {\doibase 10.1103/PhysRevD.74.084006} {\bibfield
  {journal} {\bibinfo  {journal} {\prd}\ }\textbf {\bibinfo {volume} {74}},\
  \bibinfo {eid} {084006} (\bibinfo {year} {2006})},\ \Eprint
  {http://arxiv.org/abs/astro-ph/0606569} {arXiv:astro-ph/0606569 [astro-ph]}
  \BibitemShut {NoStop}%
\bibitem [{\citenamefont {{Lai}}(2001)}]{Lai2001}%
  \BibitemOpen
  \bibfield  {author} {\bibinfo {author} {\bibfnamefont {D.}~\bibnamefont
  {{Lai}}},\ }in\ \href {\doibase 10.1063/1.1387316} {\emph {\bibinfo
  {booktitle} {Astrophysical Sources for Ground-Based Gravitational Wave
  Detectors}}},\ \bibinfo {series} {American Institute of Physics Conference
  Series}, Vol.\ \bibinfo {volume} {575},\ \bibinfo {editor} {edited by\
  \bibinfo {editor} {\bibfnamefont {J.~M.}\ \bibnamefont {{Centrella}}}}\
  (\bibinfo  {publisher} {AIP},\ \bibinfo {year} {2001})\ pp.\ \bibinfo {pages}
  {246--257},\ \Eprint {http://arxiv.org/abs/astro-ph/0101042}
  {arXiv:astro-ph/0101042 [astro-ph]} \BibitemShut {NoStop}%
\bibitem [{\citenamefont {Camelio}\ \emph {et~al.}(2021)\citenamefont
  {Camelio}, \citenamefont {Dietrich}, \citenamefont {Rosswog},\ and\
  \citenamefont {Haskell}}]{Camelio2021}%
  \BibitemOpen
  \bibfield  {author} {\bibinfo {author} {\bibfnamefont {G.}~\bibnamefont
  {Camelio}}, \bibinfo {author} {\bibfnamefont {T.}~\bibnamefont {Dietrich}},
  \bibinfo {author} {\bibfnamefont {S.}~\bibnamefont {Rosswog}}, \ and\
  \bibinfo {author} {\bibfnamefont {B.}~\bibnamefont {Haskell}},\ }\href
  {\doibase 10.1103/PhysRevD.103.063014} {\bibfield  {journal} {\bibinfo
  {journal} {Phys. Rev. D}\ }\textbf {\bibinfo {volume} {103}},\ \bibinfo
  {pages} {063014} (\bibinfo {year} {2021})}\BibitemShut {NoStop}%
\bibitem [{\citenamefont {Lattimer}\ and\ \citenamefont
  {Schutz}(2005)}]{Lattimer:2005}%
  \BibitemOpen
  \bibfield  {author} {\bibinfo {author} {\bibfnamefont {J.~M.}\ \bibnamefont
  {Lattimer}}\ and\ \bibinfo {author} {\bibfnamefont {B.~F.}\ \bibnamefont
  {Schutz}},\ }\href {\doibase 10.1086/431543} {\bibfield  {journal} {\bibinfo
  {journal} {Astrophys. J.}\ }\textbf {\bibinfo {volume} {629}},\ \bibinfo
  {pages} {979} (\bibinfo {year} {2005})},\ \Eprint
  {http://arxiv.org/abs/astro-ph/0411470} {arXiv:astro-ph/0411470} \BibitemShut
  {NoStop}%
\bibitem [{\citenamefont {Doneva}\ \emph
  {et~al.}(2013{\natexlab{b}})\citenamefont {Doneva}, \citenamefont
  {Yazadjiev}, \citenamefont {Stergioulas},\ and\ \citenamefont
  {Kokkotas}}]{Doneva:2013rha}%
  \BibitemOpen
  \bibfield  {author} {\bibinfo {author} {\bibfnamefont {D.~D.}\ \bibnamefont
  {Doneva}}, \bibinfo {author} {\bibfnamefont {S.~S.}\ \bibnamefont
  {Yazadjiev}}, \bibinfo {author} {\bibfnamefont {N.}~\bibnamefont
  {Stergioulas}}, \ and\ \bibinfo {author} {\bibfnamefont {K.~D.}\ \bibnamefont
  {Kokkotas}},\ }\href {\doibase 10.1088/2041-8205/781/1/L6} {\bibfield
  {journal} {\bibinfo  {journal} {Astrophys. J.}\ }\textbf {\bibinfo {volume}
  {781}},\ \bibinfo {pages} {L6} (\bibinfo {year}
  {2013}{\natexlab{b}})}\BibitemShut {NoStop}%
\bibitem [{\citenamefont {Lattimer}\ and\ \citenamefont
  {Prakash}(2004)}]{Lattimer:2004}%
  \BibitemOpen
  \bibfield  {author} {\bibinfo {author} {\bibfnamefont {J.~M.}\ \bibnamefont
  {Lattimer}}\ and\ \bibinfo {author} {\bibfnamefont {M.}~\bibnamefont
  {Prakash}},\ }\href {\doibase 10.1126/science.1090720} {\bibfield  {journal}
  {\bibinfo  {journal} {Science}\ }\textbf {\bibinfo {volume} {304}},\ \bibinfo
  {pages} {536} (\bibinfo {year} {2004})},\ \Eprint
  {http://arxiv.org/abs/astro-ph/0405262} {arXiv:astro-ph/0405262} \BibitemShut
  {NoStop}%
\end{thebibliography}

%

\end{document}